\documentclass[aps,prd,reprint,superscriptaddress,nofootinbib]{revtex4-2}
\usepackage{amsmath,amsfonts,amssymb,bm}
\usepackage{graphicx}
\usepackage{bbold}
\usepackage[normalem]{ulem}
\usepackage[dvipsnames,usenames]{xcolor}
\usepackage[colorlinks,citecolor=blue]{hyperref}
\usepackage{orcidlink}

\begin{document}

\title{Fast neutrino flavor conversions in a supernova: Emergence, evolution, and effects}

\author{Zewei Xiong\orcidlink{0000-0002-2385-6771}}
\email[Email: ]{z.xiong@gsi.de}
\affiliation{GSI Helmholtzzentrum {f\"ur} Schwerionenforschung, Planckstra{\ss}e 1, 64291 Darmstadt, Germany}

\author{Meng-Ru Wu\orcidlink{0000-0003-4960-8706}}
\affiliation{Institute of Physics, Academia Sinica, Taipei 11529, Taiwan}
\affiliation{Institute of Astronomy and Astrophysics, Academia Sinica, Taipei 10617, Taiwan}
\affiliation{Physics Division, National Center for Theoretical Sciences, Taipei 10617, Taiwan}

\author{Manu George\orcidlink{0000-0002-8974-5986}}
\affiliation{Institute of Physics, Academia Sinica, Taipei 11529, Taiwan}

\author{Chun-Yu Lin\orcidlink{0000-0002-7489-7418}}
\affiliation{National Center for High-performance Computing, Hsinchu 30076, Taiwan}

\author{Noshad Khosravi Largani\orcidlink{0000-0003-1551-0508}}
\affiliation{Institute of Theoretical Physics, University of Wroclaw, plac Maksa Borna 9, 50-204 Wroclaw, Poland}

\author{Tobias Fischer\orcidlink{0000-0003-2479-344X}}
\affiliation{Institute of Theoretical Physics, University of Wroclaw, plac Maksa Borna 9, 50-204 Wroclaw, Poland}

\author{Gabriel Mart{\'i}nez-Pinedo\orcidlink{0000-0002-3825-0131}}
\affiliation{GSI Helmholtzzentrum {f\"ur} Schwerionenforschung, Planckstra{\ss}e 1, 64291 Darmstadt, Germany} \affiliation{Institut f{\"u}r Kernphysik (Theoriezentrum), Fachbereich Physik, Technische Universit{\"a}t Darmstadt, Schlossgartenstra{\ss}e 2, 64289 Darmstadt, Germany}

\date{\today}

\begin{abstract}
Fast flavor conversions (FFCs) of neutrinos, which can occur in core-collapse supernovae (CCSNe), are multiangle effects.
They depend on the angular distribution of the neutrino's electron lepton number (ELN).
In this work, we present a comprehensive study of the FFCs by solving the multienergy and multiangle quantum kinetic equations with an extended set of collisional weak processes based on a static and spherically symmetric CCSN matter background profile.
We investigate the emergence and evolution of FFCs in models featuring different ELN angular distributions, considering scenarios with two and three neutrino flavors.
The spectrogram method is utilized to illustrate the small-scale spatial structure, and we show that this structure of neutrino flavor coherence and number densities in the nonlinear regime is qualitatively consistent with the dispersion relation analysis.
On the coarse-grained level, we find that different asymptotic states can be achieved following the FFCs depending on the locations and shapes of the ELN distributions, despite sharing a common feature of the elimination of the ELN angular crossing.
While equilibration among different neutrino flavors may be achieved immediately after the prompt FFCs, it is not a general outcome of the asymptotic state, as subsequent feedback effects from collisional neutrino-matter interactions come into play, particularly for cases where FFCs occur inside the neutrinosphere.
The impacts of FFCs and the feedback effect on the net neutrino heating rates, the equilibrium electron fraction of CCSN matter, and the free-streaming neutrino energy spectra are quantitatively assessed.
Other aspects including the impact of the vacuum term and the coexistence with other type of flavor instabilities are also discussed.
\end{abstract}

\maketitle
\graphicspath{{./}{figures/}}

\section{Introduction}
\label{sec:introduction}
Core-collapse supernovae (CCSNe) are cataclysmic events when massive stars reach the end of their lives.
During the evolution of a CCSN, a nascent protoneutron star (PNS) forms at the center as a prolific source of neutrinos.
These neutrinos play pivotal roles on the CCSN dynamics and the evolution of chemical composition.
They interact with the medium through both charged- and neutral-current weak interactions in the proximity of the PNS and deposit energy, facilitating the shock revival via reheating of material in the postshocked layer and leading to the eventual mass ejection.
Particularly, the charged-current interactions determine the proton-to-baryon ratio, denoted by the electron fraction $Y_e$, which is a crucial quantity for the nucleosynthesis results of CCSN explosions.
A better knowledge of the flux intensities and flavor content of neutrinos is necessary to robustly model the inner dynamics of CCSNe, to predict the elemental compositions in CCSN ejecta, and to determine the neutrino signals for the detection of the next Galactic supernova event (see e.g.,~Refs.~\cite{mirizzi2016supernova,Fischer:2023ebq} for recent reviews).

The flavor oscillations of neutrinos among $\nu_e$, $\nu_\mu$, and $\nu_\tau$ in vacuum and in medium have been well studied and confirmed by various ground-based neutrino experiments~\cite{pdg}.
When neutrino fluxes are sufficiently high in CCSNe, the forward scattering among neutrinos themselves leads to various collective phenomena of flavor instability.
Particularly, the fast flavor conversion (FFC) associated with the fast flavor instability (FFI; see e.g., Refs.~\cite{tamborra2021new,richers2023fast,capozzi2022neutrino,volpe2023neutrinos} for reviews) has attracted great interest in recent years owing to the vastly rapid conversion rate within nanoseconds and over a distance shorter than a coin.
Studies based on results from the multidimensional CCSN simulations have shown a general occurrence of FFIs in certain regions ahead of the shock wave, near the neutrinosphere, or even deep inside the PNS \cite{abbar2019occurrence,azari2019linear,azari2020fast,nagakura2019fast,morinaga2020fast,abbar2020fast,glas2020fast,nagakura2021occurrence,abbar2021characteristics,harada2022prospects}.

Since the first proposal of FFI in Ref.~\cite{sawyer2009multiangle}, the advance of both theory and methodology throughout the past decade has improved our understanding of FFCs and guided the line of research toward the ultimate goal of implementing neutrino oscillations in CCSNe.
A generic framework governing the coherent flavor evolution and collisional neutrino-matter interactions is prescribed by the neutrino quantum kinetic equation ($\nu$QKE) \cite{sigl1993general,vlasenko2014neutrino,volpe2015neutrino,blaschke2016neutrino}.
The linear stability analysis (LSA) of the $\nu$QKE provides a powerful tool to diagnose the existence of flavor instabilities \cite{banerjee2011linearized,izaguirre2017fast,capozzi2017fast,yi2019dispersion}.
It has been proved based on the LSA that the FFI requires the presence of angular zero crossings where the angular distribution of the neutrino lepton number between any two distinct flavors transitions from positive to negative values \cite{morinaga2022fast}.
The dispersion relation based on LSA implies the break of spatial homogeneity, which results in small-scale spatial structure confirmed by several dynamical simulations that numerically solved $\nu$QKE within tiny boxes~\cite{martin2020dynamic,bhattacharyya2021fast,bhattacharyya2020late,wu2021collective,richers2021particle,richers2021neutrino,zaizen2021nonlinear,abbar2022suppression,richers2022code,bhattacharyya2022elaborating,grohs2023neutrino,zaizen2023simple,xiong2023evaluating} and may affected by adopting a different boundary condition \cite{zaizen2023characterizing,cornelius2023perturbing,nagakura2023bgk}.
Quasisteady states are found to be achieved asymptotically on the coarse-grained level after neutrinos undergo the kinematic decoherence \cite{raffelt2007self,abbar2019fast,johns2020fast,bhattacharyya2021fast,xiong2023symmetry}.
It was reported by these simulations that the FFCs lead to complete flavor equilibration or partial flavor conversion depending on the ratio of neutrino fluxes and the characteristics of the angular zero crossings.
Particularly when a complete flavor equilibration or even overconversion is achieved, the neutrino flavor content can be significantly affected, which was demonstrated to be able to exert substantial impact on the neutrino-driven explosion mechanism and the nucleosynthesis of CCSNe \cite{xiong2020potential,fujimoto2023explosive,nagakura2023roles,ehring2023fast,ehring2023fast2,balantekin2023collective}, or even in the context of neutron-star mergers \cite{wu2017imprints,george2020fast,li2021neutrino,just2022fast,fernandez2023fast,nagakura2023global,froustey2024neutrino,grohs2024two}.

Recently, a growing amount of studies solving $\nu$QKE in a large spatial domain ranging from the interior of the PNS to the region around the shock wave were performed~\cite{stapleford2020coupling,shalgar2023supernova,nagakura2022time,shalgar2023neutrino,nagakura2023connecting,nagakura2023basic}.
These studies intended to capture the flavor instabilities and to evolve both collective flavor conversions as well as the neutrino transport simultaneously in a self-consistent manner.
Despite all these triumphs in the progress of FFCs, there remain several unaddressed questions.
(1)~Does the small-scale structure observed in the local simulations survive when the advection and neutrino-matter interactions are considered?
(2)~What is the general outcome of FFCs as well as its impact on the neutrino flux intensities and their ratios?
Is flavor equilibration the ultimate fate of FFCs?
(3)~What consequence could be anticipated on CCSNe and the associated neutrino signals?

In addition to the FFI, there are other flavor instabilities such as the slow (see Refs.~\cite{duan2010collective,mirizzi2016supernova} for reviews) or collisional type (see Refs.~\cite{johns2023collisional,padilla2022neutrino2,johns2022collisional,xiong2023evolution,lin2023collision,xiong2023collisional,kato2023flavor,liu2023systematic,akaho2023collisional,liu2023universality,kato2023collisional,fiorillo2023collisions,shalgar2023neutrinos}) that might coexist with the FFI.
Particularly, the collisional flavor instability (CFI), which was discovered recently, can potentially trigger flavor conversions in regions close to or inside the neutrinosphere where the FFI is known to play a role.
Certainly, there are still open issues regarding these topics.
It is important to understand whether and how different flavor instabilities interplay with each other in more sophisticated supernova models.

In this work, we aim to provide answers to the above questions (1)--(3) and to investigate the coexistence as well as impact of other flavor instabilities.
We extend our previous multienergy and multiangle neutrino flavor evolution code~\cite{george2023cosenu,xiong2023evolution} to solve the $\nu$QKE by including a more complete set of collisional weak interactions in static snapshots of background environments obtained with spherically symmetric CCSN hydrosimulations.
Notice that the multienergy spherically symmetric neutrino transport in standard CCSNe typically finds no angular crossing \cite{tamborra2017flavor,nagakura2020systematic,liu2023universality} because the imposed spherical symmetry restricts the capability of capturing the multdimensional CCSN features such as the fluid convection \cite{nagakura2020systematic} and lepton-emission self-sustained asymmetry \cite{tamborra2014self,glas2019effects}.
These multdimensional features can lead to attenuated $Y_e$ compared to the spherically symmetric model and create conditions for FFIs.
Therefore, we follow the spirit of Ref.~\cite{nagakura2023roles} and introduce radial-dependent $Y_e$ attenuation schemes in order to mimic the suitable conditions for FFCs commonly observed in multdimensional simulations.
We use schemes with different attenuation strengths to probe a variety of scenarios for the FFI close to and inside the neutrinosphere.

We adopt a similar two-step approach as in Ref.~\cite{xiong2023evolution}.
We first relax neutrino profiles into stationary states up to $\simeq 1$~ms without including coherent flavor oscillations.
Those stationary states are then used as the initial condition to simulate the evolution of neutrinos including diffusion, collisions, and flavor oscillations.
We then present a comprehensive analysis for the FFCs covering three main aspects: the small-scale structure after the FFI emerges from the linear to nonlinear regimes, the dynamical evolution of the neutrino flavor content, and the effects on CCSN physics as well as the neutrino signals.

This paper is organized as follows.
We provide the descriptions and parameters of our models for the FFCs in Sec.~\ref{sec:model}.
We use the LSA and the spectrogram method introduced in Sec.~\ref{sec:analyses} to characterize the small-scale features of FFI in Sec.~\ref{sec:emergence}.
We present a detailed study on the evolution of FFCs in Sec.~\ref{sec:evolution} and discuss their implications in the energy transport, chemical composition of the matter, and free-streaming neutrino spectra in Sec.~\ref{sec:effects}.
Further discussions and conclusions are given in Sec.~\ref{sec:discussion}.
We adopt natural units with $\hbar=c=k_B=1$ throughout the paper.

\section{Models}
\label{sec:model}
In this section we introduce our theoretical model for neutrino flavor conversions in CCSNe.
We begin with a review about the supernova model and the simulations based on which the flavor study is conducted.
We then revisit the neutrino quantum kinetic approach and introduce our computational implementation.

\subsection{Supernova model}
The CCSN simulation considered in this work is launched from the stellar progenitor star with zero-age main sequence mass of 25 $M_\odot$ and solar metallicity, from the stellar evolution series of Ref.~\cite{rauscher2002nucleosynthesis}, denoted as \texttt{s25a28}.
Our \textsc{agile-boltztran} supernova model is based on general relativistic spherical symmetric neutrino radiation hydrodynamics in comoving coordinates \cite{mezzacappa1993type,mezzacappa1993stellar,mezzacappa1993numerical,liebendorfer2001conservative,liebendorfer2004finite}, including six-species Boltzmann neutrino transport assuming ultrarelativistic particles \cite{lindquist1966relativistic,fischer2020muonization}.
The equations of radiation hydrodynamics are solved on an adaptive baryon mass mesh \cite{liebendorfer2002adaptive,fischer2010protoneutron}, for which 102 radial grid points are used for the current simulations, including about 5~$M_\odot$ of the stellar core, which covers parts of the extended silicon-sulfur layer for this progenitor model.
The neutrino distributions are solved by means of the Boltzmann neutrino transport equation in the discrete-ordinate method.
They are discretized in terms of 6 propagation angles and 36 energy bins (from 0.5 to 300~MeV) following the setup of Refs.~\cite{mezzacappa1993numerical,bruenn1985stellar}.

For the collision integral of the Boltzmann equation, a complete set of weak processes is employed (see Table~I in Ref.~\cite{fischer2020neutrino}).
This includes the charged-current neutrino emission and absorption (EA) involving heavy nuclei~\cite{juodagalvis2010improved} and unbound nucleons~\cite{fischer2020neutrino}, isoenergetic neutrino scattering on nuclei and nucleons (IS)~\cite{bruenn1985stellar,mezzacappa1993numerical}, inelastic neutrino scattering on electrons and positrons (NES)~\cite{schinder1982neutrino,mezzacappa1993stellar}, and neutrino pair reactions (PRs), including electron-positron annihilation~\cite{bruenn1985stellar}, nucleon-nucleon bremsstrahlung~\cite{thompson2001neutrino,fischer2016role} and the annihilation of electron neutrino pairs~\cite{buras2003electron,fischer2009neutrino}.
EA are implemented in the full kinematics approach at the mean-field level, taking weak-magnetism contributions self-consistently into account~\cite{guo2020charged}.
Note that muonic weak processes are omitted here, such that muon and tauon flavors are not distinguished for neutrinos and antineutrinos.
For the equation of state (EOS) used in the present supernova simulation, see details in Appendix~\ref{sec:sn_eos}.

\begin{figure}[t]
\includegraphics[width=0.48\textwidth]{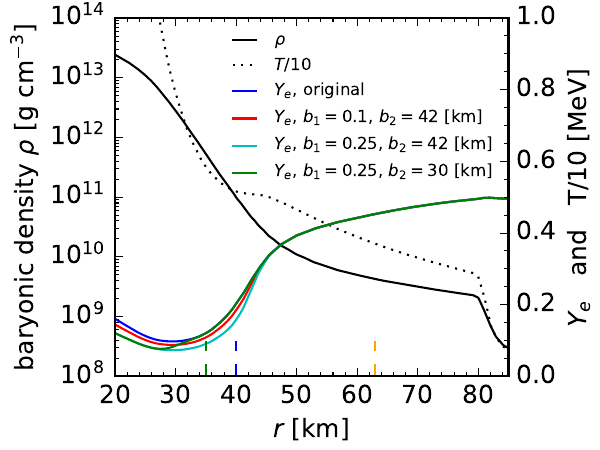}
\caption{\label{fig:ye_attenuation} Radial profiles of baryonic density, temperature $T/10$, and $Y_e$ (before and after attenuations) for a snapshot at the postbounce time $t_{\rm pb}\approx 252$~ms. Orange, blue, and green vertical dashed lines indicate the gain radius, the radii of $\nu_e$- and $\bar\nu_e$-spheres at $r\approx 63$, 40, and 35~km, respectively.
}
\end{figure}

During the early postbounce evolution, on the order of 10--25~ms after the core bounce, the $\nu_e$ deleptonization burst is released.
It is a standard feature of CCSN phenomenology that results in the rapid drop of the electron fraction $Y_e$, accordingly, from around $Y_e=0.3$--$0.4$ at core bounce to $Y_e\simeq0.1$, in the region where the neutrinos decouple from matter.
Even though neutrino decoupling is an energy- and momentum-dependent process (cf. Refs.~\cite{raffelt2002mu,keil2003monte}), often the neutrino-phase-space-averaged neutrinospheres of last elastic and inelastic scatterings are used as references~\cite{fischer2012neutrino}, inside of which the bulk part of the neutrino spectrum is trapped and outside of which it is freely streaming.
The postbounce evolution of this \texttt{s25a28} progenitor model corresponds in spherically symmetric supernova simulations to the failed branch, leading eventually to the PNS collapse and the formation of a black hole~\cite{sumiyoshi2006neutrino,fischer2009neutrino,oconnor2011black}.
Before that happens, the postbounce phase is determined by a steady state mass accretion from the gravitationally unstable layers above the stellar core.
The infalling material falls onto the standing bounce shock.
The latter is located at a radius of about 80--100~km, depending on the progenitor model and the nuclear EOS.
Thereby, the infalling heavy nuclei are dissociated into bulk nuclear matter, composed of unbound nucleons and light nuclei, that accumulate at the PNS surface.
This layer, illustrated in Fig.~\ref{fig:ye_attenuation} via the shock at $r\simeq 85$~km and the PNS surface at $r\simeq 40$~km depicting a shallow rise in density (solid curve) and temperature (dotted curve), is subject to the neutrino decoupling, spanning a range of temperatures from up to $T\simeq 10$~MeV, at densities of a few times $10^{13}$~g~cm$^{-3}$, down to $T\simeq 3$~MeV which corresponds to the shock location with densities around $10^{9}$~g~cm$^{-3}$ where the temperature increases sharply due to the shock heating from $T\simeq 0.5$ to $T\simeq 3$~MeV.

The timescale for the compression of the accumulated layer, settling at the PNS surface into the steepening gravitational potential, is determined by the high-density nuclear EOS, its temperature and isospin asymmetry dependence, and the enclosed PNS mass.
As a consequence, neutrino luminosities $L_\nu$ are obtained on the order of several $10^{52}$~erg~s$^{-1}$, after the $\nu_e$ deleptonization burst has been launched, and the average neutrino energies have values of $\langle E_\nu \rangle\simeq10$--20~MeV.
For a recent review of the role of neutrinos in the postbounce phase of core-collapse supernovae, see Refs.~\cite{Fischer2017PASA34,Fischer:2023ebq} and references therein.

Instead of surveying over different time stages of supernova, we will focus on one snapshot at a postbounce time $t_{\rm pb}\approx 252$~ms in Fig.~\ref{fig:ye_attenuation}.
Based on this snapshot, we explore different scenarios for FFCs by applying a radial-dependent attenuation to the original $Y_e$ profile to mimic suitable conditions for FFCs observed in multdimensional supernova simulations.
A radial-dependent factor,
\begin{equation}
    b_{Y_e}(r) = 1-\frac{b_1}{1+e^{r-b_2 [\mathrm{km}]}},
\end{equation}
is multiplied to the original $Y_e$ profile with $b_1$ denoting the fraction of attenuation and $b_2$ characterizing the radial range of attenuation.
With the attenuated electron fraction $Y_e$ and smaller electron chemical potential $\mu_{e^-}$, the equilibrium neutrino chemical potential $\mu_{\nu_e}^{\rm eq}=\mu_{e^-}+\mu_p-\mu_n$ is reduced to create a condition less dominated by $\nu_e$'s, or even a condition where $\bar\nu_e$'s dominate over $\nu_e$'s.
We vary $b_1$ from 10\% to 25\%, allowing different angular crossings dominated by $\nu_e$ or $\bar\nu_e$ to appear, which will be discussed later in Sec~\ref{sec:illus_quant}.
We show in Fig.~\ref{fig:ye_attenuation} three attenuated $Y_e$ profiles with different values of $(b_1, b_2[{\rm km}])=(0.1, 42), (0.25, 42)$, and $(0.25, 30)$, respectively.
Note that taking $b_2=30$~km restricts the FFI solely within the PNS, but not extending to the free-streaming regime.
Also noted is that we only attenuate $Y_e$ without modifying the neutron and proton abundances for simplification.
When the latter abundances are modified accordingly based on the EOS, $\mu_p$ decreases, and $\mu_n$ increases.
As a result, the attenuation factor, $b_1$, required to achieve a similar scenario for each model is reduced by approximately half.
In addition, although the $Y_e$ profile with the largest attenuation factor adopted here might be hard to be obtained in typical supernovae, it may mimic other physical conditions such as the second collapse in a hadron-quark phase transition supernova featuring a unique burst-like neutrino signature dominated by $\bar\nu_e$~\cite{Fischer2018NatAs2,largani2024constraining,Jakobus2022MNRAS516,Kuroda2022ApJ924,Zha2022PhRvD106}.

\subsection{Neutrino quantum kinetic equation}
We solve the neutrino quantum kinetic equation ($\nu$QKE) in both two- and three-flavor schemes.
The general equations governing the spatial and temporal evolution of density matrices $\varrho$ (for neutrinos) and $\bar\varrho$ (for anti-neutrinos) are given by
\begin{align}
    ( \partial_t + v_r \partial_r + \frac{1-v_r^2}{r}\partial_{v_r}) \varrho & = -i[\mathbf H, \varrho] +\mathbf C,
    \label{eq:eom_nu}
\end{align}
and
\begin{align}
    ( \partial_t + v_r \partial_r +\frac{1-v_r^2}{r}\partial_{v_r}) \bar\varrho & = -i[\bar{\mathbf H}, \bar\varrho] +\bar{\mathbf C},
    \label{eq:eom_nubar}
\end{align}
with
\begin{equation}
    \varrho = \begin{bmatrix} \varrho_{ee} & \varrho_{e\mu} & \varrho_{e\tau} \\ \varrho_{e\mu}^* & \varrho_{\mu\mu} & \varrho_{\mu\tau} \\ \varrho_{e\tau}^* & \varrho_{\mu\tau}^* & \varrho_{\tau\tau} \end{bmatrix}
    \text{~and~}
    \bar\varrho = \begin{bmatrix} \bar\varrho_{ee} & \bar\varrho_{e\mu} & \bar\varrho_{e\tau} \\ \bar\varrho_{e\mu}^* & \bar\varrho_{\mu\mu} & \bar\varrho_{\mu\tau} \\ \bar\varrho_{e\tau}^* & \bar\varrho_{\mu\tau}^* & \bar\varrho_{\tau\tau} \end{bmatrix}
\end{equation}
in three-flavor bases or
\begin{equation}
    \varrho = \begin{bmatrix} \varrho_{ee} & \varrho_{e\mu} \\ \varrho_{e\mu}^* & \varrho_{\mu\mu} \end{bmatrix} \text{~and~}
    \bar\varrho = \begin{bmatrix} \bar\varrho_{ee} & \bar\varrho_{e\mu} \\ \bar\varrho_{e\mu}^* & \bar\varrho_{\mu\mu} \end{bmatrix},
\end{equation}
in a simplified neutrino system of two flavors.
Their diagonal elements are normalized to the neutrino number density $n_{\nu_i}(r, t) = \int d E\, d v_r\, \varrho_{ii}$ where $i=e,\,\mu,\,\tau$.

In general, the coherent propagation Hamiltonian $\mathbf H$ on the right-hand side has three contributions.
They are the vacuum mixing term
\begin{equation}
    \mathbf H_{\rm vac}(E) = \mathbf U \mathbf M \mathbf U^\dagger,
\end{equation}
with the diagonal matrix in mass basis $\mathbf M = \mathrm{diag}[0, \delta m_{\rm L}^2, \delta m^2]/(2E)$ and the unitary mixing matrix $\mathbf U$ parametrized as in Eq.~(14.34) of Ref.~\cite{pdg}, the matter term
\begin{equation}
    \mathbf H_\mathrm{mat} = \mathrm{diag}[V_{\rm mat}, 0, 0],
\end{equation}
with the effective potential $V_{\rm mat}$ corresponding to neutrino forward scattering on ordinary matter, and the neutrino self-induced term
\begin{equation}
    \mathbf H_{\nu\nu}(v_r) = \sqrt{2} G_F \int d E'\, d v_r' (1-v_r v_r') (\varrho-\bar\varrho^*),
\end{equation}
corresponding to neutrino forward scattering on other neutrinos, respectively.
In the two-flavor scheme, the vacuum term is explicitly
\begin{equation}
    \mathbf H_{\rm vac}(E) =
    \frac{\delta m^2}{4E}
    \begin{bmatrix}
    -\cos 2\theta_V & \sin 2 \theta_V \\ \sin 2 \theta_V & \cos 2 \theta_V
    \end{bmatrix}
\end{equation}
with one mass-square difference $\delta m^2$ and one mixing angle $\theta_V$.
Although the matter term may affect the behaviors of flavor oscillations \cite{sigl2022simulations,xiong2023evolution}, we will not include it explicitly in the following discussion of this work.

The collisional term $\mathbf C$ includes all four types of weak processes in Table~\ref{tab:nu_process}.
We use the same scheme for the EA and IS processes as in Ref.~\cite{xiong2023evolution} except that muonic EA reactions are not considered.

\begin{table}[!ht]
   \centering
   \caption{\label{tab:nu_process} Set of weak processes considered in \textsc{cose$\nu$}, where $\nu$ and $\bar\nu$ are for all neutrino flavors and $N = n, p$.}
   \begin{tabular}{lcc}
   \hline\hline
   Label & Weak process & Abbreviation \\\hline
   (1a) & $\nu_e + n \leftrightarrows p + e^-$ & EA \\
   (1b) & $\bar\nu_e + p \leftrightarrows n + e^+$ & EA \\
   (1c) & $\bar\nu_e + p + e^- \leftrightarrows n$ & EA \\
   (2a) & $\nu + N \leftrightarrows \nu + N $ & IS \\
   (2b) & $\bar\nu + N \leftrightarrows \bar\nu + N $ & IS \\
   (3a) & $\nu + e^{\pm} \leftrightarrows \nu + e^{\pm} $ & NES \\
   (3b) & $\bar\nu + e^{\pm} \leftrightarrows \bar\nu + e^{\pm} $ & NES \\
   (4a) & $\nu + \bar\nu \leftrightarrows e^- + e^+ $ & PR \\
   (4b) & $\nu + \bar\nu + N+N \leftrightarrows N + N $ & PR \\
   \hline\hline
   \end{tabular}
\end{table}

Low-energy electron neutrinos and heavy-lepton neutrinos are thermalized by NES and PRs.
Because of the high computational cost of the scattering kernels associated with them, approximated schemes are employed.
For PRs, we integrate the scattering kernels in Refs.~\cite{bruenn1985stellar,hannestad1998supernova} into the emissivity $j_{\rm PR}$ and opacity $\chi_{\rm PR}$ for all flavors in a similar way as in Ref.~\cite{oconnor2015open} except assuming that the accompanied neutrinos or antineutrinos in the pair processes obey the Fermi-Dirac distribution determined by the equilibrium neutrino chemical potential (e.g., $\mu_{\nu_e}^{\rm eq}$ for $\nu_e$), in order to take the blocking effect into account.
For heavy-lepton neutrinos, the equilibrium neutrino chemical potential is assumed to be zero.
The collisional term for neutrinos can then be constructed
in an effective way similar to the EA processes as follows,
\begin{align}
    \mathbf C_\mathrm{PR}(E) = &
    \frac{1}{2} \left\{ \mathrm{diag}[j_{e,\rm PR}, j_{\mu,\rm PR}, j_{\tau,\rm PR}], \varrho_{\rm FO}-\varrho \right\} \nonumber\\
    & - \frac{1}{2} \left\{ \mathrm{diag}[\chi_{e,\rm PR}, \chi_{\mu,\rm PR}, \chi_{\tau,\rm PR}], \varrho \right\}
\end{align}
with the fully occupied differential number density $\varrho_{\rm FO}$ and the curly bracket being the anticommutator.
The full term of NES in $\nu$QKE involves channels of different Mandelstam variables \cite{blaschke2016neutrino}, which increases the complication of the treatment.
For simplicity, we take the classical treatment particularly for NES by only keeping the diagonal elements of $\mathbf C_{\rm NES}$ \cite{yueh1976scattering,yueh1976neutrino,schinder1982neutrino,mezzacappa1993stellar}.
The thermalization of the energy spectra of heavy-lepton neutrinos is guaranteed by the Kirchhoff-Planck relation (``detailed balance'') between the in- and out-scattering kernels
\begin{equation}
    R_{\rm NES}^{\rm in}(E,E',v_r,v_r') = e^{\frac{E'-E}{T}} R_{\rm NES}^{\rm out}(E,E',v_r,v_r').
    \label{eq:detailed_balance}
\end{equation}
In addition, for each flavor the NES scattering kernel is truncated at the zeroth moment following the method in \cite{schinder1982neutrino,rampp2002radiation} (see Appendix~\ref{sec:nes} for more details).

\subsection{Numerical setup}
The $\nu$QKE is solved by \textsc{cose$\nu$} under spherical symmetry within a radial range uniformly discretized by $N_r$ radial grids between $r_{\rm ib}=20$ and $r_{\rm ob}=80$~km.
At the inner boundary neutrinos of all species are in Fermi distribution functions with chemical potentials given by weak equilibrium with the processes of NES and PRs included.
The outer boundary is chosen to be around the position of the shock wave.
We employ the free-streaming boundary condition there for forward propagating neutrinos with $v_r\geq 0$.
Although neutrinos can be backscattered by heavy nuclei outside the shock through the neutrino-nucleus interactions, which leads to the presence of FFI around the shock \cite{morinaga2020fast}, a recent work suggests that their effect on the dominant neutrino outflow is negligible due to the relatively diluted flux \cite{abbar2022suppression}.
Therefore, we inject no neutrinos in the backward direction with $v_r<0$.

Because the vacuum term $\mathbf{H}_{\rm vac}$ with a nonzero mixing angle generates flavor mixing automatically, we do not employ any artificial perturbation as the seed to trigger flavor conversions in the initial condition.

For consistency across all models, the neutrino spectral and angular distributions are all discretized into $N_E=15$ energy grids and $N_{v_r}=100$ angular grids for both $\varrho$ and $\bar\varrho$.
The energy grids are taken between 1 to 100~MeV spaced uniformly in logarithmic scale, and the angular grids are linearly uniform between $v_r=-1$ and 1.

Because both EA and IS rates that generally dominate over others strongly depend on the neutrino energy and nucleonic density, taking a smaller maximum neutrino energy and a larger radius of the inner boundary compared to our previous work \cite{xiong2023evolution} reduces the maximum collisional rate that can be well resolved within our simulation domain.
Therefore, unlike in \cite{xiong2023evolution}, no attenuation is made to any collisional rate for all models in this work.

In contrast to the collisional rates, a common challenge is the tremendous size of the neutrino self-induced term $\mathbf H_{\nu\nu}$ associated with subnanosecond timescales and subcentimeter length scales.
They are much shorter than the scales related to the hydrodynamics (order of kilometers and milliseconds), which makes it computationally difficult for a simulation to well resolve both time and length scales with the unadjusted $\mathbf H_{\nu\nu}$.
A strategical mitigation in recent works \cite{nagakura2022time,xiong2023evolution} is to attenuate the values of $\mathbf H_{\nu\nu}$ while generally keeping the same ordering that the magnitude of $\mathbf H_{\nu\nu}$ dominates over collisional rates and vacuum oscillation frequency.\footnote{There are also other strategies, for example, assuming that neutrinos are homogeneous in radial blocks of a few hundred meters to reduce the needed number of the radial grids in computation and to allow for no attenuation on $\mathbf H_{\nu\nu}$ \cite{shalgar2023neutrino}.}
The oscillation term $\mathbf H$ on the right-hand side of Eq.~\eqref{eq:eom_nu} becomes $\mathbf H_{\rm vac}+a_{\nu\nu}(r)\mathbf H_{\nu\nu}$ with a radial-dependent factor
\begin{equation}
    a_{\nu\nu}(r) = \frac{a_1}{1+e^{(a_2-r)/a_3}},
\end{equation}
where $a_2=30$ and $a_3 = 2.5~{\rm km}$ for all our models.
This factor keeps almost constant $a_{\nu\nu}(r)\approx a_1$ for most of the radial range with $r>35$~km.
At inner radii, where the neutrino number densities are much larger, more attenuation is provided by the denominator in the formula by around one more order of magnitude, i.e., the attenuation factor $a_{\nu\nu}$ takes a value of $0.1 a_1$ at $r\approx 24.5$~km.
For example, when $a_1$ is taken to be $10^{-3}$, the attenuation factor is always greater than $10^{-4}$ for $r> 24.5$~km and close to $10^{-3}$ for $r> 35$~km, which is a reasonable choice compared to other works \cite{nagakura2023global,nagakura2023basic}.
We utilize this attenuation strategy and explore different choices of the attenuation parameter $a_1$, with the radial resolutions $N_r$ adapted accordingly.

We summarize all models and their parameters of this paper in Table~\ref{tab:parameters}.
There is no flavor instability in Model I.
In both two- and three-flavor models, all mixing angles in $\mathbf H_{\rm vac}$ are set to be $10^{-6}$ accounting for the effective suppression by the matter.
In three-flavor models, the CP violation phase is omitted, and the lower mass-square difference is taken as $\delta m_{\rm L}^2=7.39\times 10^{-5}~\mathrm{eV}^2$.
In order to investigate the effects of other types of flavor instabilities on the FFCs, we run simulations with different values of $\delta m^2$ in the vacuum term and introduce an auxiliary parameter $c_{\rm off}$ that is multiplied to all off-diagonal elements of $\mathbf C$.
The default value of $c_{\rm off}$ is 1.
When $c_{\rm off}$ is set to be zero, the CFI is prohibited.

\begingroup
\begin{table*}[t]
\begin{ruledtabular}
    \centering
    \caption{\label{tab:parameters} Parameters used in each model. For all models, $r_{\rm ib}=20$ to $r_{\rm ob}=80$~km, $t_\mathrm{pb}$=252~ms, $a_2=30$~km, $a_3=2.5$~km, $N_{v_r}=100$, $N_E=15$.}
    \begin{tabular}{cccc cc cc}
        model & $b_1$ & $b_2$~(km) & No. of flavors & $a_1$ & $N_r$ & $c_{\rm off}$& $\delta m^2$~(eV$^{2}$)\\\hline
        I     & 0    &    &   &                   &       &   &                   \\
        II    & 0.1  & 42 & 2 & $10^{-3}$         & 25000 & 1 & $8\times 10^{-5}$ \\
        III   & 0.15 & 42 & 2 & $10^{-3}$         & 25000 & 1 & $8\times 10^{-5}$ \\
        IV    & 0.2  & 42 & 2 & $10^{-3}$         & 25000 & 1 & $8\times 10^{-5}$ \\
        V     & 0.25 & 42 & 2 & $10^{-3}$         & 25000 & 1 & $8\times 10^{-5}$ \\
        VI    & 0.25 & 32 & 2 & $10^{-3}$         & 25000 & 1 & $8\times 10^{-5}$ \\
        VII   & 0.25 & 30 & 2 & $10^{-3}$         & 25000 & 1 & $8\times 10^{-5}$ \\\hline
        Vc0   & 0.25 & 42 & 2 & $10^{-3}$         & 25000 & 0 & $8\times 10^{-5}$ \\
        IIf3  & 0.1  & 42 & 3 & $10^{-3}$         & 25000 & 1 & $8\times 10^{-5}$ \\
        IVf3  & 0.2  & 42 & 3 & $10^{-3}$         & 25000 & 1 & $8\times 10^{-5}$ \\
        IIa2  & 0.1  & 42 & 2 & $2\times 10^{-3}$ & 50000 & 1 & $8\times 10^{-5}$ \\
        IIa4  & 0.1  & 42 & 2 & $4\times 10^{-3}$ & 50000 & 1 & $8\times 10^{-5}$ \\
        IIa10 & 0.1  & 42 & 2 & $10^{-2}$         & 50000 & 1 & $8\times 10^{-5}$ \\
        IIIa2 & 0.15 & 42 & 2 & $2\times 10^{-3}$ & 50000 & 1 & $8\times 10^{-5}$ \\
        IIIa4 & 0.15 & 42 & 2 & $4\times 10^{-3}$ & 50000 & 1 & $8\times 10^{-5}$ \\
        IVa2  & 0.2  & 42 & 2 & $2\times 10^{-3}$ & 50000 & 1 & $8\times 10^{-5}$ \\
        IVa4  & 0.2  & 42 & 2 & $4\times 10^{-3}$ & 50000 & 1 & $8\times 10^{-5}$ \\
        IVv2  & 0.2  & 42 & 2 & $10^{-3}$         & 25000 & 1 & $2.4\times 10^{-3}$ \\
    \end{tabular}
\end{ruledtabular}
\end{table*}
\endgroup

\section{Analyses}
\label{sec:analyses}
\subsection{Illustrative quantities}
\label{sec:illus_quant}
For a better illustration of our models, we define several quantities as follows.
The energy- and angle-integrated density matrices are defined by $\langle\varrho\rangle_E(v_r) = \int d E\, \varrho(E, v_r) $ and $\langle\varrho\rangle_A(E) = \int d v_r\, \varrho(E, v_r)$, respectively.
The neutrino mean energy of a given flavor is calculated as
\begin{equation}
	\langle E_{\nu_i} \rangle = \frac{\int d E\,d v_r\, E \varrho_{ii}(E,v_r)}{\int d E\,d v_r\, \varrho_{ii}(E,v_r)},
\end{equation}
where $i=e,\,\mu,\,\tau$.
A dimensionless ratio, $s_{e\mu} = |\langle \varrho_{e\mu} \rangle_E|/(|\langle \varrho_{ee} \rangle_E-\langle \varrho_{\mu\mu} \rangle_E|^2/4+|\langle \varrho_{e\mu} \rangle_E|^2)^{1/2}$, is defined to quantify the nonlinearity of flavor conversions caused by FFI.
The corresponding phase (angle in the complex plane) is defined as $\phi_{e\mu} = \arg[\langle \varrho_{e\mu} \rangle_E \, {\rm sgn}(\langle \varrho_{ee} \rangle_E-\langle \varrho_{\mu\mu} \rangle_E) ]$, where $\mathrm{sgn}$ denotes the sign function.
The electron-minus-muon lepton number of neutrinos (E-MuLN, shortened to ELN in the following discussion) is $G= \langle \varrho_{ee} \rangle_E-\langle \bar\varrho_{ee} \rangle_E-\langle \varrho_{\mu\mu} \rangle_E+\langle \bar\varrho_{\mu\mu} \rangle_E$.
When there only exists one zero crossing in the ELN angular distribution, we define two integrals $I_+=\int_{-1}^1dv_r\, G \Theta(G)$ and $I_-=\int_{-1}^1 dv_r\, G \Theta(-G)$, where $\Theta$ is the Heaviside step function.
Those two quantities allow us to evaluate the ``deepness ratio'' of the ELN angular crossing based on the value of ${\rm min}(|I_-|,|I_+|)/{\rm max}(|I_-|,|I_+|)$.

\begin{figure*}[!hbt]
\includegraphics[width=\textwidth]{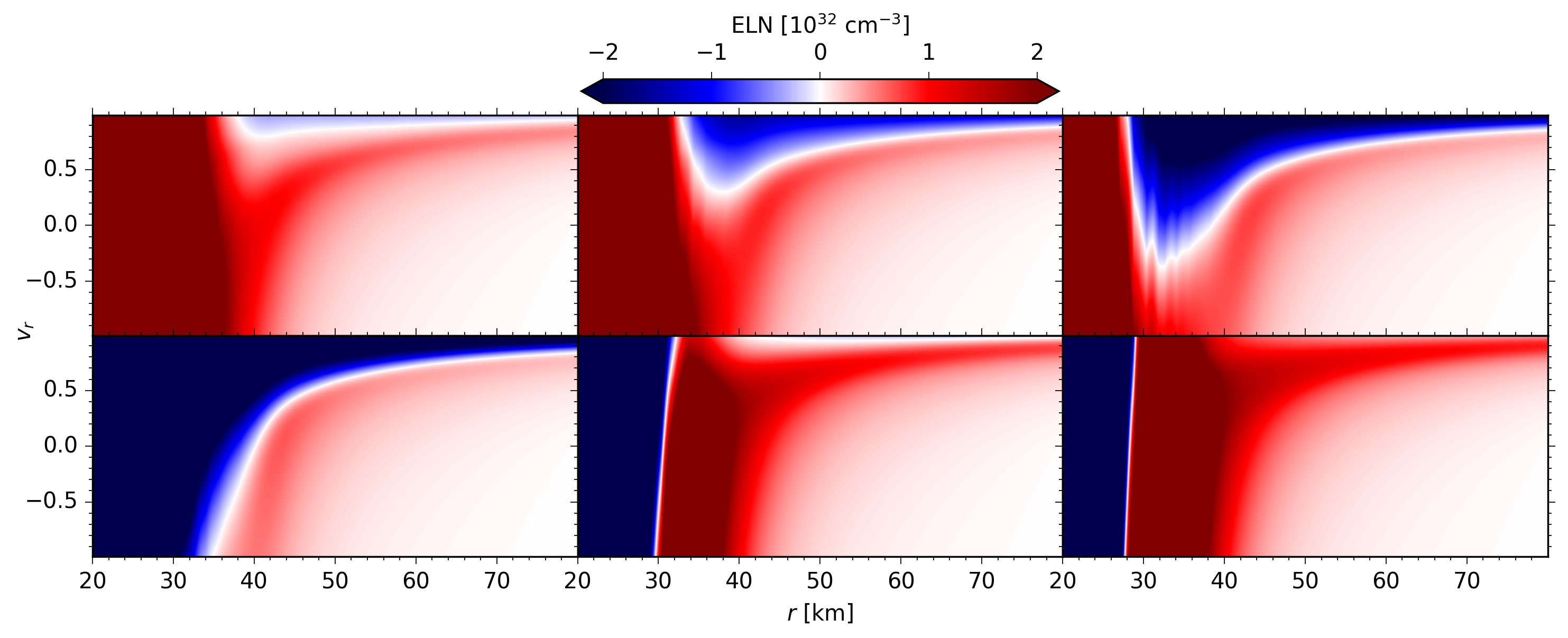}
\hspace{-0.04in}\llap{\parbox[b]{9.8in}{\small Model II \\\rule{0ex}{1.5in}}}
\hspace{-0.04in}\llap{\parbox[b]{5.4in}{\small Model III\\\rule{0ex}{1.5in}}}
\hspace{-0.04in}\llap{\parbox[b]{1.0in}{\small Model IV \\\rule{0ex}{1.5in}}}
\hspace{-0.04in}\llap{\parbox[b]{9.8in}{\small Model V \\\rule{0ex}{0.5in}}}
\hspace{-0.04in}\llap{\parbox[b]{5.4in}{\small Model VI \\\rule{0ex}{0.5in}}}
\hspace{-0.04in}\llap{\parbox[b]{1.0in}{\small Model VII\\\rule{0ex}{0.5in}}}
\caption{\label{fig:initial_ELN} Radial profiles of ELN distribution in Models II--VII that have different attenuated $Y_e$ profiles. The angular crossings of ELN are indicated by the location of curves with the white color. Model I (without $Y_e$ attenuation) is not shown as it contains no ELN angular crossing.
}
\end{figure*}

\begin{figure}[!hbt]
\includegraphics[width=\columnwidth]{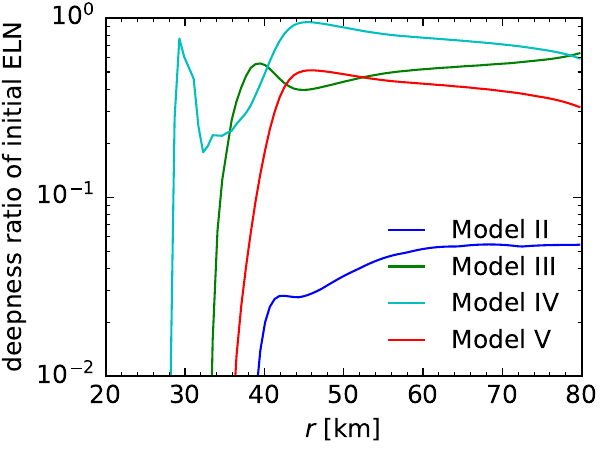}
\caption{\label{fig:I_ratio} Deepness ratios of the initial ELN distribution, ${\rm min}(|I_-|,|I_+|)/{\rm max}(|I_-|,|I_+|)$, in Models II--V (see text for the definition of $I_\pm$).
}
\end{figure}

Except the original Model I without $Y_e$ attenuation and angular crossings, Fig.~\ref{fig:initial_ELN} shows the initial ELN distributions for Models II--VII in the stationary state obtained by solving Eqs.~\eqref{eq:eom_nu} and \eqref{eq:eom_nubar} without including the commutators on the right-hand side.
Clearly, these six models show a significant variation in the radial shapes of their ELN angular crossings due to different $Y_e$ attenuation, while each of them only contains at most one angular crossing at any specific radius.
In particular, the ELN crossings in Models II--V extend all the way to the outer boundary where neutrinos freely stream outward.
With the least $Y_e$ attenuation imposed in Model II (corresponding to the red curve in Fig.~\ref{fig:ye_attenuation}), a relatively shallow angular crossing begins to appear with crossing radial velocity $v_{r,c}\lesssim 1$ at $r\approx 38$~km, and extends to smaller $v_{r,c}$ as low as $v_{r,c}\approx 0.8$ at $r \approx 41$~km.
For the entire domain, the deepness ratios are less than 0.06 as shown in Fig.~\ref{fig:I_ratio}.
When we apply more attenuation on $Y_e$, which results in more production of $\bar\nu_e$'s, the starting radii where the angular crossing appears shift to $r\approx 33$~km and $28$~km in Model III and IV, respectively.
The minimum values of $v_{r,c}$ are $v_{r,c}\approx 0.3$ at $r\approx 38$~km and $v_{r,c}\approx -0.4$ at $r\approx 32$~km accordingly.
For Model V (corresponding to the cyan curve in Fig.~\ref{fig:ye_attenuation}), the inner boundary becomes so $\bar\nu_e$-dominated that the crossing starts from $v_{r,c}=-1$ at $\approx 35$~km and monotonically approaches $v_{r,c}\approx 0.85$ as the radius increases.
The deepness ratios in Models III--V are generally greater than 0.3, particularly reaching $\approx 1$ at $r\simeq 30$~km and 45~km in Model IV.
In the last two models (VI and VII), ELN crossings are only confined inside the neutrino spheres as they originate from the change of sign of $\mu^{\rm eq}_{\nu_e}$ due to the $Y_e$ attenuation.
The crossing in Model VI spans from $r\approx 40$~km to $r\approx 42$~km, while it is narrower in Model VII (corresponding to the green curve in Fig.~\ref{fig:ye_attenuation}), ranging from $r\approx 38$~km to $r\approx 39$~km.
These different radial shapes of ELN crossings lead to very different FFI and the evolution of FFC, which will be discussed in the following sections.

\subsection{Linear stability analysis}
We perform the LSA for models with two flavors.
The analyses on three-flavor models are expected to be similar when neutrinos of muon and tauon flavors are identical, although a nontrivial interplay among flavors may happen when difference in $\mu$ and $\tau$ neutrinos exist \cite{capozzi2020mu,capozzi2022supernova}.
Assuming $|\varrho_{e\mu}|/|\varrho_{ee}-\varrho_{\mu\mu}|\ll 1$ and $|\bar\varrho_{e\mu}|/|\bar\varrho_{ee}-\bar\varrho_{\mu\mu}|\ll 1$ before any instability leads to flavor conversion, $\nu$QKE can be linearized to find unstable normal modes if flavor instability exists \cite{banerjee2011linearized,izaguirre2017fast}.
We follow our previous treatment in \cite{xiong2023evolution}, neglecting the terms associated with $\partial_{v_r}$ and IS but including the vacuum term.
We further assume that in a local region of a width $\sim \mathcal O($km) near a radius $r$, $\varrho_{ee}$ and $\varrho_{\mu\mu}$ are homogeneous and can be characterized by the values at $r$, while a collective mode of the perturbation $\varrho_{e\mu} = Q(\Omega,K_r,E,r,v_r)e^{-i[\Omega t-K_r (r'-r)]}$ and $\bar\varrho_{e\mu}^* = \bar Q(\Omega,K_r,E,r,v_r)e^{-i[\Omega t-K_r (r'-r)]}$ can develop locally with $r'$ denoting the spatial dependence locally around $r$.
The off-diagonal parts of Eqs.~\eqref{eq:eom_nu} and \eqref{eq:eom_nubar} become\footnote{The dependence of $Q$ and $\bar Q$ on $\Omega$, $K_r$, and $r$ is not shown explicitly.}
\begin{align}\label{eq:LEQ1}
& \left[ \Omega - K_r v_r - \Phi(v_r) + i C_{e\mu}(E) + \delta m^2/2E \right]  Q(E,v_r) \nonumber\\
= & -\sqrt{2}G_F[\varrho_{ee}(E,v_r)-\varrho_{\mu\mu}(E,v_r)] \int dE'\, dv_r'\,\times \nonumber\\
&  (1- v_r v_r') [Q(E',v_r')-\bar Q(E',v_r')],
\end{align}
and
\begin{align}\label{eq:LEQ2}
& \left[\Omega - K_r v_r - \Phi(v_r) + i\bar C_{e\mu}(E) - \delta m^2/2E \right]  \bar Q(E,v_r) \nonumber\\
= & -\sqrt{2}G_F[\bar\varrho_{ee}(E,v_r)-\bar\varrho_{\mu\mu}(E,v_r)] \int dE'\, dv_r'\,\times \nonumber\\
& (1-v_r v_r') [Q(E',v_r')-\bar Q(E',v_r')],
\end{align}
where $\Phi(v_r)=\sqrt{2}G_F\int dE'\, dv_r'\, (1-v_r v_r') [\varrho_{ee}(E',v_r')-\varrho_{\mu\mu}(E',v_r')-\bar\varrho_{ee}(E',v_r')+\bar\varrho_{\mu\mu}(E',v_r')]$, $C_{e\mu}(E)$ and $\bar C_{e\mu}(E)$ are the damping rates attributed to EA and the approximate PR terms.
For example, $C_{e\mu}(E) = (j_{e,\rm EA}+\chi_{e,\rm EA}+j_{e,\rm PR}+\chi_{e,\rm PR}+j_{\mu,\rm PR}+\chi_{\mu,\rm PR})/2$.
To account for the attenuation scheme used for $\mathbf{H}_{\nu\nu}$, we modify Eqs.~\eqref{eq:LEQ1} and \eqref{eq:LEQ2} by multiplying $G_F$ with the attenuation factor $a_{\nu\nu}$, which affects the right-hand side and the term $\Phi$ on the left-hand side.

For a given wave number $K_r$, the eigenvalues of $\Omega$ and their corresponding eigenvectors can be numerically solved.
The contribution of $\Phi$ is often absorbed into $\Omega$ and $K_r$ in literature \cite{izaguirre2017fast,yi2019dispersion}, but we do not follow this procedure in order to keep the original meaning of $K_r$ as the wave number in the local frame.
When $K_r=0$, it represents the homogeneous mode.
We highlight that we search for nonzero $K_r$ modes in addition to the homogeneous one to avoid missing the most unstable mode.

We note that there may exist some other aspects that cannot be captured by the linearized equations \eqref{eq:LEQ1} and \eqref{eq:LEQ2}.
For instance, the behaviors of flavor instability may be affected by the angular advection term associated with $\partial_{v_r}$ and the fact that neutrino number density is not homogeneous.
Although the pure FFI without collisions and with vanishing vacuum oscillation frequency is free from the contamination of spurious modes~\cite{capozzi2019fast}, which are associated with the use of an insufficient number of angular grids in the discretization, those spurious modes may appear in multiangle slow instabilities when the vacuum term is taken into account \cite{sarikas2012spurious,abbar2015flavor,morinaga2018linear} or in the CFI when collisions are included~\cite{martin2021fast,kato2023flavor}.
Their growth rates should approach zero as the adopted angular grid number increases toward the continuous limit.

\subsection{Spectrogram}\label{sec:ana-meth-spec}
In addition to the LSA, the onset of specific unstable modes as well as the small-scale spatial structure can be simply analyzed by the Fourier transform of relevant quantities over the entire simulation domain in the periodic-box setup.
In contrast, the inhomogeneous neutrino and matter profiles in the spherical symmetric supernova model do not allow such a simple approach.
Nonetheless, the short-time Fourier transform (or spectrogram analysis) provides a powerful tool in signal processing to diagnose the characteristic features in the frequency domain over a short-time window for a time-varying system, which has been generically used in the engineering and science (see, e.g., \cite{flanagan1972speech,shen2018natural,abbott2016observation}).
In the same spirit, we can perform similar spectrogram analysis to the neutrino flavor evolution outcomes except that here we calculate the spectra in terms of wave number $K_r$ over a short-length window.

The spectrogram analysis can be performed in various ways.
In order to mitigate the spectral leakage, we use the Tukey window function in the spectrogram analysis~\cite{harris1978use,bloomfield2004fourier}.
The flat top of the Tukey window also naturally provides less amplitude attenuation than in others such as a Gaussian window.
We divide the whole radial range of our simulation domain into $N_r^{\rm sp}$ blocks with each block centered at a radius $r$, and calculate discretized Fourier transform for the neutrino off-diagonal mixing at a snapshot of simulation in each block
\begin{align}
    & \mathcal F_{r,K_r} = \frac{1}{\Delta r^\mathrm{sp}} \int_{r-\frac{5}{8}\Delta r^\mathrm{sp}}^{r+\frac{5}{8}\Delta r^\mathrm{sp}} dr'\, e^{-i K_r r'} W(r'-r) \times \nonumber\\
    & \int dv_r\, \langle \varrho_{e\mu} \rangle_E(v_r,r') \mathrm{sgn}[\langle \varrho_{e\mu} \rangle_E(v_r,r')-\langle \varrho_{e\mu} \rangle_E(v_r,r')],
\end{align}
where $\Delta r^\mathrm{sp}=(r_{\rm ob}-r_{\rm ib})/N_r^{\rm sp}$ is the radial width of the nonoverlapping part of each block, $K_r$ takes discretized values $z \pi/\Delta r^\mathrm{sp}~\mathrm{km}^{-1}$ with integers $z$, and $W(r)$ is the Tukey window function with a cosine lobe
\begin{equation}
    W(r) = \begin{cases} 1 & \mathrm{if}~|r|\leq \Delta r^\mathrm{sp}/2, \\ \cos^2 \left(\frac{4\pi |r|}{\Delta r^\mathrm{sp}}-2\pi \right) & \mathrm{if}~|r|> \Delta r^\mathrm{sp}/2. \end{cases}
\end{equation}
Depending on $N_r$ used in the simulations, we take $N_r^{\rm sp}$ as 50 or 100 to ensure sufficient resolution in $K_r$ space.

\section{Emergence of flavor instability}
\label{sec:emergence}
\subsection{Scalability of the dispersion relation}
Perturbed by the vacuum flavor mixing (or potentially quantum fluctuation), the flavor waves of the neutrino fields can be launched and lead to flavor transformation driven by the dispersion relation and the unstable collective modes in LSA.
Ignoring the contributions from other aspects such as advection, collisions and the vacuum term, the dispersion relation branches for the axisymmetric unstable mode can have distinct structures for different ELN distributions~\cite{yi2019dispersion}.
For a shallow crossing, the dispersion relation typically yields two separated unstable branches with respect to the wave number of the perturbed flavor wave $K_r$.
For a deep crossing, only one unstable branch exists, and the maximum growth rate of the branch is enhanced.

All LSA results shown below are based on the stationary neutrino profiles obtained without including oscillations (see e.g., Fig.~\ref{fig:initial_ELN} for the energy-integrated profiles).
The labels \texttt{v0}, \texttt{v1}, \texttt{v2}, and \texttt{v-2} represent results using four different values of $\delta m^2=0~\mathrm{eV}^2$, $8\times 10^{-5}~\mathrm{eV}^2$,  $2.4\times 10^{-3}~\mathrm{eV}^2$ and $-2.4\times 10^{-3}~\mathrm{eV}^2$, respectively.
The labels \texttt{c0} and \texttt{c1} denote settings with $c_{\rm off}=0$ and 1, respectively.
Note that a positive (negative) $\delta m^2$ value represents the normal (inverted) neutrino mass ordering.

\begin{figure*}[!hbt]
\includegraphics[width=0.32\textwidth]{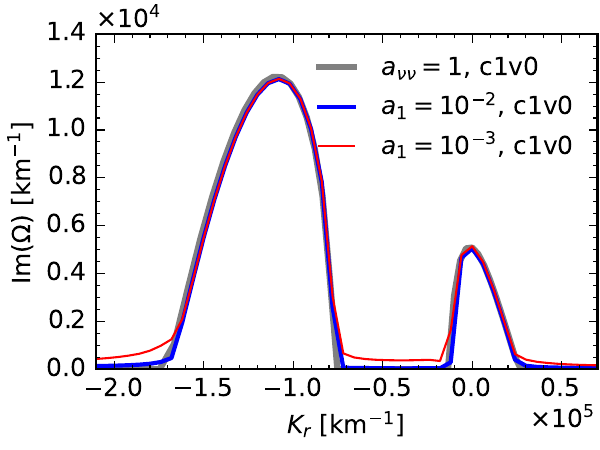}
 \llap{\parbox[b]{3.5in}{\small (a)\\\rule{0ex}{1.3in}}}
\includegraphics[width=0.32\textwidth]{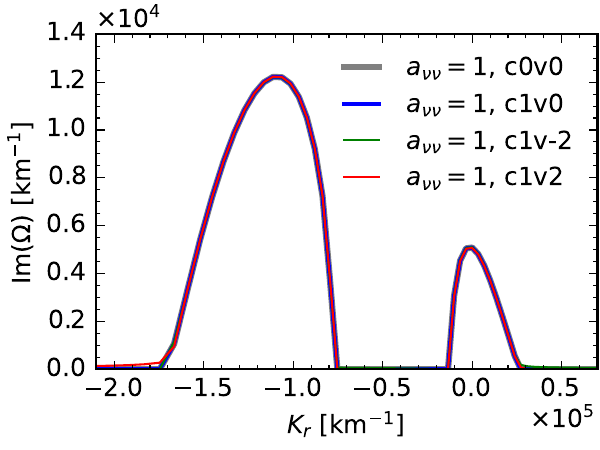}
 \llap{\parbox[b]{3.5in}{\small (b)\\\rule{0ex}{1.3in}}}
\includegraphics[width=0.32\textwidth]{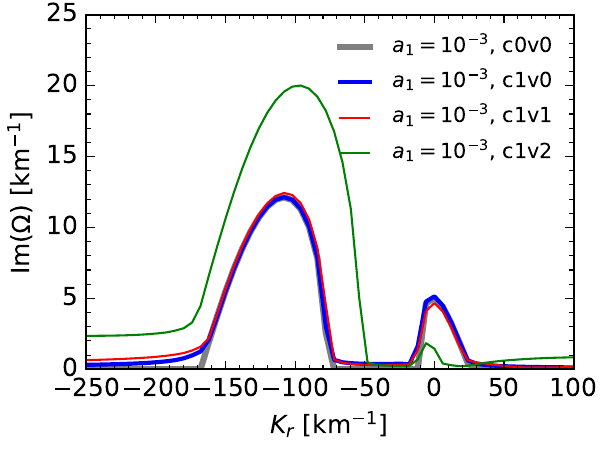}
 \llap{\parbox[b]{3.5in}{\small (c)\\\rule{0ex}{1.4in}}}

\includegraphics[width=0.32\textwidth]{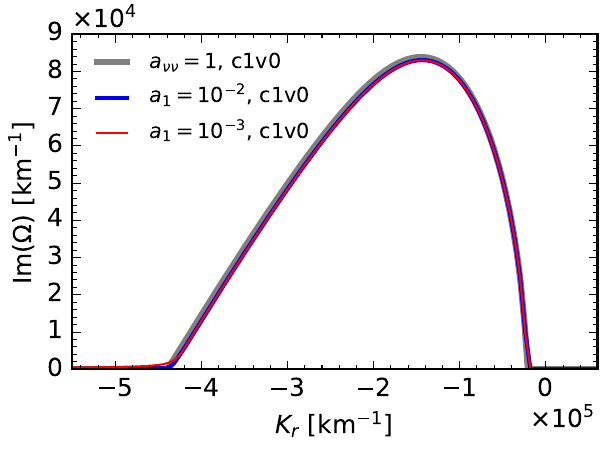}
 \llap{\parbox[b]{3.6in}{\small (d)\\\rule{0ex}{1.0in}}}
\includegraphics[width=0.32\textwidth]{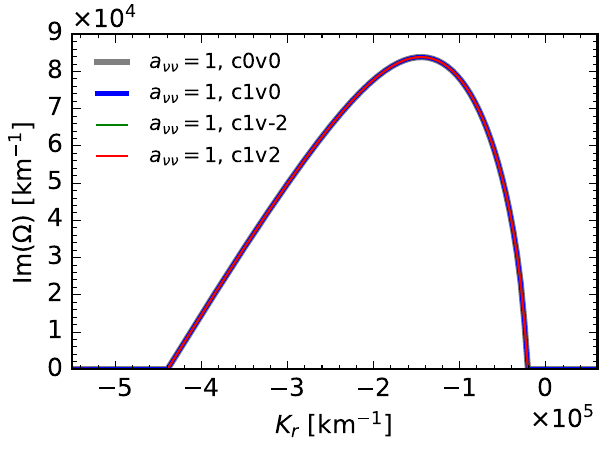}
 \llap{\parbox[b]{3.6in}{\small (e)\\\rule{0ex}{0.9in}}}
\includegraphics[width=0.32\textwidth]{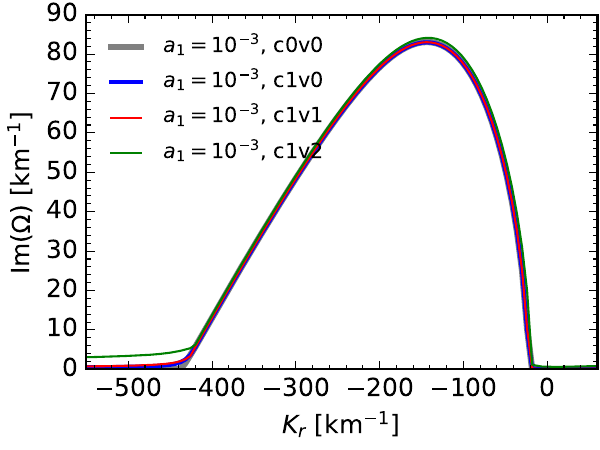}
 \llap{\parbox[b]{3.6in}{\small (f)\\\rule{0ex}{0.9in}}}
\caption{\label{fig:DR_other} Growth rates Im$(\Omega)$ as functions of $K_r$ at $r=41.6$~km in Models II (a--c), and V (d--f). The values of Im$(\Omega)_{\rm max}$ for the cases with $a_1=10^{-2}$ and $a_1=10^{-3}$ in (a) and (d) are multiplied by $10^2$ and $10^3$, respectively, to show the scalability of the dispersion relation discussed in the text.
}
\end{figure*}

We begin with Model II that has the most shallow angular crossing compared to other models.
Figures~\ref{fig:DR_other}(a--c) show the growth rates Im$(\Omega)$ of the unstable mode in Model II with different choices of $\delta m^2$, $c_{\rm off}$, and $a_{\nu\nu}$.
From Eqs.~\eqref{eq:LEQ1} and \eqref{eq:LEQ2}, the dispersion relation of the FFI is clearly scalable when the vacuum and collision terms are omitted.
In this case, applying the attenuation factor $a_{\nu\nu}$ simply reduces the magnitude of $\Omega$ and $K_r$ for the same amount but does not affect the shape of the dispersion relation branches.

\begin{figure*}[!hbt]
\includegraphics[width=0.32\textwidth]{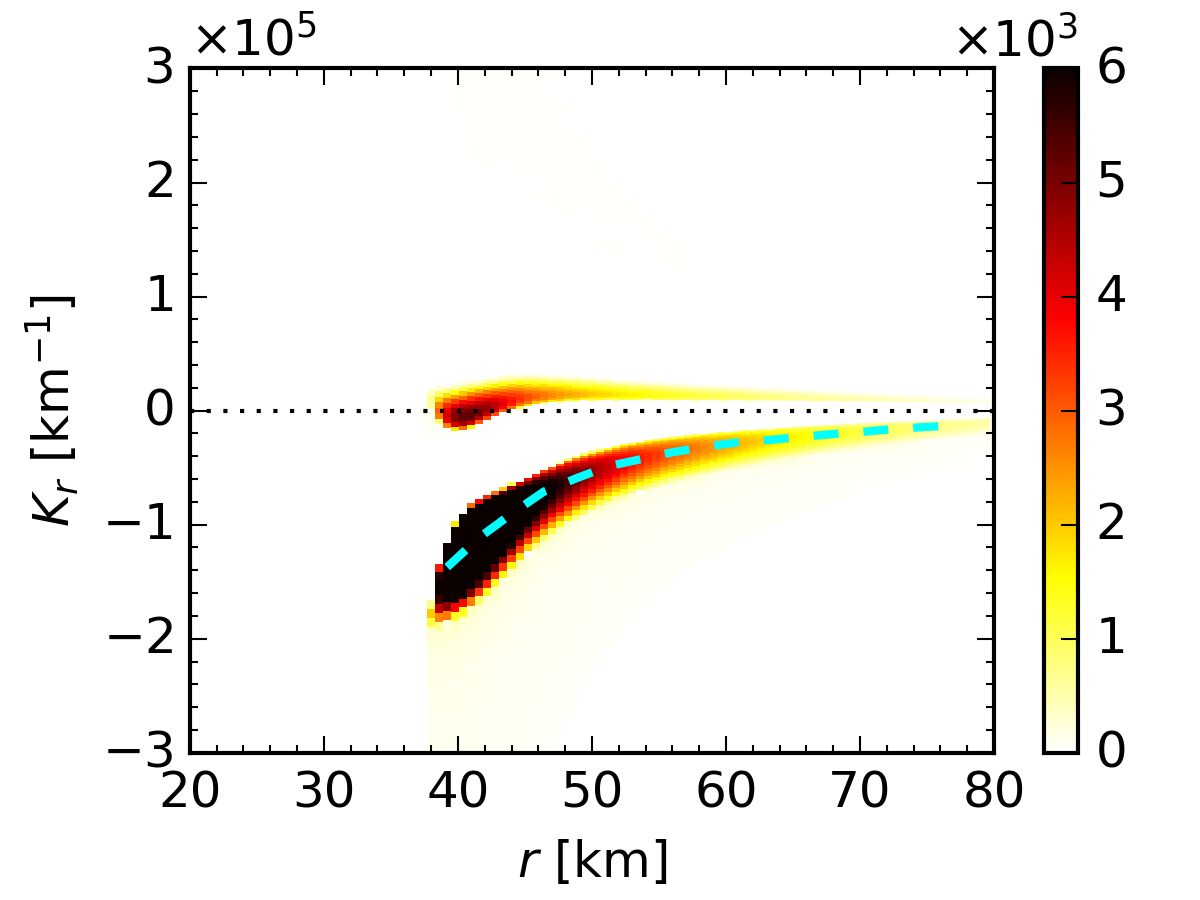}
\llap{\parbox[b]{2.7in}{\small (a) $a_{\nu\nu}=1$, \texttt{c1v2}\\\rule{0ex}{1.3in}}}
\includegraphics[width=0.32\textwidth]{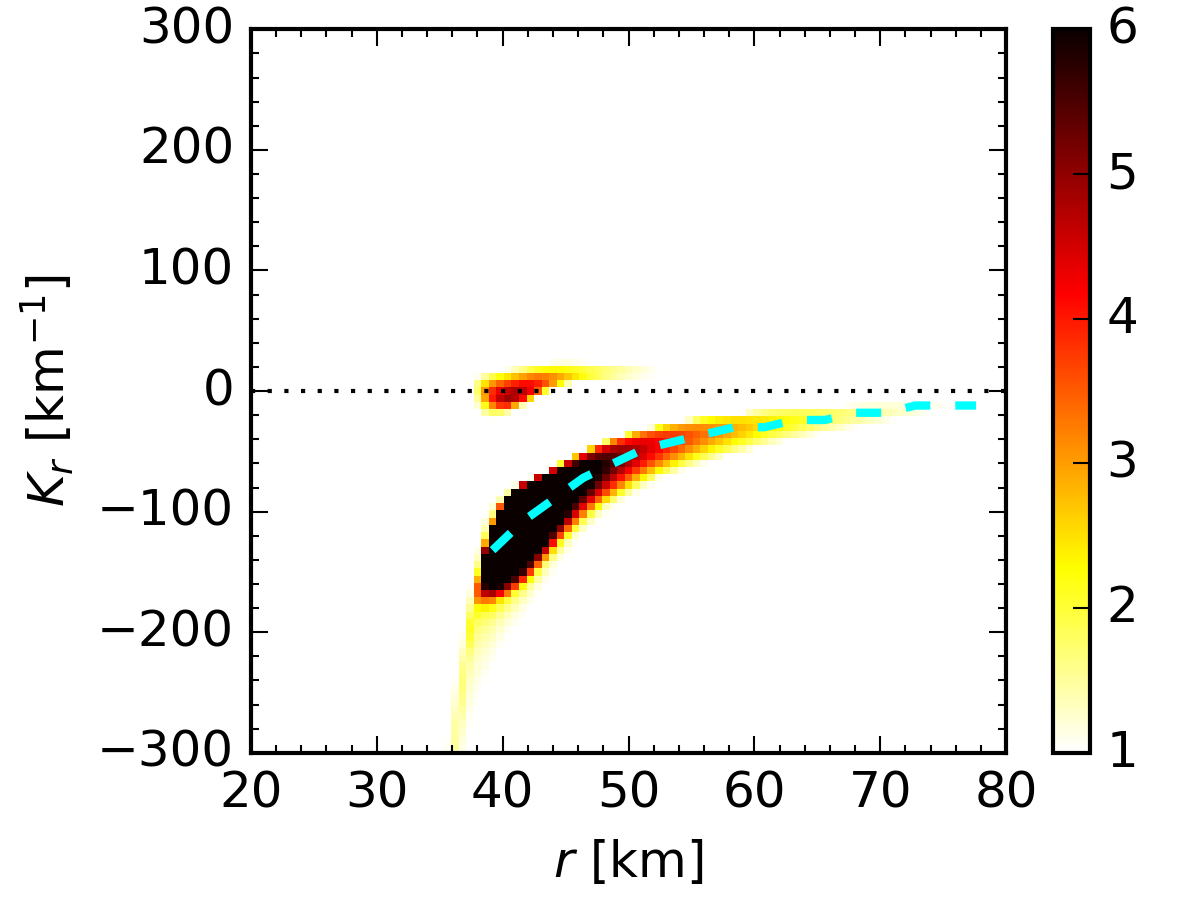}
\llap{\parbox[b]{2.4in}{\small (b) $a_1=10^{-3}$, \texttt{c1v1}\\\rule{0ex}{1.4in}}}
\includegraphics[width=0.32\textwidth]{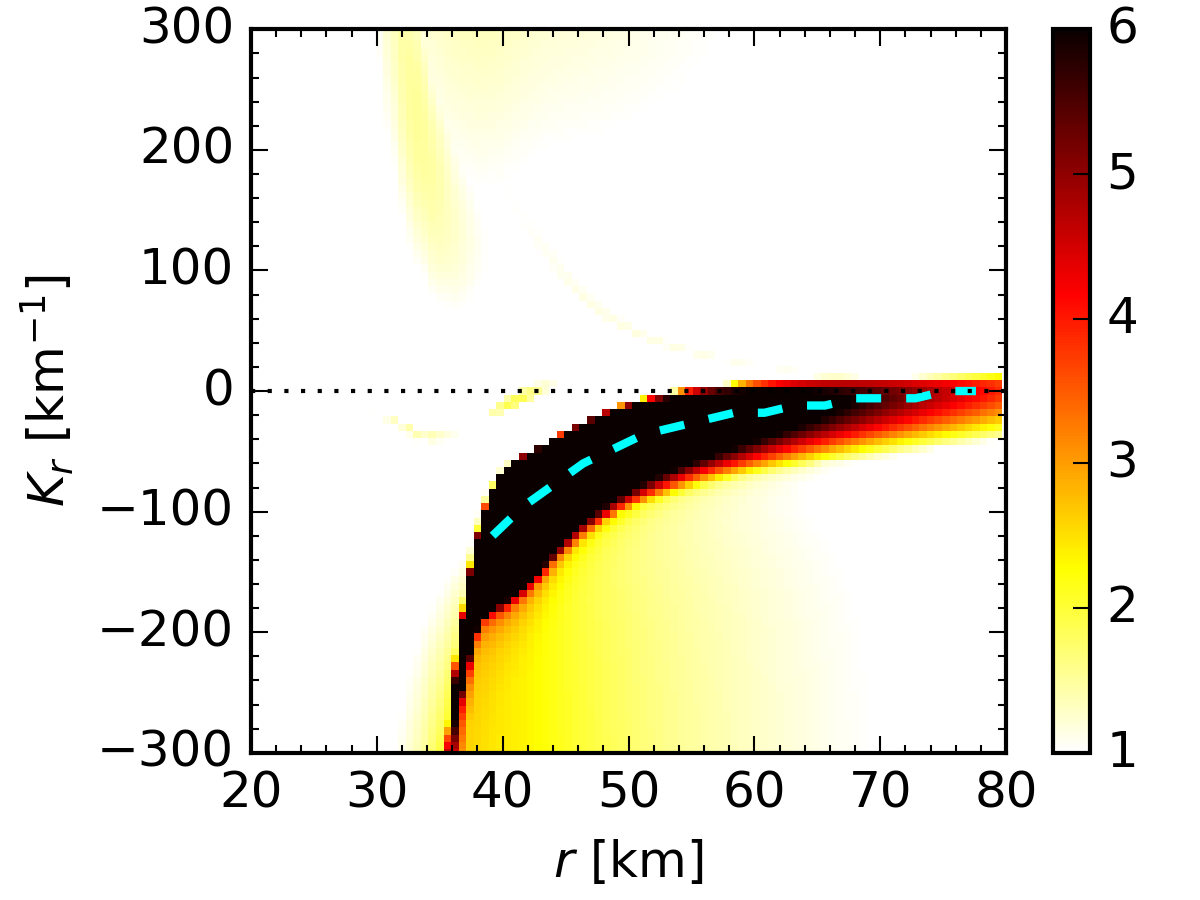}
\llap{\parbox[b]{2.4in}{\small (c) $a_1=10^{-3}$, \texttt{c1v2}\\\rule{0ex}{1.4in}}}

\includegraphics[width=0.32\textwidth]{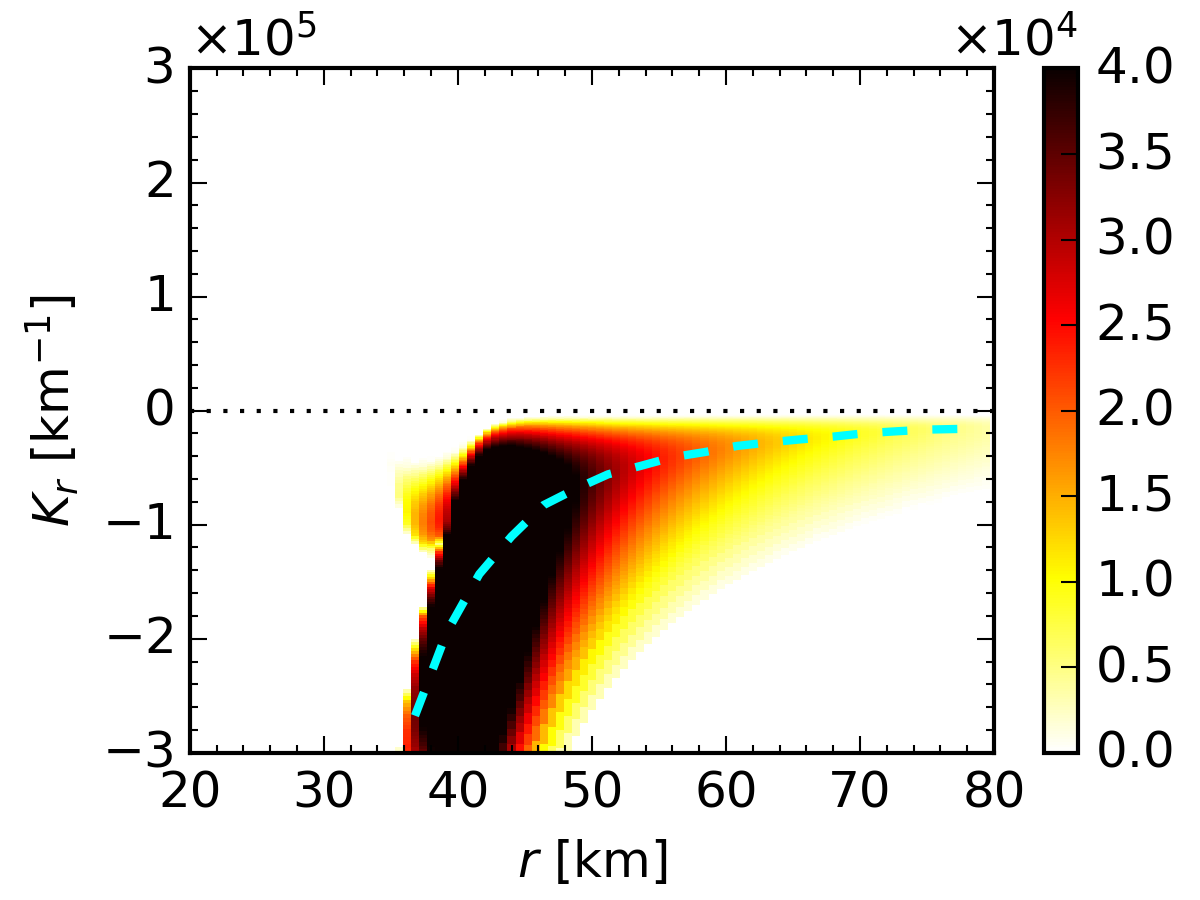}
\llap{\parbox[b]{2.7in}{\small (d) $a_{\nu\nu}=1$, \texttt{c1v2}\\\rule{0ex}{1.3in}}}
\includegraphics[width=0.32\textwidth]{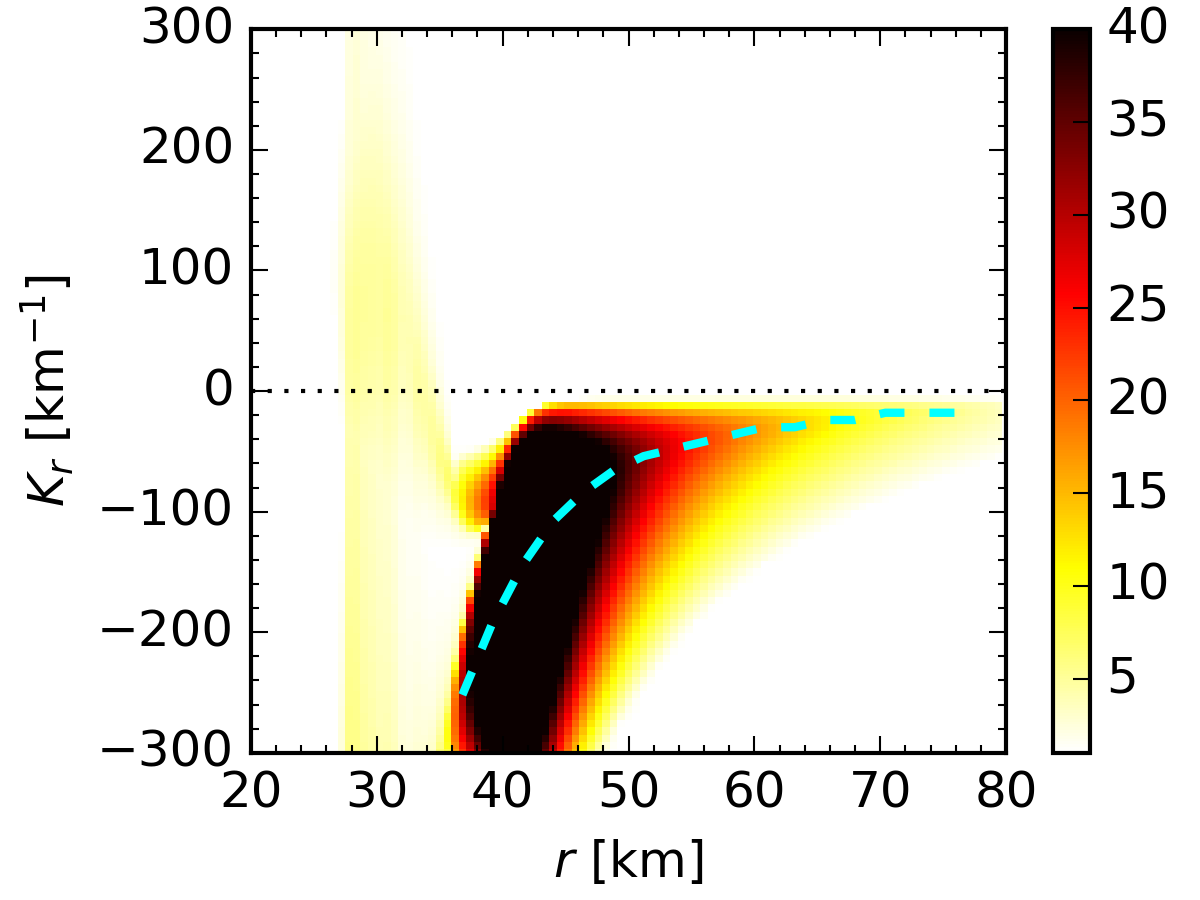}
\llap{\parbox[b]{2.4in}{\small (e) $a_1=10^{-3}$, \texttt{c1v1}\\\rule{0ex}{1.4in}}}
\includegraphics[width=0.32\textwidth]{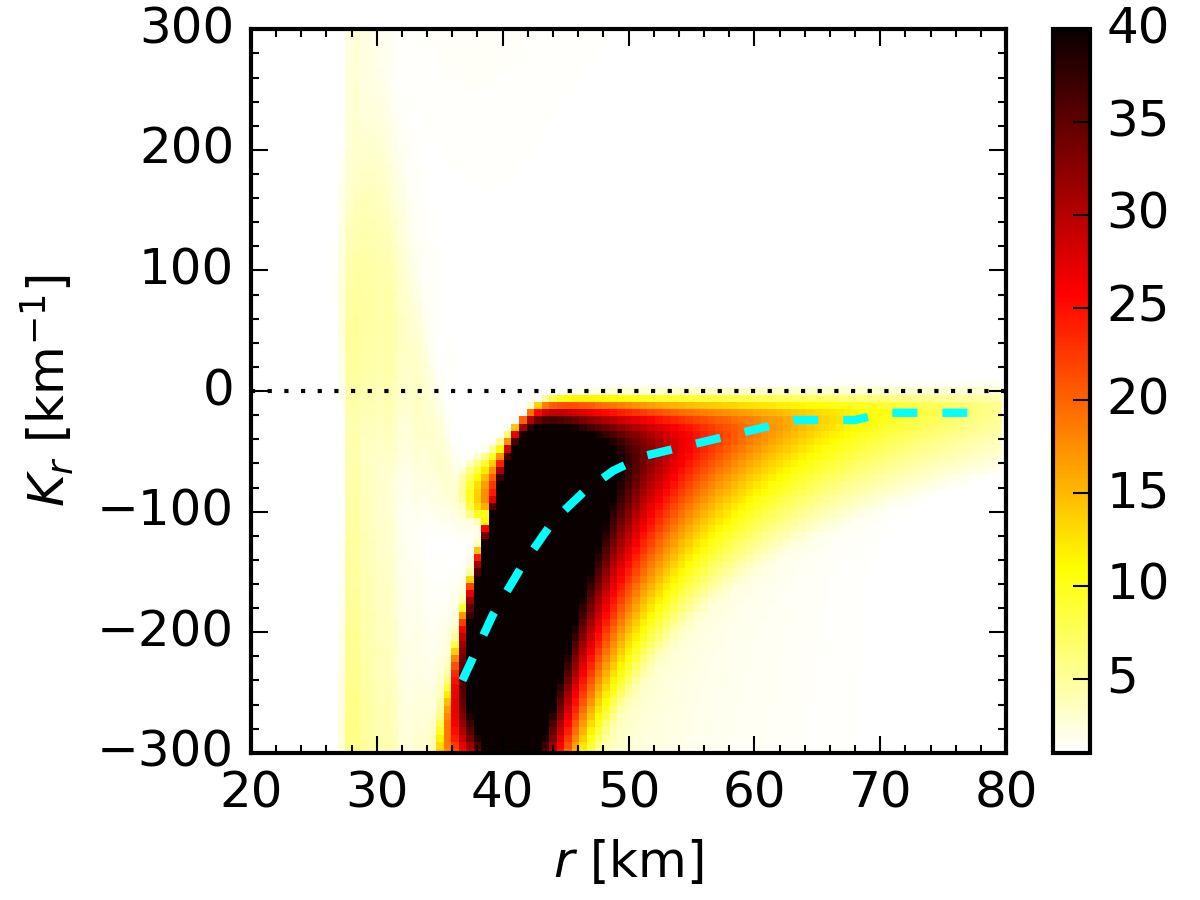}
\llap{\parbox[b]{2.4in}{\small (f) $a_1=10^{-3}$, \texttt{c1v2}\\\rule{0ex}{1.4in}}}
\caption{\label{fig:Kr} Growth rates Im$(\Omega)~[{\rm km}^{-1}]$ as functions of $K_r$ and $r$ in Models II (a--c) and V (d--f). The cyan dashed curve marks $K_r$ with maximum growth rates.
}
\end{figure*}

The scalability approximately holds even if the collisional terms are included.
Figure~\ref{fig:DR_other}(a) compares the unstable branches of Im$(\Omega)$ at $r\approx 41.6$~km in Model II from the case without attenuation ($a_{\nu\nu}=1$) and from cases with attenuation factors $a_1=10^{-2}$ and $a_1=10^{-3}$.
For the latter two with attenuation, their unstable branches are shown by multiplying factors of $10^2$ and $10^3$ to both $K_r$ and Im$(\Omega)$, respectively.
Among those three cases, there is good agreement on the rescaled growth rate in the dominant parts of the two separated unstable branches.

We further examine the effects of collisions and vacuum terms on the dispersion relation in Model II for the case with $a_{\nu\nu}=1$ and $a_1=10^{-3}$ in Fig.~\ref{fig:DR_other}(b) and (c) separately.
Figure~\ref{fig:DR_other}(b) shows that without attenuation, the dispersion relation of FFI is hardly changed in the presence of collisions and the vacuum term.
This suggests that at this radius, the neutrino number density is so high that other types of flavor instabilities are overwhelmed by the large growth rate from the fast mode.

When taking the attenuation factor $a_1=10^{-3}$ in Fig.~\ref{fig:DR_other}(c), the shape of the dispersion relation remains quantitatively similar when including the collisions in the case of \texttt{c1v0} or a vacuum term with smaller $\delta m^2$ in \texttt{c0v1}.
A slight difference is that small nonzero growth rates Im$(\Omega)\simeq 1~{\rm km}^{-1}$, which connect the two otherwise separated branches of pure fast mode when only the attenuated $\mathbf H_{\nu\nu}$ is included (case \texttt{c0v0}), become visible in comparison to Fig.~\ref{fig:DR_other}(b).
This is because with the attenuation, $\mathbf H_{\nu\nu}$ becomes closer to the $\mathbf H_{\rm vac}$ in magnitude.\footnote{These extended nonzero growth rates may also originate from the spurious mode associated with insufficient number of grids. However, here we do not further clarify its exact origin because the flavor conversion at this radius is dominated by the FFI.}
When we take a larger vacuum term with $\delta m^2=2.4\times 10^{-3}~{\rm eV}^2$, the slow flavor instability becomes so strong that it mixes with the FFI and drastically affects the structure of the dispersion relation branches.
The growth rate of the left unstable branch is enhanced by a factor of $\simeq 2$, and that of the right one is suppressed.

For Model V with a much deeper ELN angular crossing at the same radius than that in Model II, we show the corresponding LSA results in Fig.~\ref{fig:DR_other}(d-f).
In this case, the dispersion relation has a single unstable branch with larger growth rates.
With unattenuated $\mathbf H_{\nu\nu}$, the maximum growth rate in Model V shown in Fig.~\ref{fig:DR_other}(d) is $\approx 8.4\times 10^4~{\rm km}^{-1}$, which is about seven times larger than that in Model II, and is nearly independent of the presence of the collisional and vacuum terms [Fig.~\ref{fig:DR_other}(e)].
Moreover, due to the enhanced growth rate, even when an attenuation with $a_1=10^{-3}$ is taken, the shape of the dispersion relation branch remains hardly affected by the inclusion of collisional terms and different sizes of vacuum terms as shown in Fig.~\ref{fig:DR_other}(f).

The scalability of dispersion relation discussed above also holds for a wider radial range.
Figure~\ref{fig:Kr}(a) shows the maximum growth rate Im$(\Omega)$ as a function of $K_r$ and $r$ in Model II without attenuation for the case \texttt{c1v2}.
Similarly, two separate unstable branches span over the whole radial region where there are ELN angular crossings.
The maximum growth rate sits on the unstable branch with smaller (more negative) $K_r$ for all radii, with the corresponding $K_r$ value increases from $K_r\simeq -1.5\times 10^5~{\rm km}^{-1}$ at $r= 38$~km to $K_r\simeq 0$~km$^{-1}$ at the outer boundary as the neutrino flux decreases.
The other branch is located in the range of $0~{\rm km}^{-1}\lesssim K_r\lesssim 2\times 10^4~{\rm km}^{-1}$.
For the case with parameters of $\texttt{c1v1}$ and with $a_1=10^{-3}$ shown in Fig.~\ref{fig:Kr}(b), one sees that the shapes of the unstable branches remain largely similar to those shown in Fig.~\ref{fig:Kr}(a), when both the $K_r$ range and the color scheme of Im$(\Omega)$ are also rescaled by the same factor $a_1$.
For the case $\texttt{c1v2}$ with a larger $\delta m^2$ and with the attenuation parameter $a_1=10^{-3}$, Fig.~\ref{fig:Kr}(c) shows that the shape of the dispersion relation for the whole radial range is affected by the vacuum term.
The values of Im$(\Omega)$ in the lower branch are enhanced while those in the upper one are suppressed, consistent with the result discussed earlier at $r\approx 41.6$~km.
Considering that this change is related to the attenuation of $\mathbf H_{\nu\nu}$ and does not appear in the original dispersion relation without attenuation, we mainly adopt $\delta m^2=8\times 10^{-5}~{\rm eV}^2$ in our simulation models to avoid this artifact.

Figure~\ref{fig:Kr}(d--f) compares the radial profiles of dispersion relation in Model V for the case without attenuation to the attenuated cases with \texttt{c1v1} and \texttt{c1v2}.
For all cases, two unstable branches only appear at $r\simeq 37$~km where a very shallow angular crossing exists.
Those two branches merge into one at larger radii.
In the original case without attenuation in Fig.~\ref{fig:Kr}(d), the value of $K_r$ corresponding to the maximum growth rate increases from $K_r\simeq -2.5\times 10^5~{\rm km}^{-1}$ at $r= 37$~km to close to zero at the outer boundary.
The other two cases in Fig.~\ref{fig:Kr}(e) and (f) also follow this trend with a factor of $10^3$ smaller in terms of $K_r$ and the growth rate, showing again very good scalability as discussed earlier.

\begin{figure}[!hbt]
\includegraphics[width=0.48\textwidth]{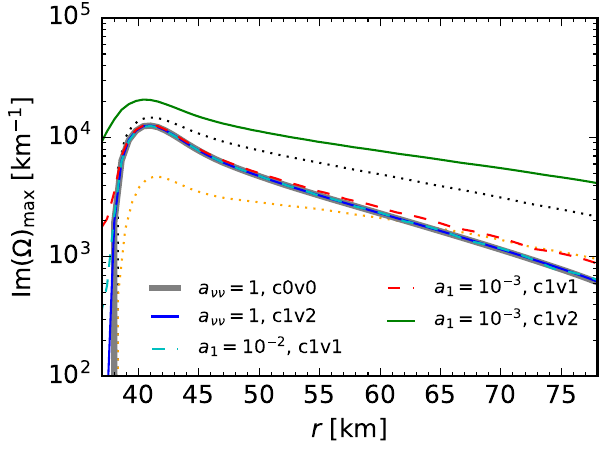}
\hspace{-0.03in}\llap{\parbox[b]{5.3in}{\small (a)\\\rule{0ex}{2.2in}}}
\includegraphics[width=0.48\textwidth]{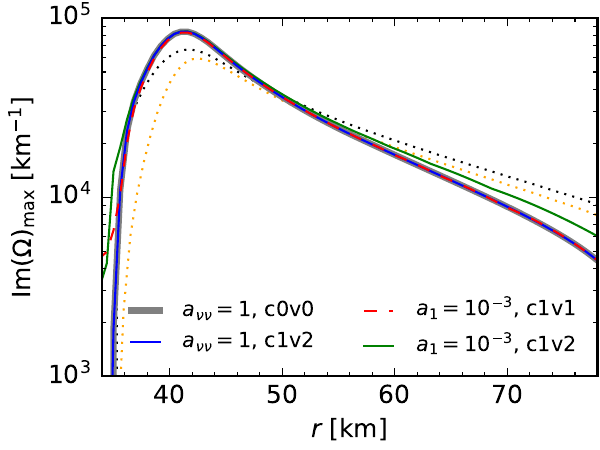}
\hspace{-0.03in}\llap{\parbox[b]{5.3in}{\small (b)\\\rule{0ex}{2.2in}}}
\caption{\label{fig:gr} Maximum growth rates Im$(\Omega)_{\rm max}$ as functions of radius $r$ in Models II (a) and V (b). For the cases with $a_1=10^{-2}$ and $a_1=10^{-3}$, the values of Im$(\Omega)_{\rm max}$ are multiplied by $10^2$ and $10^3$, respectively, for easier comparison. Black and orange dotted curves are obtained using the simple analytical formulas in Refs.~\cite{nagakura2019fast} and \cite{padilla2021multi}, respectively.
}
\end{figure}

Quantitatively, we compare in Fig.~\ref{fig:gr}(a) and (b) the maximum growth rates in Model II and V for cases with different attenuation factors as well as with different collision and vacuum terms.
Once again, the values of Im$(\Omega)_{\rm max}$ for the attenuated cases with $a_1=10^{-2}$ and $a_1=10^{-3}$ are multiplied by $10^2$ and $10^3$, respectively.
For Model II, Fig.~\ref{fig:gr}(a) clearly shows that a good scalability can be reached without attenuation or when the attenuated cases take the smaller $\delta m^2$ (case \texttt{c1v1}).
When the larger $\delta m^2$ (case \texttt{c1v2}) is adopted for the attenuated case with $a_1=10^{-3}$, it artificially enhances the maximum growth rates for the whole region with FFI.
For Model V shown in Fig.~\ref{fig:gr}(b), because the angular crossing is deeper and the maximum growth rate is generally larger than in Model II, the impact due to the vacuum term is relatively smaller.
Even for the case with the larger $\delta m^2$ (\texttt{c1v2}) and $a_1=10^{-3}$, the rescaled maximum growth rates closely follow the original ones, with deviations only up to $\lesssim 20\%$ at $r\gtrsim 60$~km.

Also shown in Fig.~\ref{fig:gr} are the maximum growth rates estimated by two previously proposed analytical formulas $\sqrt{2} G_F \sqrt{|I_- I_+|}$ and $\sqrt{2} G_F (2|I_- I_+|)/(|I_-|+|I_+|)$, which are proportional to the geometric and harmonic averages of $|I_-|$ and $|I_+|$, respectively~\cite{nagakura2019fast,padilla2021multi}.
These estimated rates qualitatively agree with each other, as well as with the values computed by the LSA.

With all the discussions above, we conclude that the overall behavior of the dispersion relation in the region where the ELN angular crossings exist is predominantly determined by the FFI.
The attenuated schemes can retrieve those structures of unstable modes to a good approximation by taking appropriate values of $\delta m^2$ for specific cases.

\subsection{Other type of flavor instability}
\label{sec:otherFI}

\begin{figure}[t]
\includegraphics[width=0.48\textwidth]{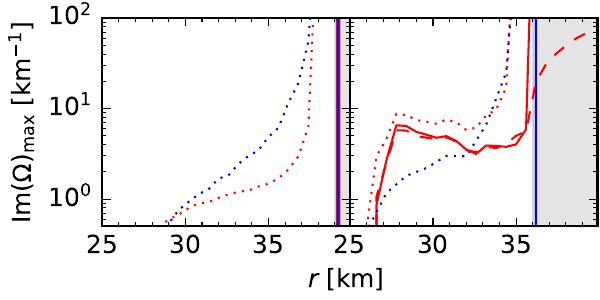}
\hspace{-0.03in}\llap{\parbox[b]{5.3in}{\small (a)\\\rule{0ex}{1.4in}}}
\hspace{-0.03in}\llap{\parbox[b]{2.5in}{\small (b)\\\rule{0ex}{1.4in}}}
\caption{\label{fig:gr_other} Maximum growth rates Im$(\Omega)_{\rm max}$ as functions of radius $r$ for in Models II (a) and V (b). The gray shaded areas indicate the radial regions where ELN angular crossings exist in the initial ELN distribution without oscillations. Blue and red curves represent results with parameters in the cases \texttt{c0v2} and \texttt{c1v0} respectively. Dotted and solid curves represent the cases using $a_{\nu\nu}=1$ without and with excluding the spurious noncollective modes respectively. The dashed curve in (b) is with $a_1=10^{-3}$ and with all the spurious modes excluded.
}
\end{figure}

For regions without angular crossings, it is curious to see whether other types of flavor instabilities such as the CFI or the slow type may exist.
In Model IV where the angular crossing extends to more inner radius, the FFI essentially dominates all radii of concern.
While for other models, we find that CFI exists around the neutrinosphere in Model V with the strongest $Y_e$ attenuation leading to the dominance of $\bar\nu_e$'s over $\nu_e$'s, but it is suppressed in the $\nu_e$-dominated Models I--III.
For Models VI and VII with FFI only appearing in limited range inside the neutrinosphere, their ELN angular profiles outside the neutrinosphere are similar to Model I and thus not discussed in detail below.
We do not find any slow instabilities that are strong enough to be distinguished from spurious modes in any of the models.

As two representative instances, we show the maximum growth rates calculated from the LSA for Models II and V in Fig.~\ref{fig:gr_other}(a) and (b) for $r<40$~km by the dotted lines.
Although in radii without angular crossings (nonshaded region), nonzero small growth rates are found for cases \texttt{c0v2} and \texttt{c1v0} in both Models II and V, some of them turn out to be spurious modes that contain noncollective features.
When applying a practical selection rule detailed in Appendix~\ref{sec:selection_rule_spurious} to preclude the spurious modes, none of those in Model II survive [see the solid lines in Fig.~\ref{fig:gr_other}(a)].
In Model V, the unstable mode due to the nonzero vacuum term is also spurious, whereas the mode due to CFI persists [solid red line in Fig.~\ref{fig:gr_other}(b)].
When the $\mathbf{H}_{\nu\nu}$ attenuation with $a_1=10^{-3}$ is applied, the growth rates due to CFI are nearly the same as the original case without attenuation in regions without ELN crossings [dashed red line in Fig.~\ref{fig:gr_other}(b)].
Notice that this CFI mode also appears visibly around $r\simeq 30$~km over a wide range of $K_r$ in Fig.~\ref{fig:Kr}(e) and (f).

Some further comments regarding the spurious modes and the CFI are in order.
First, the spurious modes, in principle, should also exist in simulations since the LSA results presented here are based on the same number of energy and angular grids as in simulations.
However, we find that these noncollective spurious modes do not grow in the linear regime but are suppressed by the angular derivative in the advection term $(1-v_r^2)r^{-1}\partial_{v_r}$.
Thus, their presence does not affect the numerical outcome discussed in the next section.

Second, in contrast to our previous work \cite{xiong2023evolution} where CFI were found to be present in snapshots that are similar to the unattenuated Model I here, the CFI now only exists in Model V in which $\bar\nu_e$'s dominate over $\nu_e$'s due to the significant $Y_e$ attenuation.
We find that the main distinction is related to the newly added NES process in this work.
Without the NES process, the heavy-lepton neutrinos are thermalized less efficiently and have higher mean energy than the electron flavor neutrinos (see the comparison shown in Fig.~\ref{fig:boltztran_cosenu} in Appendix~\ref{sec:nes}).
Given that all flavors of neutrinos are still approximately isotropic at this radius, the distinct behaviors can be understood by examining the quantity
\begin{equation}
    \gamma_{\rm coll} = \frac{\int dE\,dv_r\, [C_{e\mu} (\varrho_{ee}-\varrho_{\mu\mu})-\bar C_{e\mu} (\bar\varrho_{ee}-\bar\varrho_{\mu\mu})]}{\int dE\,dv_r\,(\varrho_{ee}-\varrho_{\mu\mu}-\bar\varrho_{ee}+\bar\varrho_{\mu\mu})},
\end{equation}
which is the imaginary part of the simple formula, Eq.~(13) obtained in Ref.~\cite{xiong2023collisional}.
The denominator, the total ELN, is positive in Models I and II, as well as in all cases of the Ref.~\cite{xiong2023evolution}, but is negative in Model V.
For the numerator, it is mainly dominated by the first term $C_{e\mu} (\varrho_{ee}-\varrho_{\mu\mu})$ due to the larger EA rate associated with $\nu_e$.
In Ref.~\cite{xiong2023evolution}, the heavy-lepton neutrinos dominate over $\nu_e$'s at higher energy in the region where the CFI can develop.
Weighted by the collisional rate $C_{e\mu}$, which is, in general, proportional to $E^2$, the high-energy part of the neutrinos contributes predominantly to the total integral of the numerator and leads to a negative $\gamma_{\rm coll}$.
With a positive denominator, the associated CFI corresponds to the minus type defined in Ref.~\cite{xiong2023collisional}, whose growth rate can be well approximated as $|\gamma_{\rm coll}|$ and as large as a few km$^{-1}$, mainly determined by the collisional rates of the corresponding high-energy part.
On the other hand, in Models I and II considered here, the NES process thermalizes the spectra of heavy-lepton neutrinos so that the mean energy of $\nu_\mu$ is close to $\nu_e$.
As a result, $\gamma_{\rm coll}$ is positive, which implies that the corresponding CFI would be the plus type of Ref.~\cite{xiong2023collisional} if it exists.
However, unlike the minus type solution, the CFI may not be assured for the plus type.
The actual growth rate can be less than $\gamma_{\rm coll}$ or even completely vanish, depending on the neutrino spectral properties and the total ELN (see, e.g., Fig.~2 in Ref.~\cite{xiong2023collisional}).
In Model V, $\bar\nu_e$'s dominate over $\nu_e$'s with a negative total ELN.
With the numerator being positive as in Models I and II, $\gamma_{\rm coll}$ is negative.
As a result, the CFI is of the minus type,
for which $|\gamma_{\rm coll}|$ is a good approximation to indicate the corresponding growth rate.

The above discussion suggests the following intriguing fact.
Although the EA processes in Table~\ref{tab:nu_process} only involve the electron flavor neutrinos from the classical viewpoint, it can damp the coherent state between $\nu_e$ and $\nu_\mu$ in $\nu$QKE so that the CFI does depend on the properties of heavy-lepton neutrinos.
We also note that although CFI is suppressed in our models dominated by $\nu_e$'s, it does not imply that CFI is generally not present in all $\nu_e$-dominated conditions.
Multidimensional CCSN simulations with convection may potentially result in conditions favoring the minus type CFI when the dynamical evolving heavy-lepton neutrinos are less efficiently thermalized by the NES process.

\subsection{Spectrogram}
In this subsection we perform the analysis based on the simulation results using the spectrogram defined in Sec.~\ref{sec:ana-meth-spec} when flavor conversions reach the nonlinear regime and compare the structures to the dispersion relation obtained by the LSA in the previous subsections.
Note that here the model parameters are given in Table~\ref{tab:parameters}, and the $\mathbf{H}_{\nu\nu}$ attenuation is applied to all the models.

\begin{figure}[!hbt]
\includegraphics[width=0.48\textwidth]{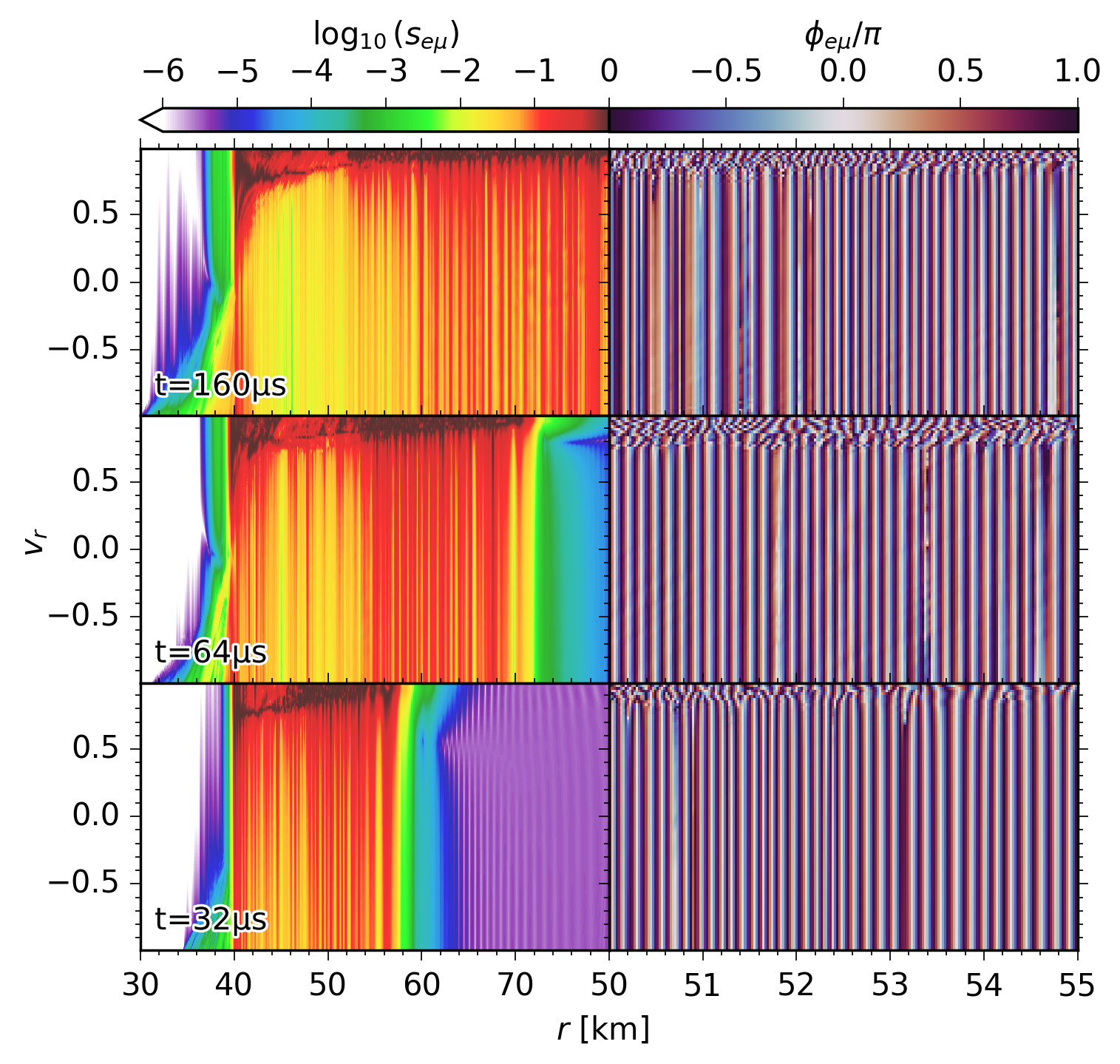}
\caption{\label{fig:2d_model_II} Evolution of the dimensionless ratio of off-diagonal mixing $s_{e\mu}$ in $30$~km~$<r<80$~km (left panel), and its associated phase angle $\phi_{e\mu}$ within a smaller radial range  $50$~km~$<r<55$~km (right panel) for Model II.
}
\end{figure}

Before presenting the spectrogram analysis, we first illustrate the evolution of FFC from the onset of FFI in Fig.~\ref{fig:2d_model_II} for Model II as an example.
Since the maximum growth rate of FFI, Im$(\Omega)_{\rm max}$, in this model peaks at $r\simeq 42$~km, FFI grows faster to reach the nonlinear regime around that radius first.
The left side of Fig.~\ref{fig:2d_model_II} shows that $s_{e\mu}$, the dimensionless ratio representing the off-diagonal flavor mixing introduced in Sec.~\ref{sec:illus_quant}, reaches the nonlinear regime over a wide range from $r\simeq 40$~km to $r\simeq 58$~km by the time of $t=32~\mu$s.
The range over which the nonlinear FFC takes place further widens to $r\simeq 70$~km at $t=64~\mu$s and extends to the outer boundary at a much later time $t=160~\mu$s.

\begin{figure*}[!hbt]
\includegraphics[width=0.32\textwidth]{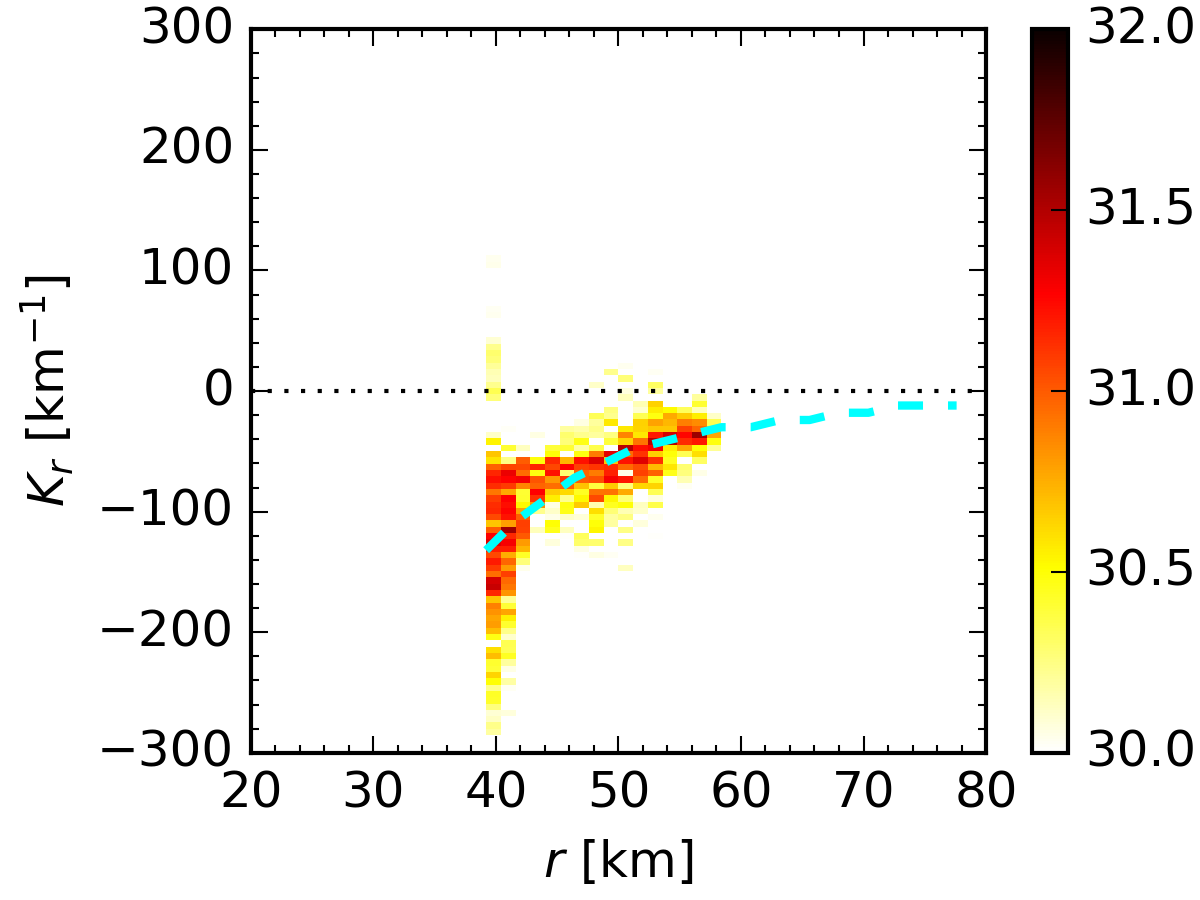}
 \llap{\parbox[b]{2.3in}{\small (a) Model II, $32~\mu$s\\\rule{0ex}{1.4in}}}
\includegraphics[width=0.32\textwidth]{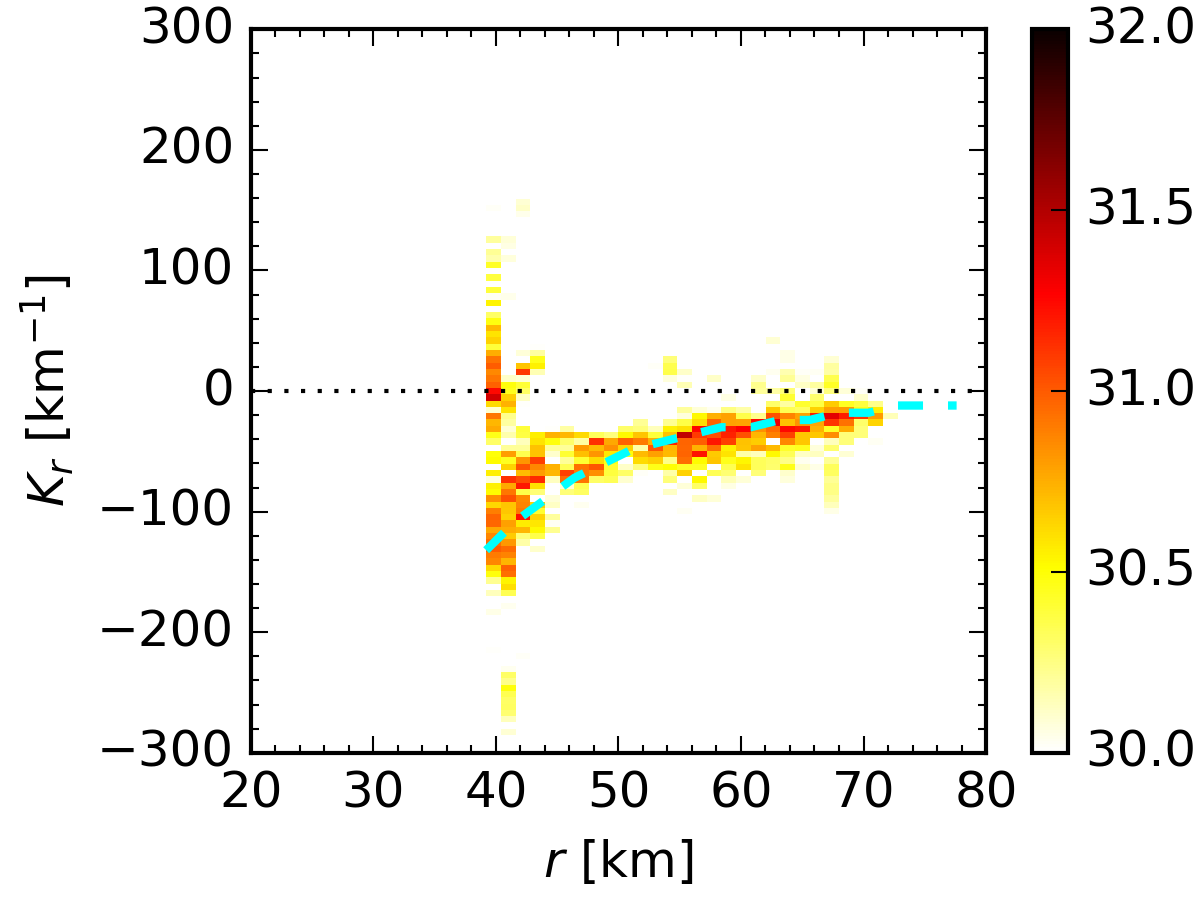}
 \llap{\parbox[b]{2.3in}{\small (b) Model II, $64~\mu$s\\\rule{0ex}{1.4in}}}
\includegraphics[width=0.32\textwidth]{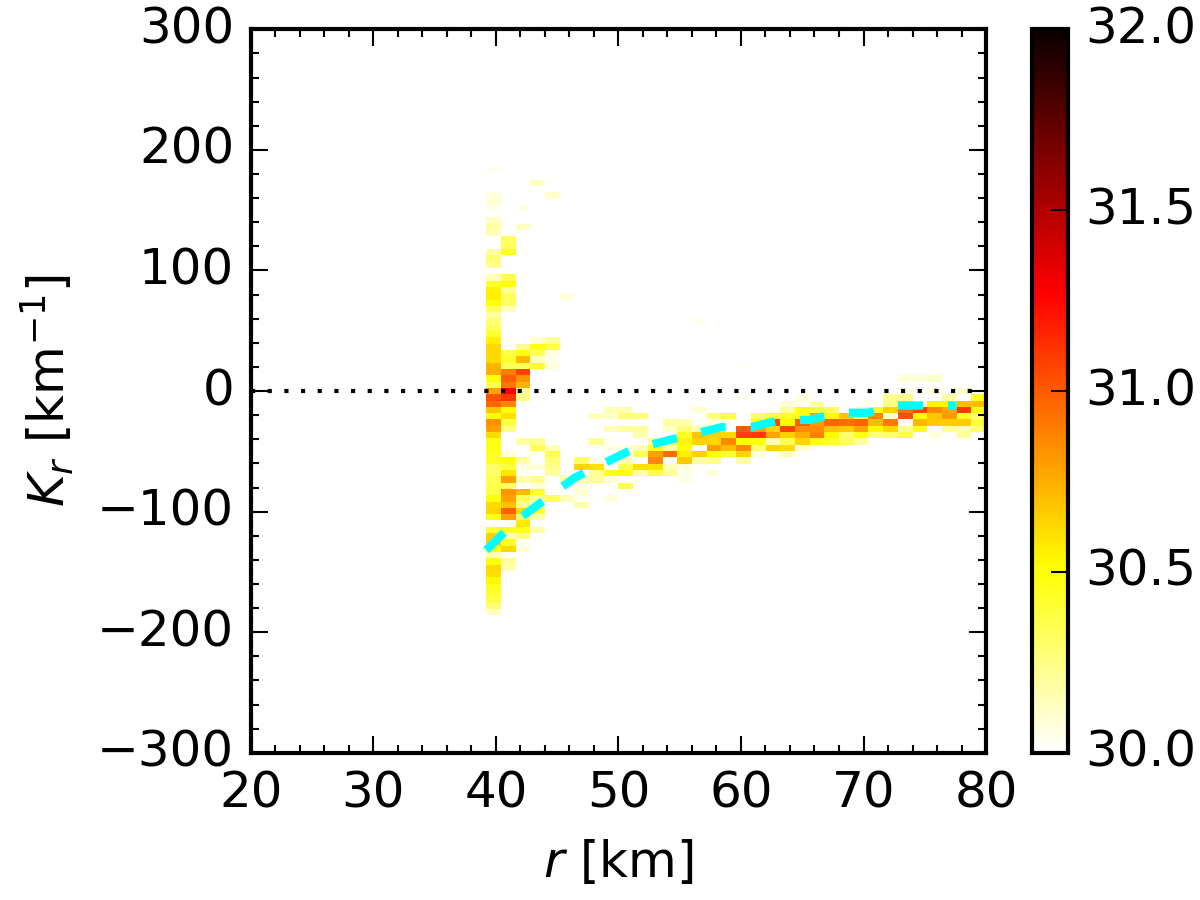}
 \llap{\parbox[b]{2.3in}{\small (c) Model II, $160~\mu$s\\\rule{0ex}{1.4in}}}
\includegraphics[width=0.32\textwidth]{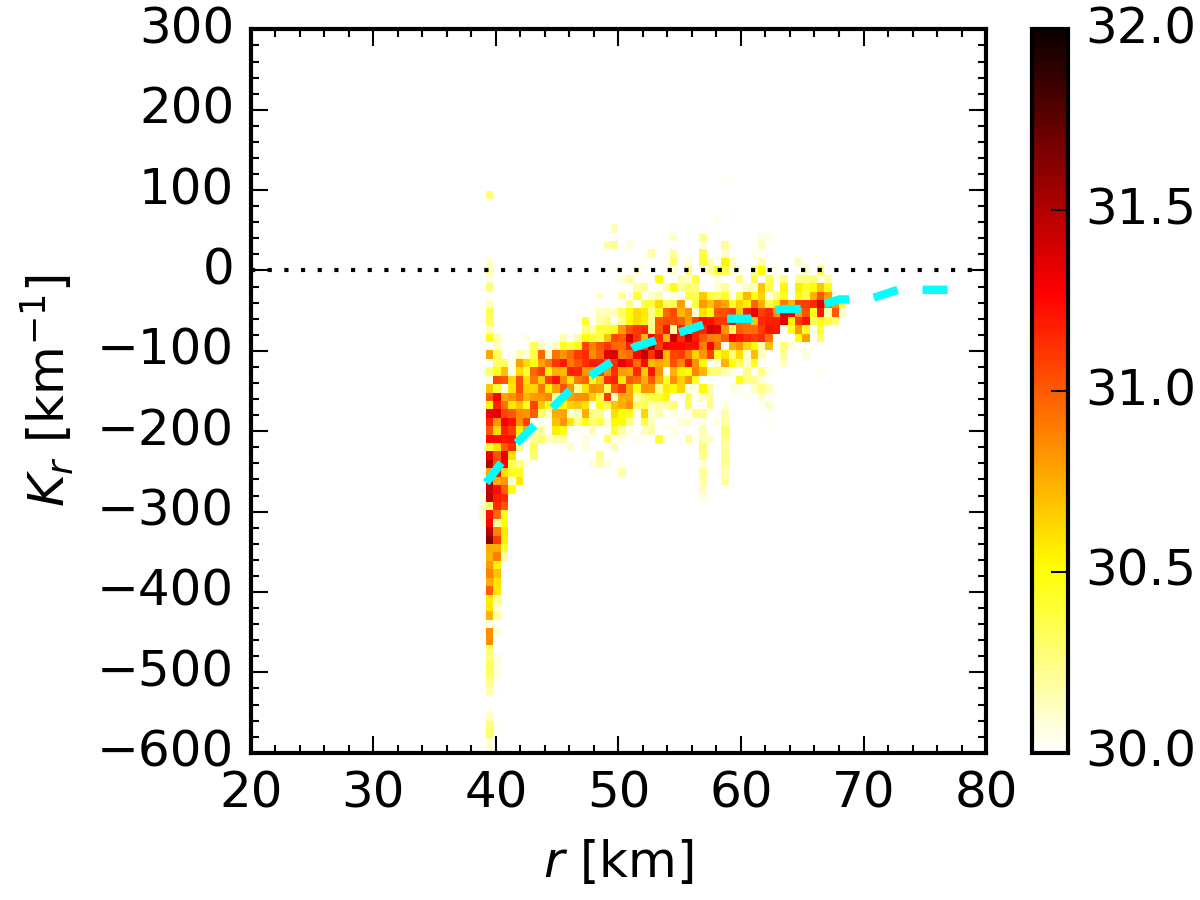}
 \llap{\parbox[b]{2.3in}{\small (d) Model IIa2, $32~\mu$s\\\rule{0ex}{1.4in}}}
\includegraphics[width=0.32\textwidth]{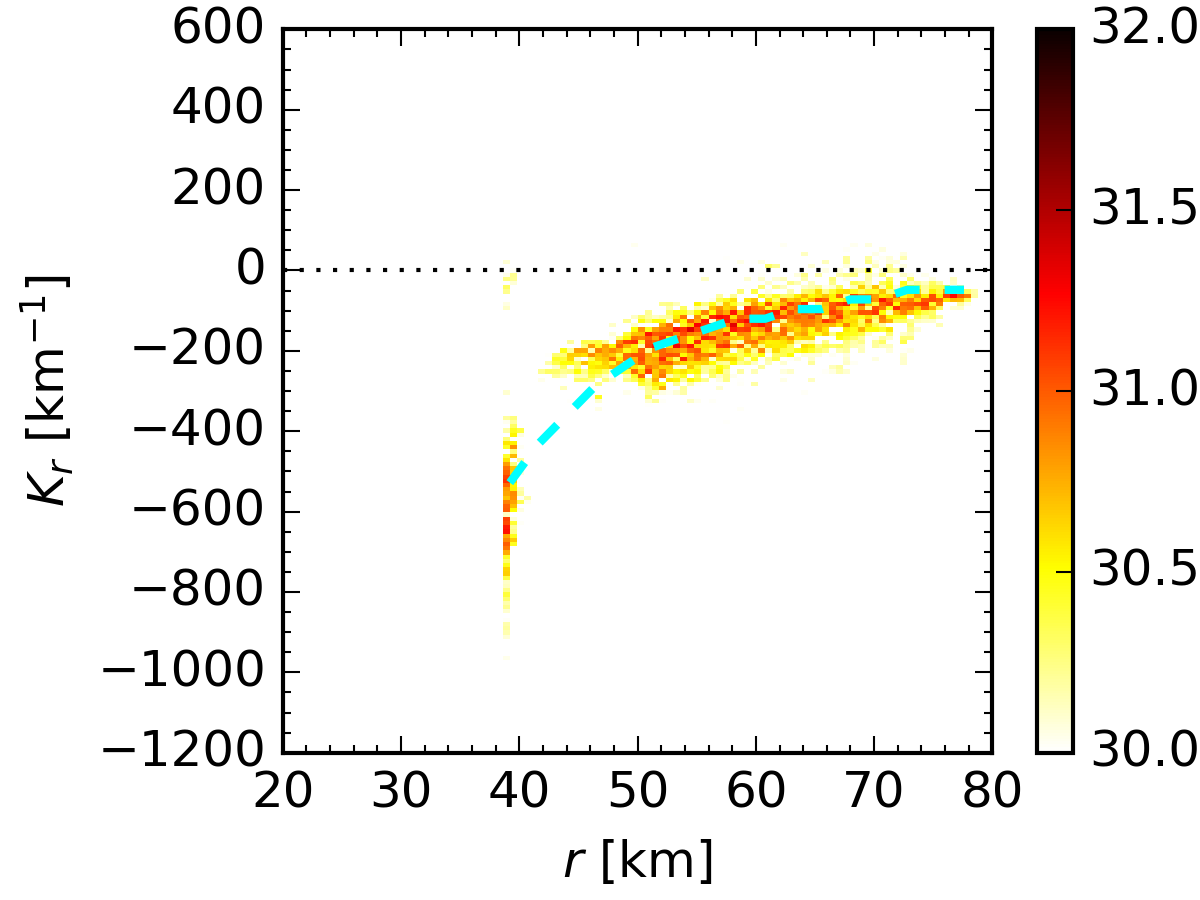}
 \llap{\parbox[b]{2.2in}{\small (e) Model IIa4, $32~\mu$s\\\rule{0ex}{1.4in}}}
\includegraphics[width=0.32\textwidth]{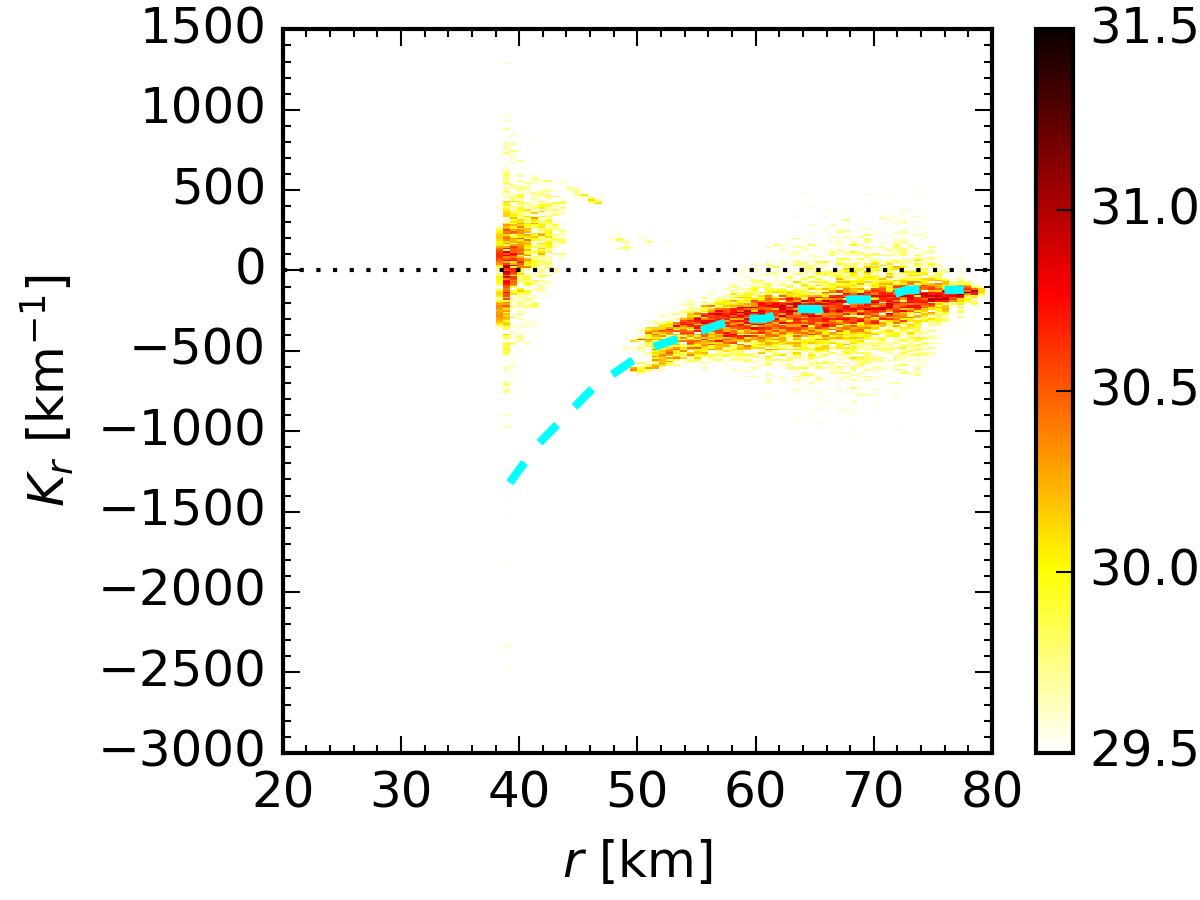}
 \llap{\parbox[b]{2.2in}{\small (f) Model IIa10, $32~\mu$s\\\rule{0ex}{1.4in}}}
\includegraphics[width=0.32\textwidth]{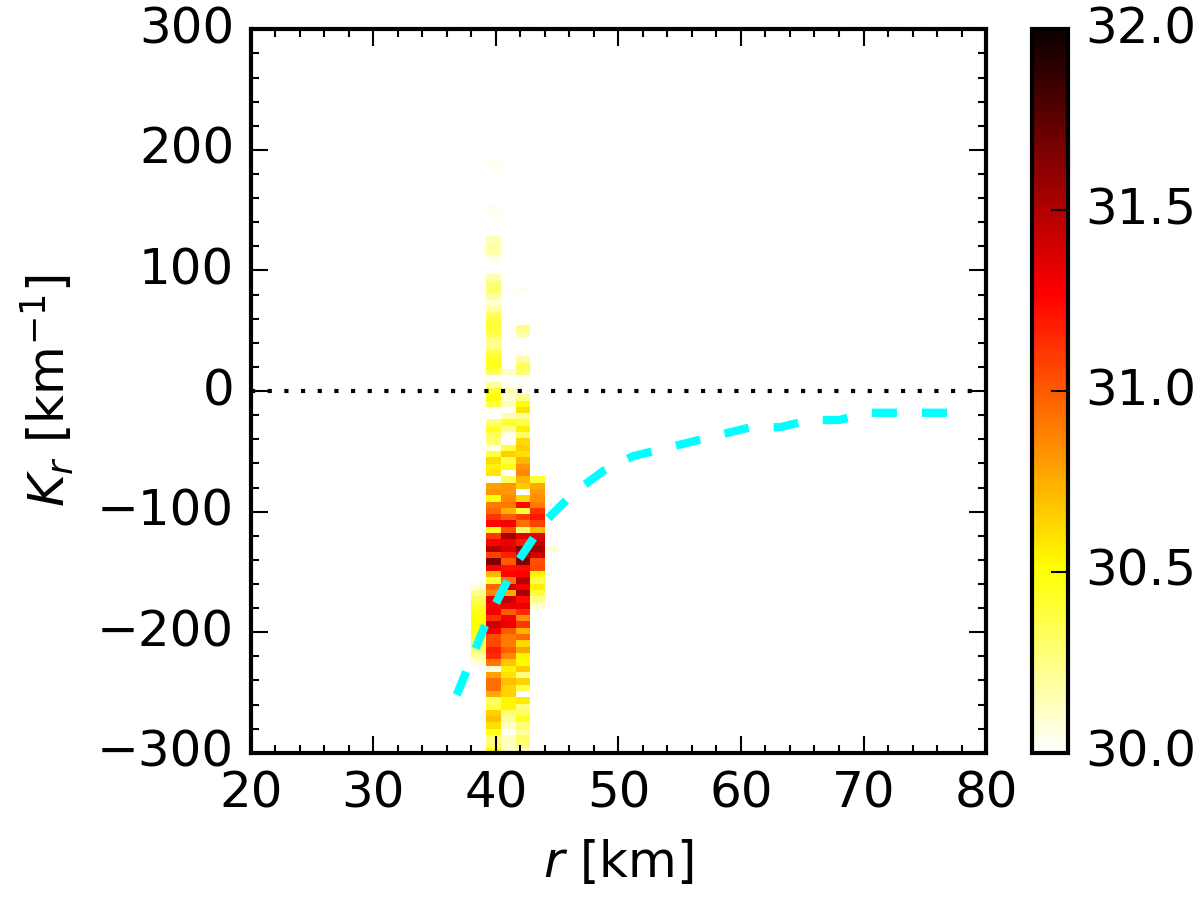}
 \llap{\parbox[b]{2.3in}{\small (g) Model V, $1.6~\mu$s\\\rule{0ex}{1.4in}}}
\includegraphics[width=0.32\textwidth]{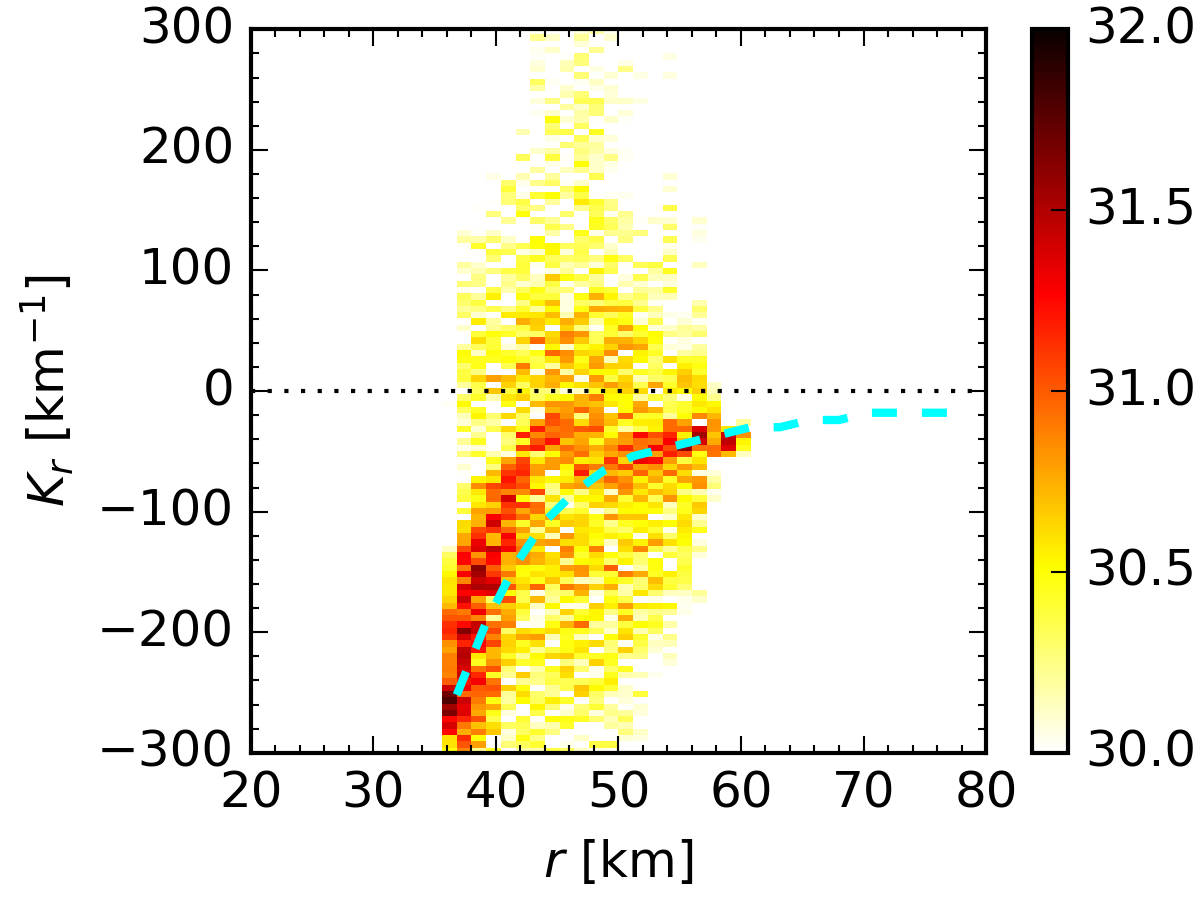}
 \llap{\parbox[b]{2.3in}{\small (h) Model V, $6.4~\mu$s\\\rule{0ex}{1.4in}}}
\includegraphics[width=0.32\textwidth]{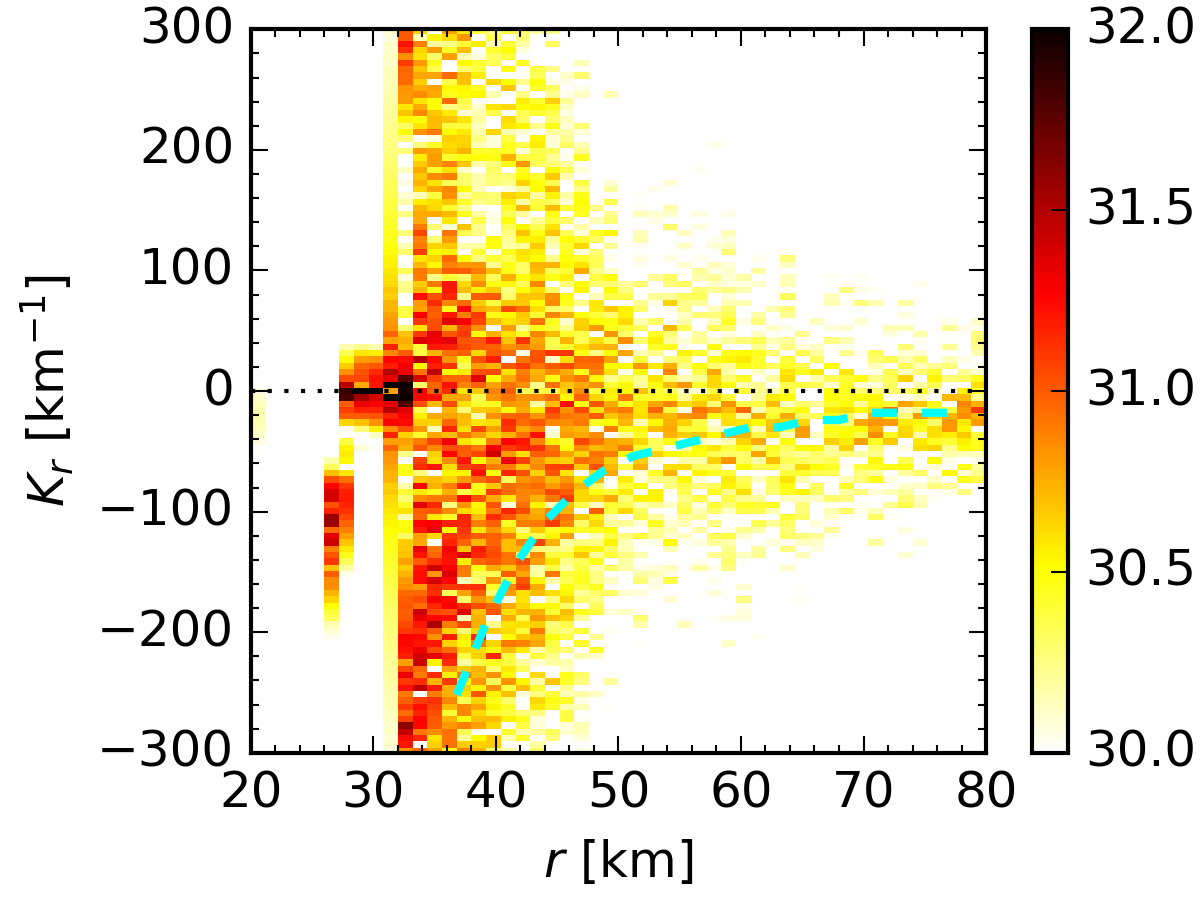}
 \llap{\parbox[b]{2.3in}{\small (i) Model V, $32~\mu$s\\\rule{0ex}{1.4in}}}
\includegraphics[width=0.32\textwidth]{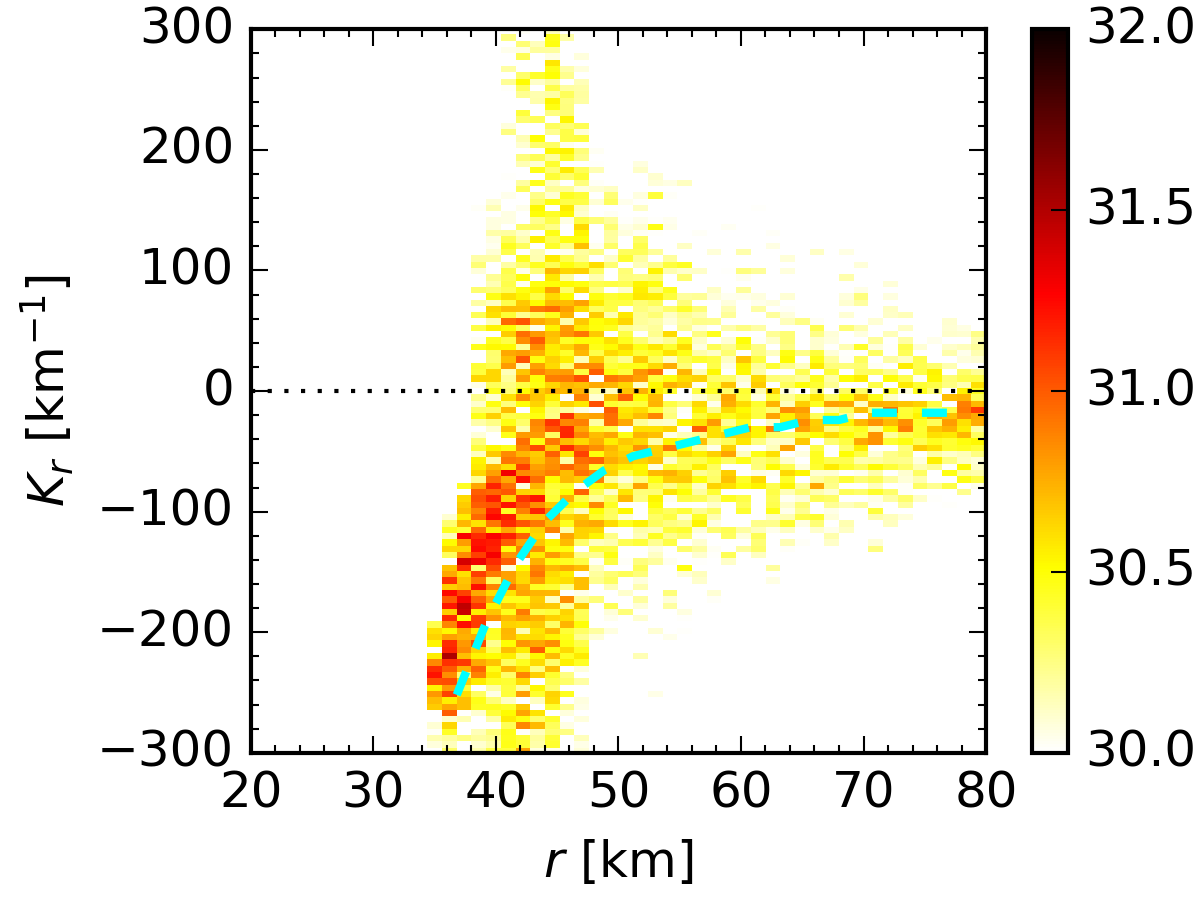}
 \llap{\parbox[b]{2.3in}{\small (j) Model Vc0, $32~\mu$s\\\rule{0ex}{1.4in}}}
\includegraphics[width=0.32\textwidth]{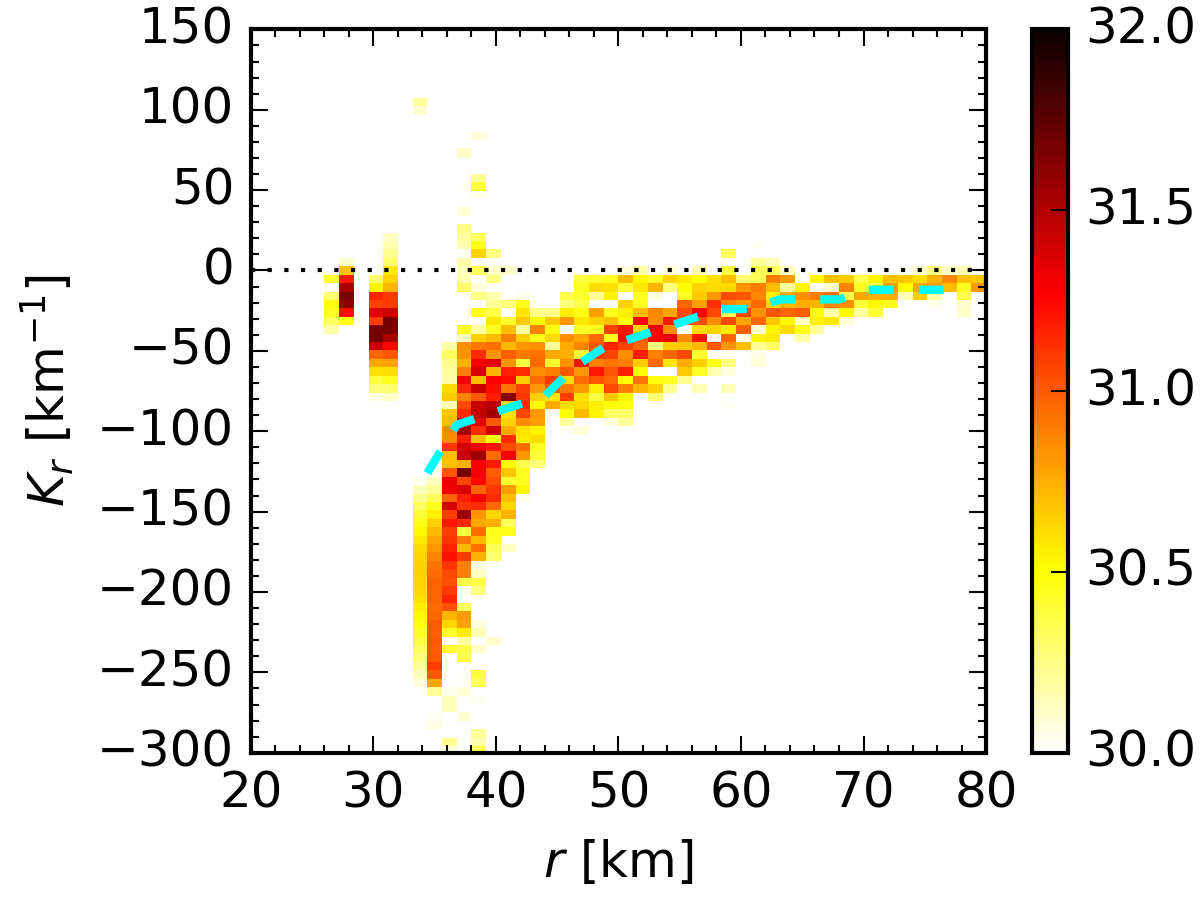}
 \llap{\parbox[b]{2.3in}{\small (k) Model III, $160~\mu$s\\\rule{0ex}{1.4in}}}
\includegraphics[width=0.32\textwidth]{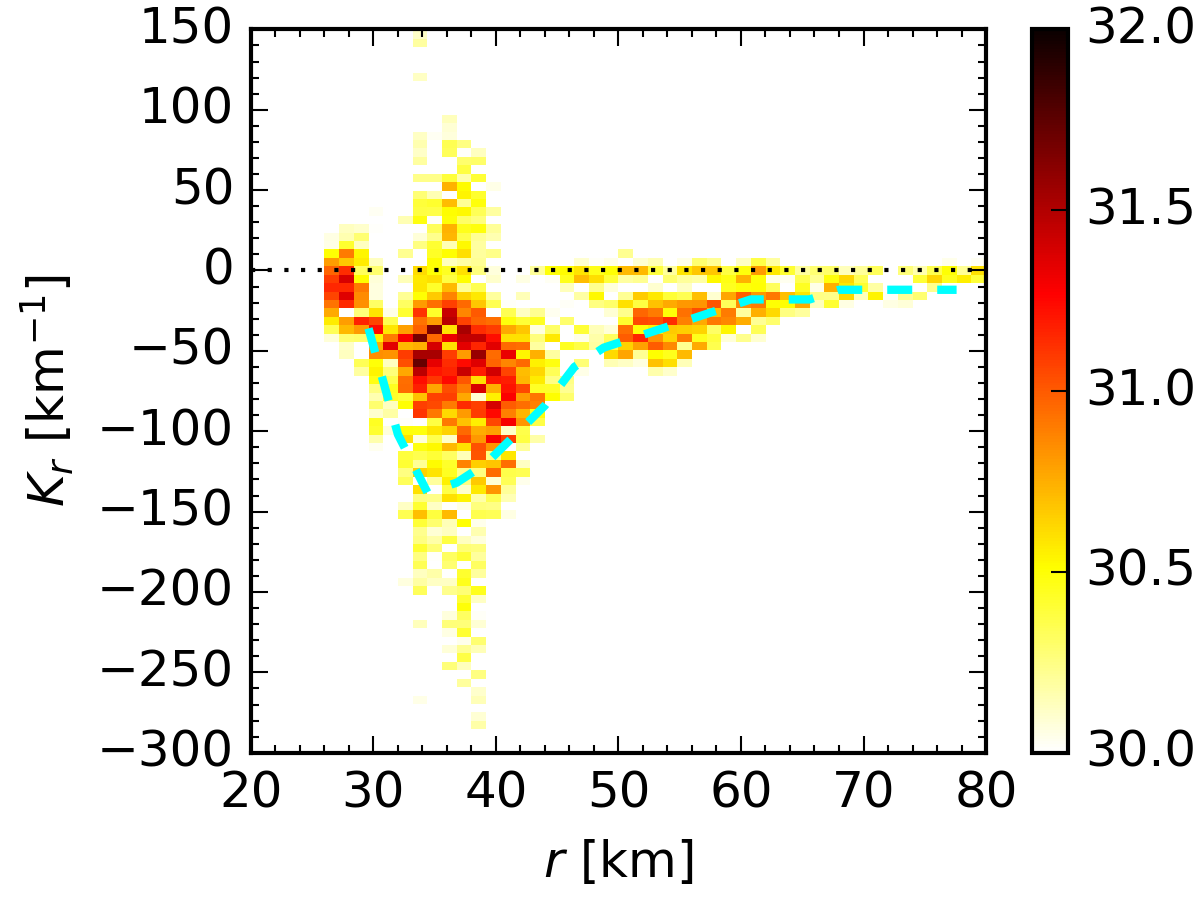}
 \llap{\parbox[b]{2.3in}{\small (l) Model IV, $160~\mu$s\\\rule{0ex}{1.4in}}}
\caption{\label{fig:sp} Spectrograms with respect to $K_r$ and $r$ for different models at different times (a)--(l). The color bar is for the Fourier transformed magnitude $\log_{10}\big|\mathcal F_{r,K_r} [{\rm cm}^{-3}]\big|$. The cyan dashed curve shows the $K_r$ with maximum growth rates calculated from LSA.
}
\end{figure*}

On the right side in Fig.~\ref{fig:2d_model_II}, we also show the radial shape of the complex quantum phase $\phi_{e\mu}$ defined in Sec.~\ref{sec:illus_quant} in $50$~km~$< r< 55$~km to illustrate the coherent feature of FFC.
The phase angle $\phi_{e\mu}$ not only oscillates rapidly in time during FFCs, but also develops coherent spatial-dependent oscillatory patterns.
The oscillatory pattern appears to be irregular for $v_r\gtrsim 0.8$ close to the angular crossing.
On the other side, the pattern is highly correlated for a wide range of $v_r$ due to the collective nature of FFC.
The typical oscillation length scale of the synchronized phase angle is $\lesssim 0.2$~km, broadly consistent with the scale determined by the maximally unstable $K_r$ shown in Fig.~\ref{fig:Kr}(b) in the same radial range.
This simple comparison suggests that the characteristic length scale determined by the dispersion relation in the linear regime correlates with the coherent length scale at late times when FFCs are in the nonlinear or asymptotic regime~\cite{duan2021flavor,richers2021neutrino}, which will be further discussed below with the spectrograms.
We also note that this spatial variation can lead to a substantial contribution to the flavor evolution equation through the advection term $v_r \partial_r$ and thus participate in determining the outcome of flavor evolution.

The spectrograms shown in Fig.~\ref{fig:sp}(a--c) compare the Fourier transformed magnitudes $\mathcal F_{r,K_r}$ at the same three time frames in Fig.~\ref{fig:2d_model_II} with the curve of $K_r$ corresponding to the maximum growth rate from LSA.
At $t=32~\mu$s, the dominant wave number $K_r$ of the off-diagonal flavor mixing in $\mathcal F_{r,K_r}$ generally aligns well with the maximally unstable $K_r$ values derived with LSA.
Relatively small differences of $\simeq 20$--30~km$^{-1}$ that exist are mainly related to the change of ELN distribution due to flavor conversions in simulation, while the LSA results are based on unoscillated neutrino profiles.
We note that near $r=40$~km, the range of dominant $K_r$ widens over time to enclose the homogeneous modes with $K_r\approx 0~{\rm km}^{-1}$.
The amplitude of the off-diagonal flavor mixing between $r\approx 42$~km and 50~km decreases at a later time $t=160~\mu$s, with $\mathcal F_{r,K_r}\lesssim 10^{30}~{\rm cm}^{-3}$.
For $r\gtrsim 50$~km, similar to the case with a shallow angular crossing studied under the periodic boundary \cite{richers2022code}, our result suggests that a steady wavelike pattern along the radial direction can also survive in a more realistic setting without adopting the periodic boundary condition.

The radial-dependent coherent pattern of the flavor mixing remains qualitatively similar when different attenuation factors are used.
Figure~\ref{fig:sp}(d--f) compares the spectrograms for Models IIa2, IIa4, and IIa10 at the same time $t=32~\mu$s.
The features of the spectrograms in all three models again agree well with the results derived from the LSA.
The patterns also show a good scalability as discussed before.
Because models with less attenuation evolve faster, they reach the asymptotic pattern at earlier times as shown in Fig.~\ref{fig:sp}(e) and (f).

In Model V, because the maximum growth rates are generally larger than those in Model II, flavor conversions occur faster.
Figure~\ref{fig:sp}(g) shows that a significant flavor mixing with $\mathcal F_{r,K_r}>10^{31}~{\rm cm}^{-3}$ is attained for $r\approx 42$~km at $t=1.6~\mu$s.
At $t=6.4~\mu$s shown in Fig.~\ref{fig:sp}(h), the FFC in the region from $\approx 36$~km to $60$~km evolves to the nonlinear regime.
The dominant wave numbers in $r\lesssim 40$~km and $r\gtrsim 50$~km still qualitatively follow  the curve of maximum-growth-rate $K_r$ from the LSA.
However, in the region between 40 and 50~km, the spectrogram covers a much wider wave number mode, spreading from $K_r\simeq -300~{\rm km}^{-1}$ to $300~{\rm km}^{-1}$.
This becomes even more evident at $t=32~\mu$s shown in Fig.~\ref{fig:sp}(i).
Also at $t=32~\mu$s, the emergence of CFI leads to flavor mixing between $r\approx 27$~km and 33~km with the unstable modes centered on $K_r\simeq 0~{\rm km}^{-1}$.

We use Model Vc0 to disentangle the effect due to the interplay between the CFI and the FFI.
Figure~\ref{fig:sp}(j) shows that the asymptotic spectrogram of Model Vc0 at the same time $t=32~\mu$s generally follows the trend of $K_r$ from LSA.
The dominant $K_r$ of the spectrogram spans from $K_r\simeq -250~{\rm km}^{-1}$ at $r\approx 36$~km to $K_r\simeq -30~{\rm km}^{-1}$ at $r\approx 46$~km.
Between $r\simeq 40$~km and $60$~km, the range of $K_r$ in the spectrogram becomes wider, consistent with similar features shown in Fig~\ref{fig:sp}(h) and (i) in the same radial range.

Finally, we also show in Fig.~\ref{fig:sp}(k) and (l) the asymptotic spectrograms in Models III and IV, respectively.
For both models, the radial shape of the dominant $K_r$ modes again agree qualitatively with the LSA results, although quantitative differences do exist.
In any case, our spectrogram analysis reveals that asymmetric patterns with respect to $K_r$ generally exist, which suggests that the outcome of FFCs and the flavor waves are not necessarily homogeneous.
The overall development of small-scale structures reported here also qualitatively agrees with the findings in local simulations adopting periodic boundary conditions.

\begin{figure}[!hbt]
\includegraphics[width=0.48\textwidth]{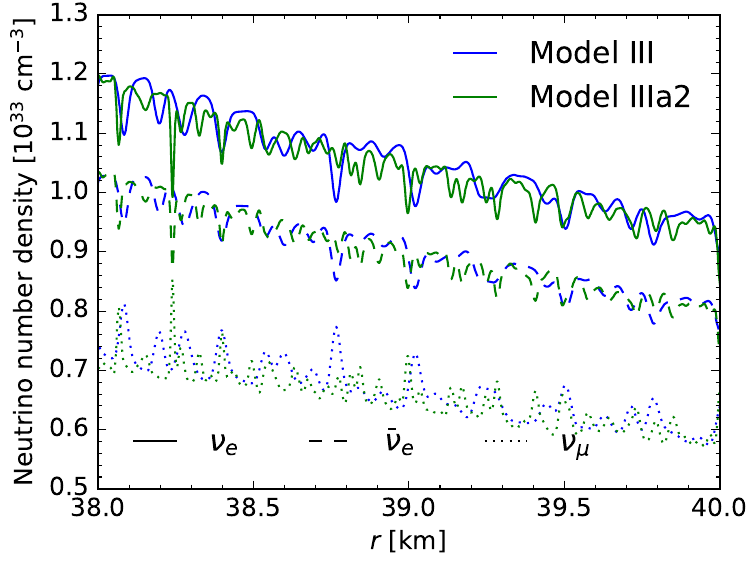}
\caption{\label{fig:small_scale_structure} Radial profiles of neutrino number densities between $r=38$~km and 40~km at $t=64~\mu$s in Models III and IIIa2.
}
\end{figure}

In addition to the off-diagonal flavor mixing, the diagonal elements of the density matrices associated with the neutrino number density also contain the same small-scale structures.
In Fig.~\ref{fig:small_scale_structure}, we show the neutrino number densities of Models III and IIIa2 between $r=38$~km and 40~km for illustration.
Clearly, oscillatory patterns exist, in general, for all three different neutrino species.
In Model III, its typical oscillatory wavelength is $\simeq 0.07$~km, corresponding to a $K_r\simeq 90~{\rm km}^{-1}$, which is consistent with the dominant mode of spectrogram in Fig.~\ref{fig:sp}(k).
With a less attenuation in Model IIIa2, the oscillatory length scale becomes correspondingly smaller.
The presence of small-scale structures may lead to implications associated with the refractive energy and the potential self-thermalization of the neutrino gas~\cite{johns2023thermodynamics,fiorillo2024inhomogenous}.

\section{Evolution of fast flavor conversions}
\label{sec:evolution}

\begin{figure*}[!hbt]
\includegraphics[width=0.48\textwidth]{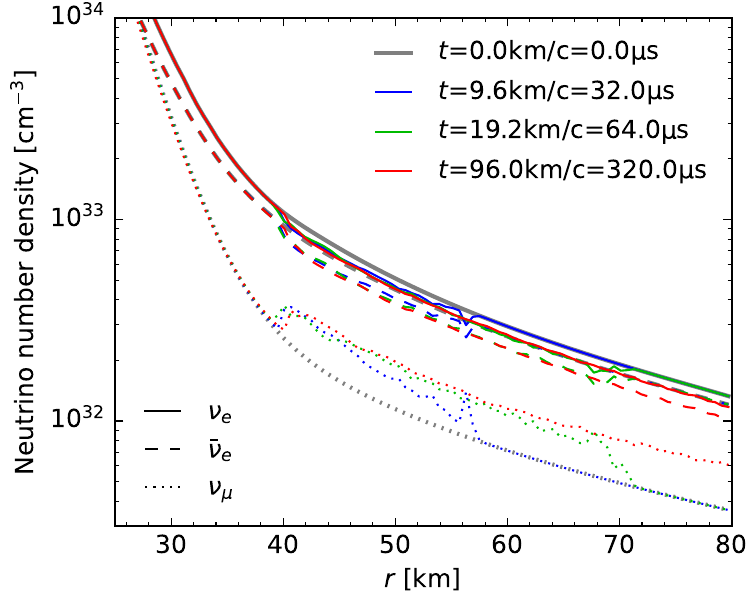}
\llap{\parbox[b]{5.0in}{\small (a)\\\rule{0ex}{2.4in}}}
\includegraphics[width=0.48\textwidth]{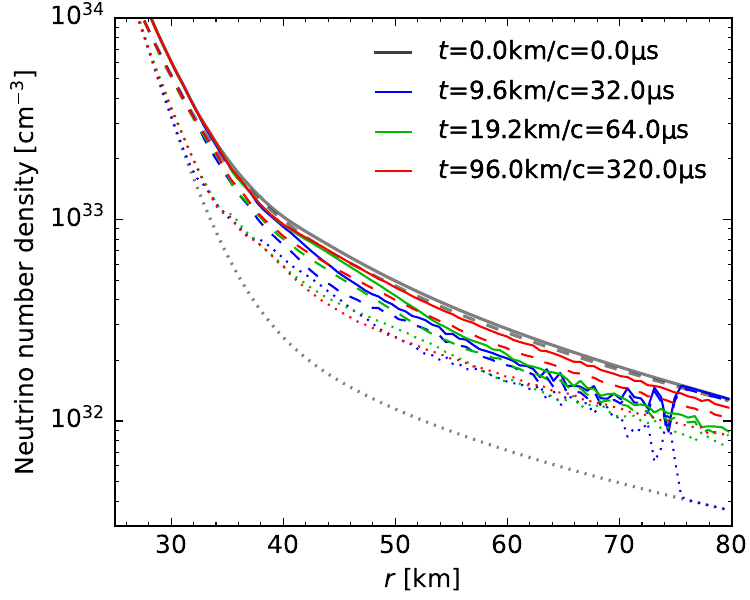}
\llap{\parbox[b]{5.0in}{\small (b)\\\rule{0ex}{2.4in}}}
\includegraphics[width=0.48\textwidth]{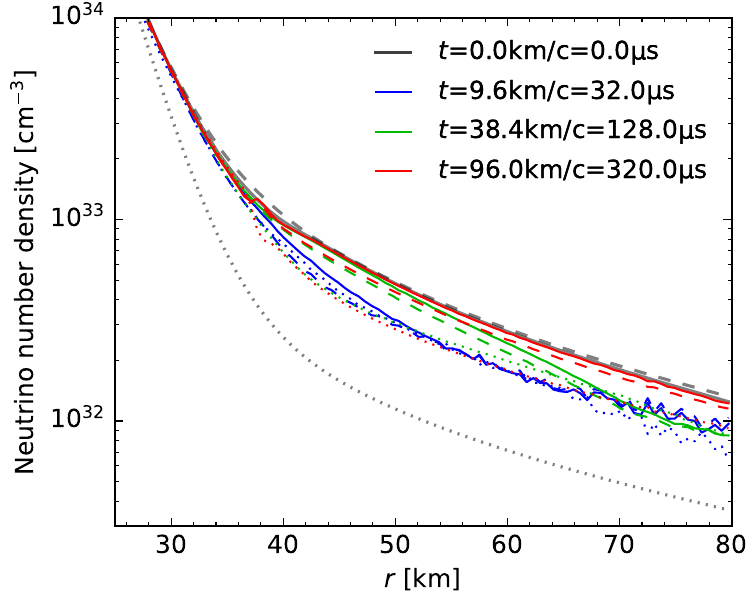}
\llap{\parbox[b]{5.0in}{\small (c)\\\rule{0ex}{2.4in}}}
\includegraphics[width=0.48\textwidth]{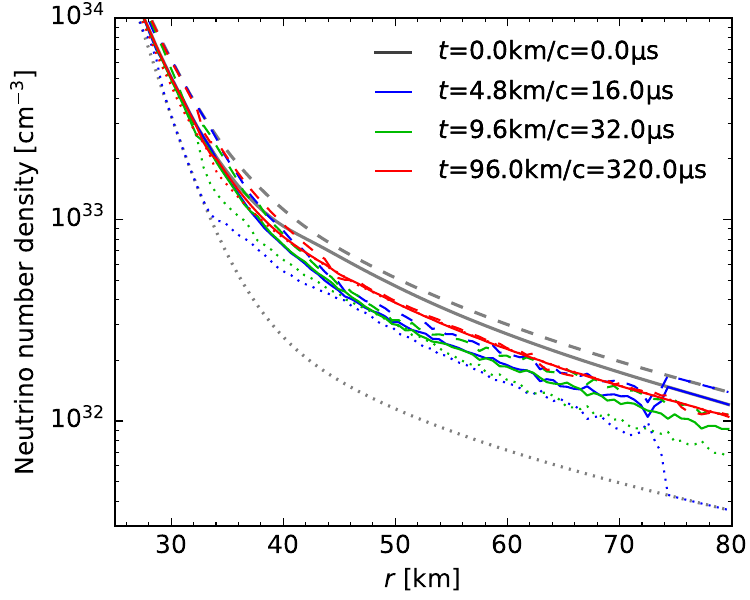}
\llap{\parbox[b]{5.0in}{\small (d)\\\rule{0ex}{2.4in}}}
\includegraphics[width=0.48\textwidth]{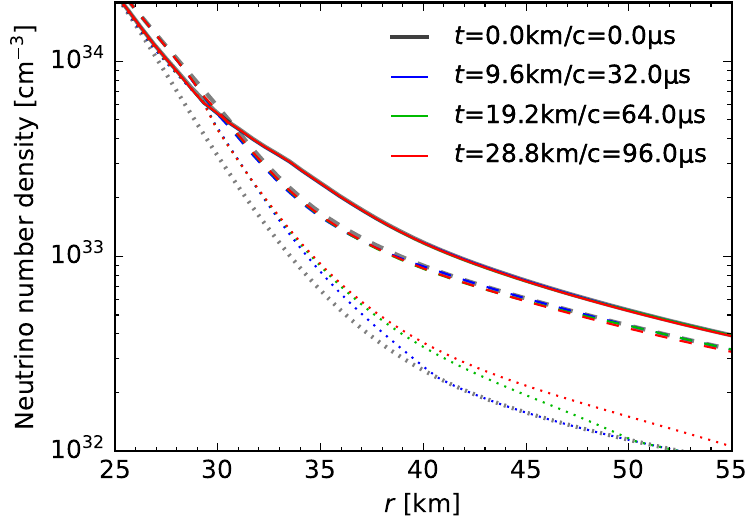}
\llap{\parbox[b]{5.0in}{\small (e)\\\rule{0ex}{2.1in}}}
\includegraphics[width=0.48\textwidth]{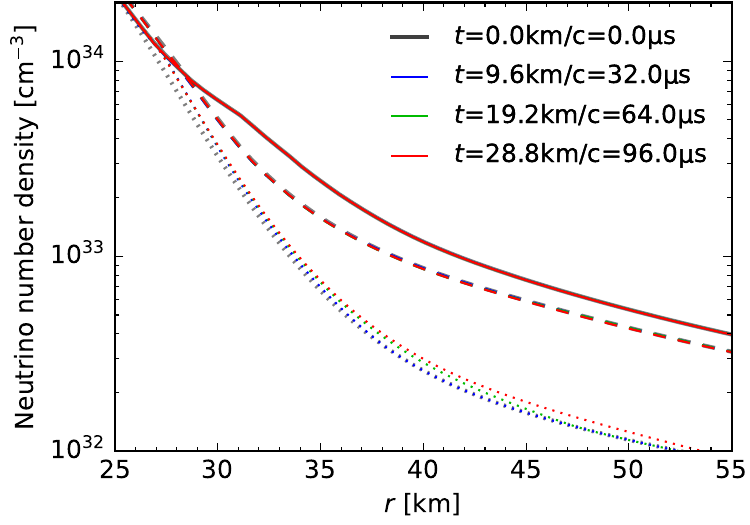}
\llap{\parbox[b]{5.0in}{\small (f)\\\rule{0ex}{2.1in}}}
\caption{\label{fig:rn} Radial profiles of neutrino number densities for $\nu_e$, $\bar\nu_e$, and $\nu_\mu$ in Models II--VII (a--f).
}
\end{figure*}

\begin{figure*}[!hbt]
\includegraphics[width=0.32\textwidth]{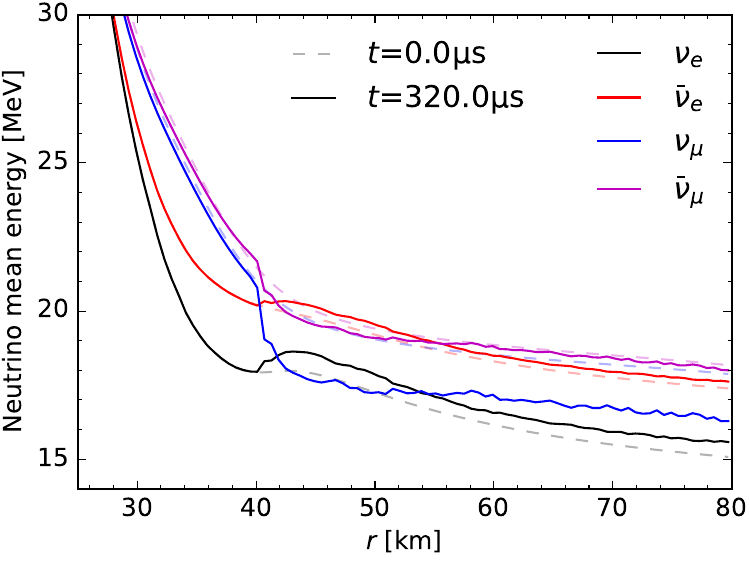}
\llap{\parbox[b]{3.5in}{\small (a)\\\rule{0ex}{1.4in}}}
\includegraphics[width=0.32\textwidth]{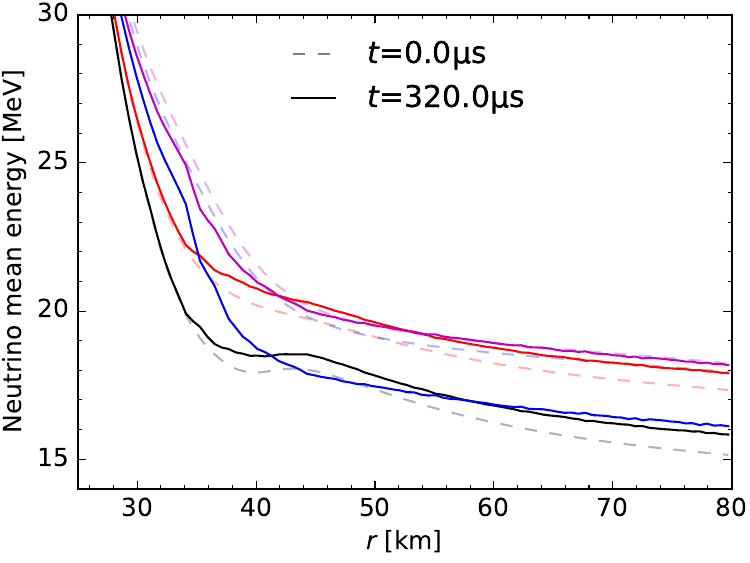}
\llap{\parbox[b]{3.5in}{\small (b)\\\rule{0ex}{1.4in}}}
\includegraphics[width=0.32\textwidth]{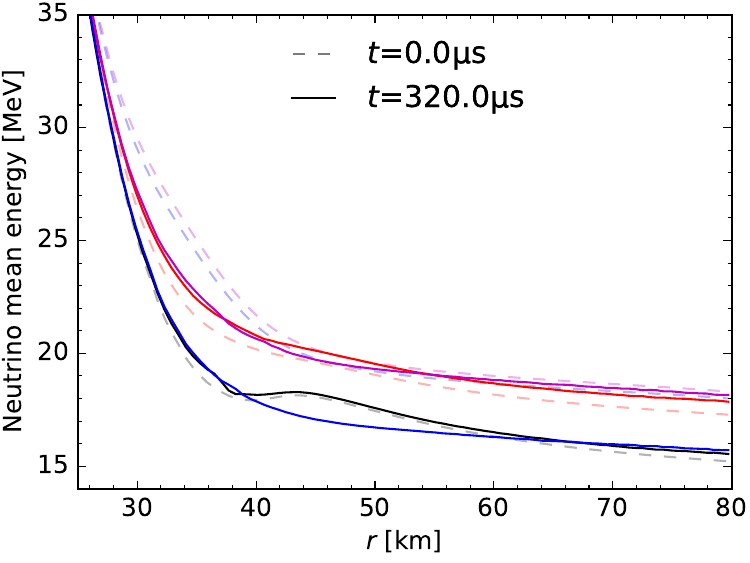}
\llap{\parbox[b]{3.5in}{\small (c)\\\rule{0ex}{1.4in}}}
\includegraphics[width=0.32\textwidth]{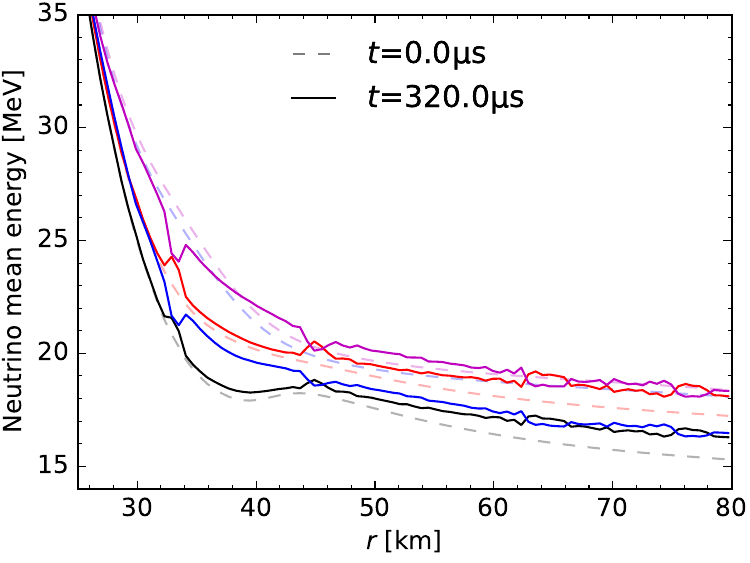}
\llap{\parbox[b]{3.5in}{\small (d)\\\rule{0ex}{1.4in}}}
\includegraphics[width=0.32\textwidth]{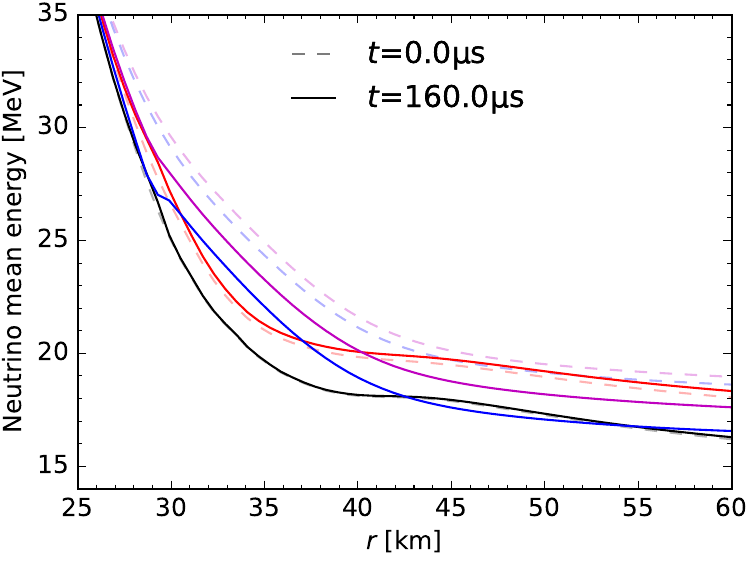}
\llap{\parbox[b]{3.5in}{\small (e)\\\rule{0ex}{1.4in}}}
\includegraphics[width=0.32\textwidth]{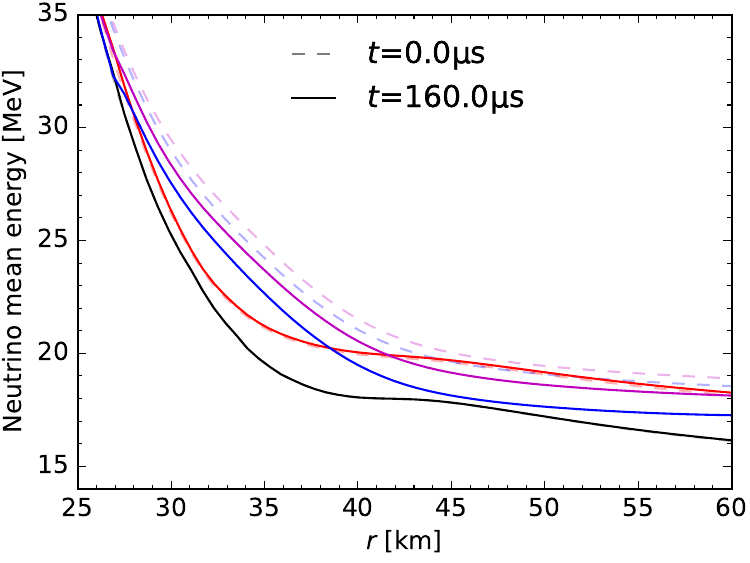}
\llap{\parbox[b]{3.5in}{\small (f)\\\rule{0ex}{1.4in}}}
\caption{\label{fig:re} Radial profiles of neutrino mean energies for $\nu_e$, $\bar\nu_e$, $\nu_\mu$, and $\bar\nu_\mu$ in Models II--VII (a--f).
}
\end{figure*}

In this section, we will present more details regarding the evolution of the FFC and its impact on key properties of neutrinos.
Figures~\ref{fig:rn} and \ref{fig:re} show the neutrino number density and mean energy of different species for Models II--VII at different times.
For the number densities, because the initial values of $\nu_\mu$ and $\bar\nu_\mu$ are rather close to each other, and given that the fast pairwise flavor conversion produces equal amount of $\nu_\mu$ and $\bar\nu_\mu$, only the $\nu_\mu$ ones are shown in Fig.~\ref{fig:rn}.
For the average energies, all fours species are shown explicitly in Fig.~\ref{fig:re}.
Also noted is that the values shown in these two figures are coarse-grained ones, computed by averaging over a size of 0.6~km from the original simulation outputs, for the purpose of capturing the global trend.
Below, we will first compare the similarities and differences between those models.
Different adopted schemes such as the number of neutrino flavors, the amount of attenuation on $\mathbf H_{\nu\nu}$, and the size of the vacuum term will then be discussed.

\subsection{Model II}
We start with Model II in which the angular crossing is the shallowest.
Prior to the occurrence of the FFC, the number densities of all neutrino species decrease smoothly with radius, following a hierarchy of $n_{\nu_e}\gtrsim n_{\bar\nu_e}> n_{\nu_\mu}$ in the radial range shown in Fig.~\ref{fig:rn}(a).
Because of the FFI discussed earlier, the FFC takes place and leads to the pairwise conversion from $\nu_e$ and $\bar\nu_e$ to muon flavor neutrinos.
At $t=32~\mu$s, the region between $r\approx 40$~km and $r\approx 57~$km is influenced by FFC.
The impact of FFC extends to $r\approx 70$~km at $t=64~\mu$s and eventually reaches the outer boundary at a later time.
Notice that there are ``spiky'' features just behind the outermost range affected by FFC, e.g., at $r\approx 56$~km at $t=32~\mu$s and $r\approx 68$~km at $t=64~\mu$s.
Those are consistent with what is reported in local simulations adopting periodic boundaries where a larger amount of flavor conversions can exist as a transient phenomenon before the system settles down to the asymptotic state.

In Model II, $n_{\nu_e}$ and $n_{\bar\nu_e}$ in the region affected by FFC are reduced by $\simeq 10\%$--$20\%$, while $n_{\nu_\mu}$ increases by $\simeq 50\%$--$70\%$.
Once the asymptotic state in a certain radial range is reached, the number densities do not vary much anymore, illustrated by comparing the curves at $t=64~\mu$s and at $t=320~\mu$s in $40$~km$\lesssim r\lesssim 68$~km.
Clearly, flavor equilibration is not achieved in Model II.
This is because the FFC here is restricted within a limited angular range (see e.g., Fig.~\ref{fig:2d_model_II} and later in Fig.~\ref{fig:an_iz50}).

For the mean energies of $\nu_\mu$ and $\bar\nu_\mu$, different from the similar evolution of $n_{\nu_\mu}$ and $n_{\bar\nu_\mu}$, they can be affected very differently by FFCs.
Figure~\ref{fig:re}(a) shows that $\langle E_{\nu_\mu} \rangle$ is reduced substantially to $\approx 16$--17~MeV for $r>40$~km, while $\langle E_{\bar\nu_\mu} \rangle$ remains around $18$--20~MeV.
For $\nu_e$ and $\bar\nu_e$, their mean energies increase by less than 1~MeV.

\subsection{Models III and IV}
\label{sec:III_IV}
For Models III and IV, which possess deeper angular crossings in terms of $v_r$, the ELN crossings also appear at radii deeper inside the neutrinospheres (see Fig.~\ref{fig:initial_ELN}).
We show in Fig.~\ref{fig:rn}(b) and (c) the evolution of the neutrino number densities.
At $t=32~\mu$s, FFCs take place between $r\approx 32$ and $r\approx 76$~km in Model III, and from $r\approx 28$~km utill the outer boundary in Model IV.
The deeper angular crossings in both models lead to an increased amount of the converted neutrinos from the electron to the muon flavor than that in Model II.
Particularly, the nearly equal number densities in 50~km$\lesssim r\lesssim 70$~km of all different flavors indicate that flavor equilibration is approximately achieved in that radial range in both models.
The zigzag pattern around $r=70$--72~km in Model III is associated with the transient phenomenon from the linear to nonlinear regime aforementioned in Model II.

Different from Model II, the existence of the ELN angular crossings at deeper radii inside the neutrino spheres of $\nu_e$ enables an intriguing feedback effect between the FFC and the collisional EA processes after the initial phase of FFC in both models.
In Model III, the number densities of $\nu_e$ and $\bar\nu_e$ around $r\simeq 45$~km at $t=64~\mu$s increase by a similar amount from those at $t=32~\mu$s as shown in Fig.~\ref{fig:rn}(b).
This is because the EA collisions around the neutrinospheres of both $\nu_e$ and $\bar\nu_e$ act to restore their number densities to the stationary values prior to the occurrence of FFC.
The increase of $n_{\nu_e}$ and $n_{\bar\nu_e}$ continues and also affects regions outside the neutrinosphere.
For instance, at $t=160~\mu$s, $n_{\nu_e}$ and $n_{\bar\nu_e}$ become closer to their initial values for $r\gtrsim 40$~km.
Similar behaviors also appear in Model IV shown in Fig.~\ref{fig:rn}(c).
Near flavor equilibration is achieved at $t=32~\mu$s initially, followed by the increase of $n_{\nu_e}$ and $n_{\bar\nu_e}$ locally in $40$~km$\lesssim r\lesssim 66$~km at $t=128~\mu$s, and later at $t=320~\mu$s the restoration of $n_{\nu_e}$ and $n_{\bar\nu_e}$ back to the initial values for $r\gtrsim 40$~km.
The restoration is stronger than in Model III, particularly for $n_{\nu_e}$, which almost overlaps with the unoscillated level.
We note that because this feedback involves mostly the collisional processes and advection, the associated timescale of the number density increase is not related to the conversion rate of FFC.
For instance, in Model IV, the range affected by this feedback effect extends from $r\simeq 40$~km at $t=32~\mu$s to $r\approx 66$~km at $t=128~\mu$s, consistent with the corresponding neutrino propagation time, which is a few tens of km/$c$.

In addition to the restoration of $\nu_e$ and $\bar\nu_e$ number densities discussed above, the feedback effect contains an interesting twist in reshaping the ELN angular distribution, which then leads to different local ELN angular distributions compared to those after the prompt flavor conversions.
One major difference is related to the conservation of the neutrino lepton number.
In the prompt FFCs, in addition to the elimination of the ELN angular crossings, the neutrino and antineutrino lepton numbers, as well as the total ELN are locally conserved to a good approximation.
However, all these three numbers are no longer conserved quantities on a longer timescale when the collision and neutrino diffusion play a role.
Although the FFC still works to erase the ELN angular crossings, the collisional weak processes such as EA can change those lepton numbers as well as the total ELN.
Consequently, the angular distributions of the ELN and each neutrino species keep evolving on timescales determined by the collision rates, while the FFC functions to ensure that no new ELN angular crossings appear during the evolution.
The system can then settls down to a quasisteady state determined by the FFC, collisions, and advection all together.

\begin{figure*}[!hbt]
\includegraphics[width=0.38\textwidth]{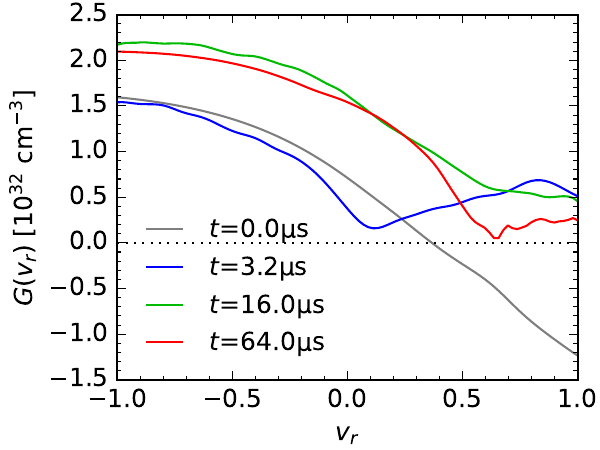}
\llap{\parbox[b]{1.9in}{\small (a) Model III\\\rule{0ex}{1.7in}}}
\includegraphics[width=0.38\textwidth]{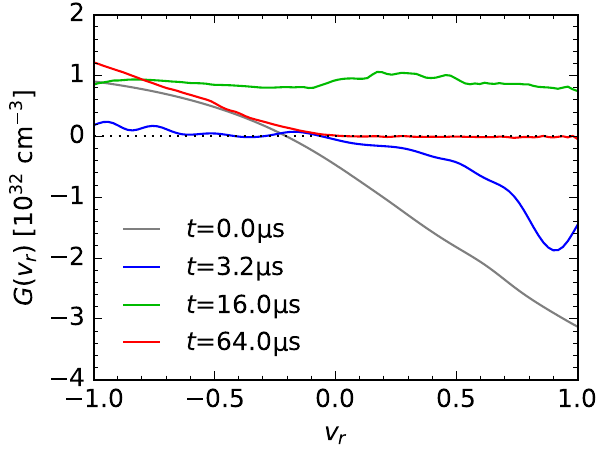}
\llap{\parbox[b]{3.2in}{\small (c) Model IV\\\rule{0ex}{1.7in}}}
\includegraphics[width=0.38\textwidth]{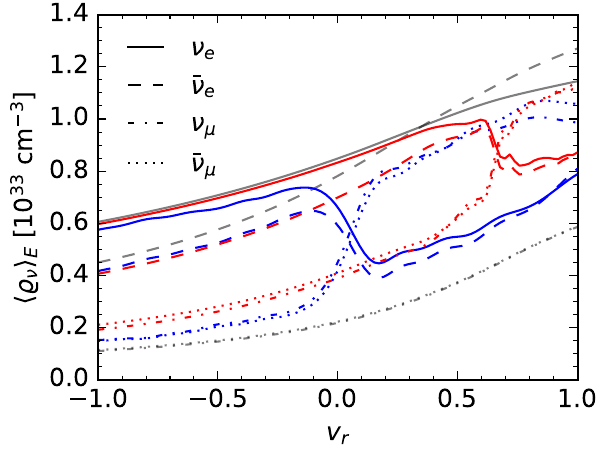}
\llap{\parbox[b]{2.6in}{\small (b) Model III\\\rule{0ex}{1.7in}}}
\includegraphics[width=0.38\textwidth]{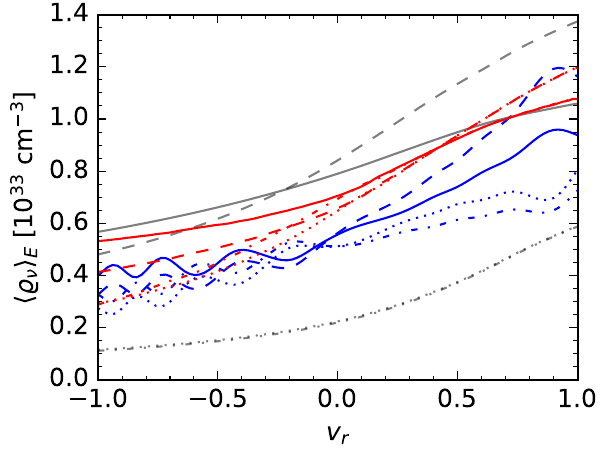}
\llap{\parbox[b]{3.6in}{\small (d) Model IV\\\rule{0ex}{1.7in}}}
\caption{\label{fig:an_iz26} Evolution of the angular distributions of the ELN, $G(v_r)$ (upper row), and that of each neutrino species at $r\approx 36$~km (lower rows) in Models III (a--b) and IV (c--d).
}
\end{figure*}

To illustrate this effect, we first show in Fig.~\ref{fig:an_iz26} the neutrino angular distributions for different times at $r\approx 36$~km, which is inside the $\nu_e$-sphere in both Model III and IV.
In Model III, the initial ELN distribution transitions from positive to negative at $v_{r,c} \approx 0.37$ [Fig.~\ref{fig:an_iz26}(a)].
With the condition $|I_+|>|I_-|$, the prompt FFC between the electron and muon flavors erases the ELN crossing by making it positive in the entire range of $v_r$, accompanied with significant flavor swap for $v_r>0$ at $t=3.2~\mu$s [Fig.~\ref{fig:an_iz26}(b)].
Since this radius is inside the neutrinosphere, the EA processes of $\nu_e$ and $\bar\nu_e$ with matter tend to restore their angular distributions back to the initial state prior to FFC.
Because of the higher opacity for $\nu_e$'s than $\bar\nu_e$'s, the angular distribution of $\nu_e$'s increases faster, which leads to a more positive ELN distribution at $t=16~\mu$s [Fig.~\ref{fig:an_iz26}(a)].
On a longer timescale, the restoration of $\nu_e$'s slows down as its angular distribution becomes closer to the initial state.
Because of the restoration of $\bar\nu_e$'s, the ELN distribution starts to decrease, which potentially leads to a new angular crossing.
However, the FFC can smear any newly formed angular crossing and  keep the ELN angular distribution positive during this process.
Finally, the neutrino distributions settle into a quasisteady state at $t\approx 64~\mu$s where the ELN angular distribution is nearly flat for $v_r\gtrsim 0.5$ [Fig.~\ref{fig:an_iz26}(a)].
Meanwhile, the $\nu_e$ and $\bar\nu_e$ distributions are restored to values closer to their initial states for $v_r\lesssim 0.5$ [Fig.~\ref{fig:an_iz26}(b)].

\begin{figure*}[!hbt]
\includegraphics[width=0.32\textwidth]{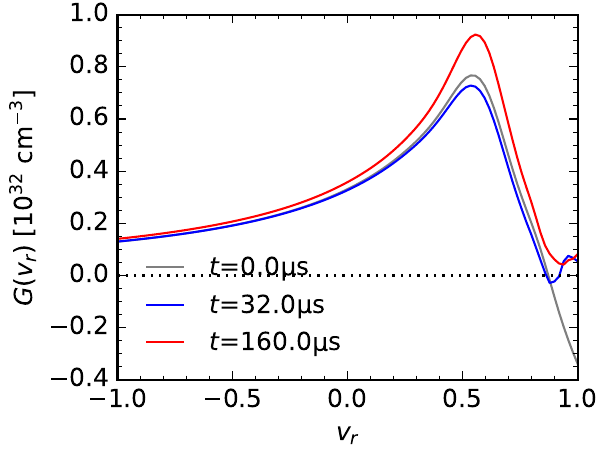}
\llap{\parbox[b]{2.8in}{\small (a) Model II\\\rule{0ex}{1.4in}}}
\includegraphics[width=0.32\textwidth]{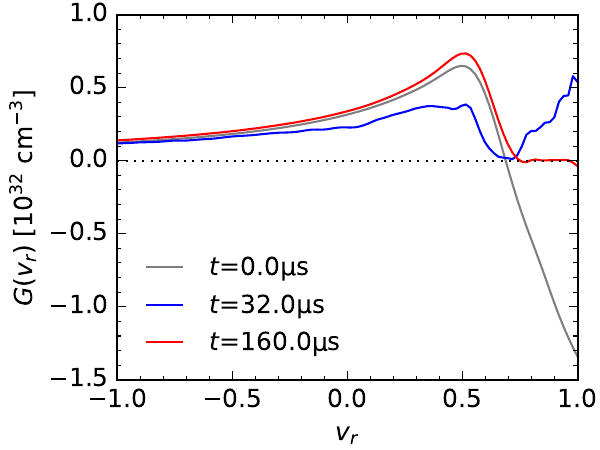}
\llap{\parbox[b]{2.8in}{\small (c) Model III\\\rule{0ex}{1.4in}}}
\includegraphics[width=0.32\textwidth]{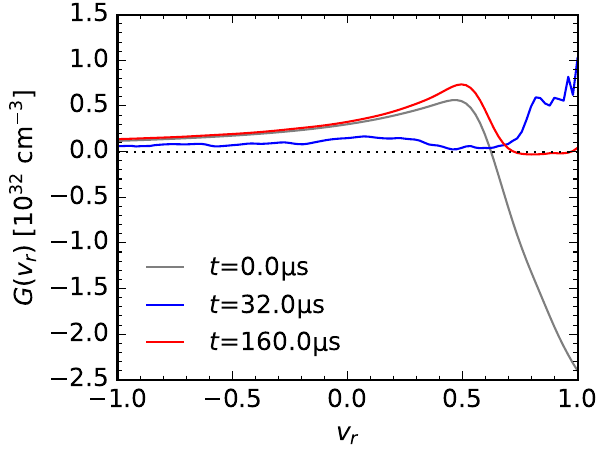}
\llap{\parbox[b]{2.8in}{\small (e) Model IV\\\rule{0ex}{1.4in}}}
\includegraphics[width=0.32\textwidth]{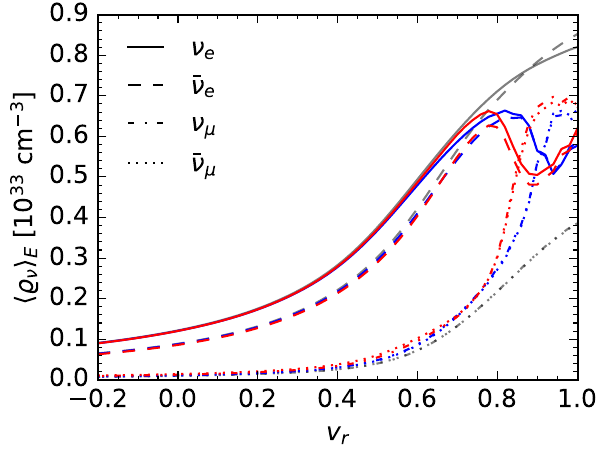}
\llap{\parbox[b]{2.2in}{\small (b) Model II\\\rule{0ex}{1.4in}}}
\includegraphics[width=0.32\textwidth]{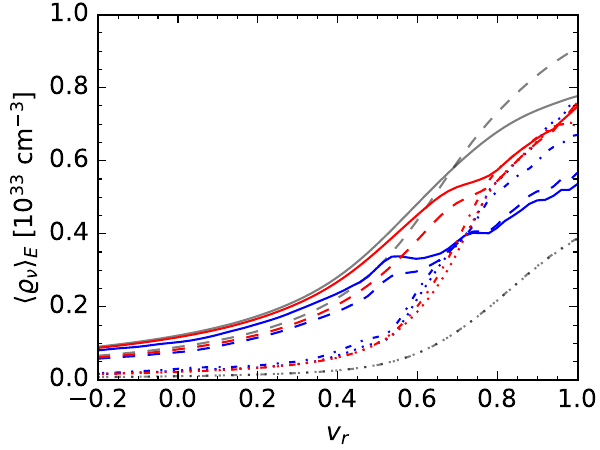}
\llap{\parbox[b]{3.0in}{\small (d) Model III\\\rule{0ex}{1.4in}}}
\includegraphics[width=0.32\textwidth]{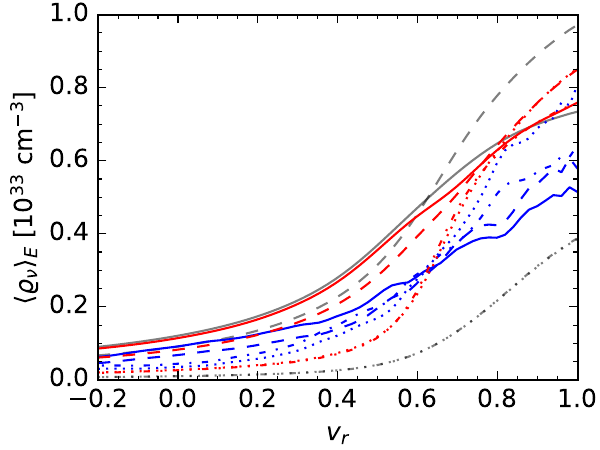}
\llap{\parbox[b]{3.0in}{\small (f) Model IV\\\rule{0ex}{1.4in}}}
\caption{\label{fig:an_iz50} Evolution of the angular distributions of the ELN, $G(v_r)$ (upper panels), and that of each neutrino species at $r\approx 50$~km (lower panels) in Models II (a--b), III (c--d), and IV (e--f). Note that $\langle \varrho_\nu \rangle_E$ are shown in $-0.2\leq v_r\leq 1$ because most of neutrinos propagate outward.
}
\end{figure*}

For Model IV, the neutrino distribution is dominated by $\bar\nu_e$'s with $|I_+|<|I_-|$ at the same radius $r\simeq 36$~km with an initial crossing at $v_r\simeq -0.2$.
Thus, flavor equilibration takes place mainly for negative $v_r$ at $t=3.2~\mu$s [Fig.~\ref{fig:an_iz26}(c)], and the ELN angular distribution becomes essentially negative for all $v_r$'s.
However, the faster EA rates for $\nu_e$'s increase the ELN distribution and turn it into positive values at $t=16~\mu$s [Fig.~\ref{fig:an_iz26}(c)].
Later, the effect of slower $\bar\nu_e$ rates takes effect and lowers the values of ELN distribution for a major $v_r$ range.
At $t=64~\mu$s, the distributions also settle into an approximately steady state, with nearly flat ELN distribution around zero for $v_r\gtrsim 0$ [Fig.~\ref{fig:an_iz26}(c)], as well as the $\langle \varrho_{\nu_e}\rangle_E \simeq \langle \varrho_{\bar\nu_e}\rangle_E$ and $\langle \varrho_{\nu_\mu}\rangle_E \simeq \langle \varrho_{\bar\nu_\mu}\rangle_E$ [Fig.~\ref{fig:an_iz26}(d)].

The collisional feedback effect inside the neutrinosphere also affects the flavor evolution in the outer part.
Figure~\ref{fig:an_iz50}(c--f) shows that, in both Model III and IV, the prompt FFC at $r=50$~km and $t=32~\mu$s turns the ELN distribution to the same sign, with significant amount of $\nu_e$'s and $\bar\nu_e$'s being converted.
However, as neutrinos continuously stream out, what are observed at $r=50$~km at later times reflect the state of neutrinos escaping from inside the neutrinosphere at earlier times.
At $t=160~\mu$s, the angular distributions are restored back to $\simeq 80\%$--90\% of the initial values for $\nu_e$'s and $\bar\nu_e$'s in $v_r\gtrsim 0.8$ in Model III, or even beyond that for $\nu_e$'s in $v_r\gtrsim 0.9$ in Model IV.
The asymptotic ELN distributions are almost zero for $v_r\gtrsim 0.7$ in both Models, showing very different shapes from the prompt FFC.
This feedback is also confirmed in \cite{xiong2024robust} by using analytical prescriptions for the flavor evolution.

For comparison, we also show in Fig.~\ref{fig:an_iz50}(a--b) the corresponding distributions in Model II.
From $t=32$ to $t=160~\mu$s, the angular distributions of $\nu_e$'s and $\bar\nu_e$'s are only slightly changed.
This is because, in Model II, the FFCs only start at $r\approx 40$~km located just outside the $\nu_e$-sphere.
No significant amount of $\nu_e$'s and $\bar\nu_e$'s are produced additionally from within the neutrinosphere.

In spite of the intriguing feedback effect on the neutrino distributions discussed above, the main impact of FFC on the radial profiles of the neutrino mean energies in Model III and IV are relatively similar to that in Model II, as shown in Fig.~\ref{fig:re}(b) and (c).
FFC slightly enhances $\langle E_{\nu_e}\rangle$ and $\langle E_{\nu_e}\rangle$ by $\lesssim 1$~MeV in all radii, while it largely reduces $\langle E_{\nu_\mu}\rangle$ by $\simeq 2$~MeV in regions for $r\gtrsim 35$~km.
One visible difference from Model II here is that for $r\lesssim 35$~km, both $\langle E_{\nu_\mu}\rangle$ and $\langle E_{\bar\nu_\mu}\rangle$ are reduced due to the backwardpropagating neutrinos from outer radii.

\subsection{Model V}
\begin{figure}[!hbt]
\includegraphics[width=0.48\textwidth]{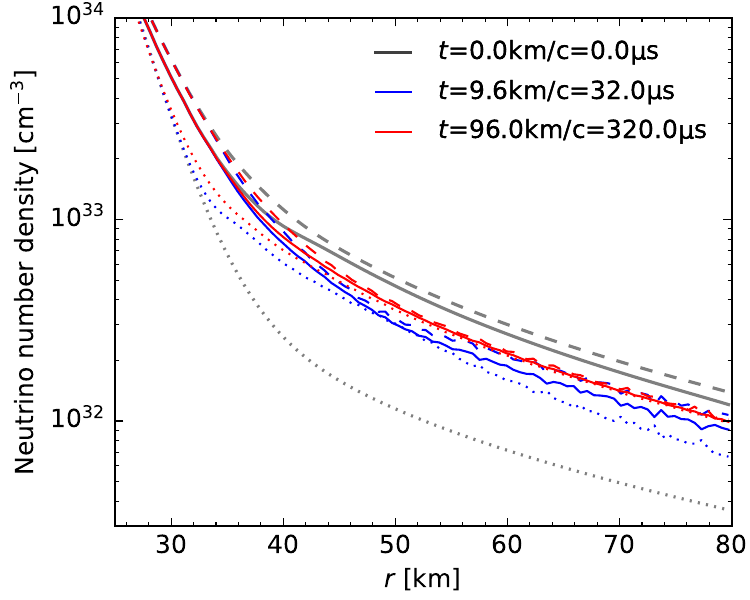}
\hspace{-0.03in}\llap{\parbox[b]{4.8in}{\small (a)\\\rule{0ex}{2.4in}}}
\includegraphics[width=0.48\textwidth]{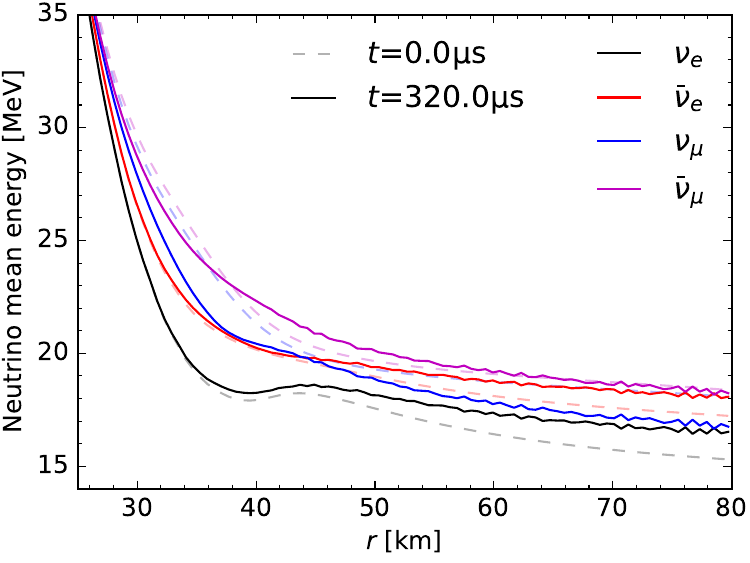}
\hspace{-0.03in}\llap{\parbox[b]{5.3in}{\small (b)\\\rule{0ex}{2.2in}}}
\caption{\label{fig:rn_Vc0} Radial profiles of neutrino number densities (a) and mean energies (b) in Model Vc0.
}
\end{figure}

Model V is special because of the coexistence of CFI and FFI.
Before discussing the combined effect originating from both flavor instabilities, we first discuss the results in the corresponding Model Vc0 where the CFI is switched off.
Figure~\ref{fig:rn_Vc0} shows the radial profiles of the neutrino number densities and their mean energies in Model Vc0.
The impact of FFCs here is qualitatively similar to those discussed earlier.
At $t=32~\mu$s, the prompt FFC affects the radial range from $r\approx 35$~km to the outer boundary and leads to approximate flavor equilibration in $40\lesssim r\lesssim 54$~km where the deepness ratio of the initial ELN, rewritten as $|I_+/I_-|$, is larger (see Fig.~\ref{fig:I_ratio}).
As neutrinos continue propagating outward, their flavor content is replaced by the emission and oscillation properties close to the neutrinosphere.
Together with the feedback mechanism related to the EA processes, the number densities of $\nu_e$ and $\bar\nu_e$ are also enhanced for $r\gtrsim 40$~km at $t=320~\mu$s as in Models III and IV.
For $r\gtrsim 50$~km, the mean energies of $\nu_\mu$ and $\bar\nu_\mu$ become much closer to those of $\nu_e$ and $\bar\nu_e$, respectively.
Because the angular crossings for radii inside the $\nu_e$ neutrinosphere in Model Vc0 is located at negative $v_r$, the FFC affects more the backward propagating $\nu_\mu$'s in that region, which leads to a minor enhancement of $n_{\nu_\mu}$ near $r=32$~km as well as moderate reductions of the mean energies for heavy-lepton neutrinos there.

For Model V where the collisional flavor conversions are also present, Fig.~\ref{fig:rn}(d) shows that in addition to the same FFCs in $r\gtrsim 35$~km, the CFI also leads to flavor transformations and thus the enhanced $n_{\nu_\mu}$ in $r\simeq 28$--35~km inside the neutrinosphere.
As a result, $n_{\nu_\mu}$ becomes almost the same as $n_{\nu_e}$ already at $t=32~\mu$s and remains utill $t=320~\mu$s.
For the mean energies, once again, the muon flavor neutrinos become closer to those of electron flavor as shown in Fig.~\ref{fig:re}(f).

We notice that, in both Models V and Vc0 at $t=320~\mu$s, the neutrino number densities in regions for $r\gtrsim 50$~km become $n_{\nu_e}\simeq n_{\bar\nu_e}$, different from the initial hierarchy of $n_{\nu_e}< n_{\bar\nu_e}$.
Because the flavor content is still dominated by $\bar\nu_e$'s after the prompt conversions, this effect cannot be attributed to the pairwise FFC.
It suggests that the collisional feedback effect inside the neutrinosphere leads to a larger net change of flavor content in $\bar\nu_e$'s than $\nu_e$'s.

\subsection{Models VI and VII}
For Models VI and VII, the initial ELN angular crossings only reside in a limited radial range at 29~$\lesssim r\lesssim 33$~km and 27~$\lesssim r\lesssim 29$~km (see Fig.~\ref{fig:initial_ELN}), respectively.
Figure~\ref{fig:rn}(e) and (f) shows that the FFC affects mostly the number density of heavy-lepton flavors outside the regions where the crossings exist initially.
For $\nu_e$ and $\bar\nu_e$, their number densities remain nearly unchanged for all radii due to the following reasons.
First of all, since these crossings are inside the neutrinospheres, the neutrino-matter interaction outside the FFI region plays an important role in resetting the properties of $\nu_e$ and $\bar\nu_e$.
Second, as the neutrino angular distributions are more isotropic inside the neutrinosphere, neutrinos not affected by FFCs can also diffuse back into the region of FFC to mitigate the impact of flavor conversions.
Moreover, even inside the region of FFC, the dynamical feedback from the collisions mentioned above in Models III and IV also acts to restore the distributions of $\nu_e$ and $\bar\nu_e$ back to the level close to those before oscillations.
For the heavy-lepton neutrinos, their spectra at inner radii become closer to those of $\nu_e$ and $\bar\nu_e$ when the system reaches the quasisteady state, which limits further flavor conversions.
Since the FFI region in Model VI is closer to the neutrinosphere than in Model VII, the overall impact of flavor conversions is larger in Model VI.

Similar behaviors are found in terms of neutrino mean energies as well.
In both models, the mean energy for heavy-lepton neutrinos are reduced significantly, while the mean energies of $\nu_e$ and $\bar\nu_e$ only increase very little.
The impact in Model VI is larger than in Model VII.

\subsection{Two flavors vs three flavors}
\begin{figure*}[!hbt]
\includegraphics[width=0.32\textwidth]{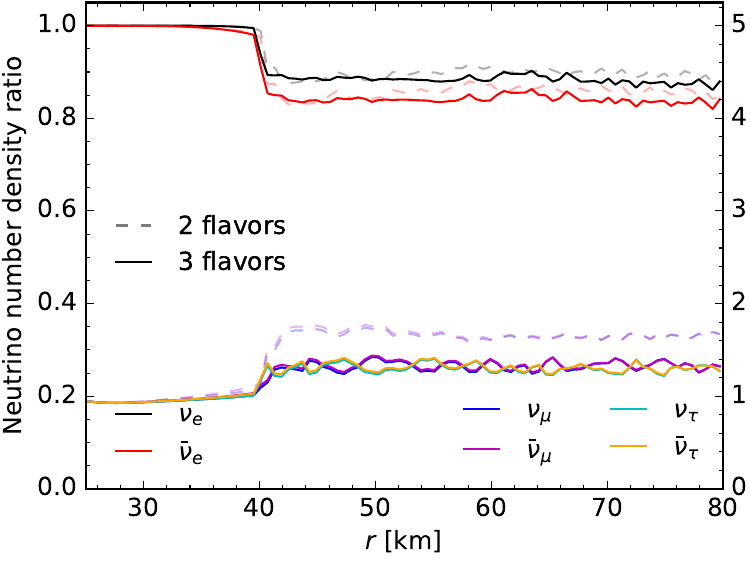}
\llap{\parbox[b]{3.8in}{\small (a)\\\rule{0ex}{1.4in}}}
\includegraphics[width=0.32\textwidth]{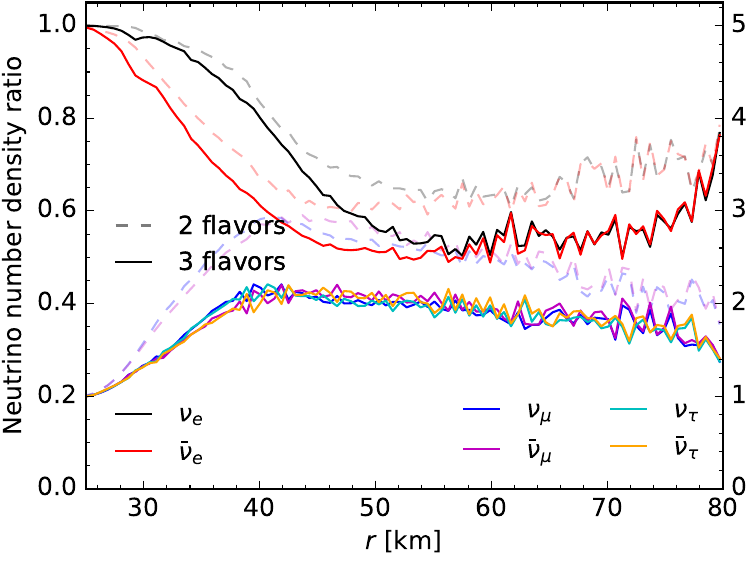}
\llap{\parbox[b]{3.8in}{\small (b)\\\rule{0ex}{1.4in}}}
\includegraphics[width=0.32\textwidth]{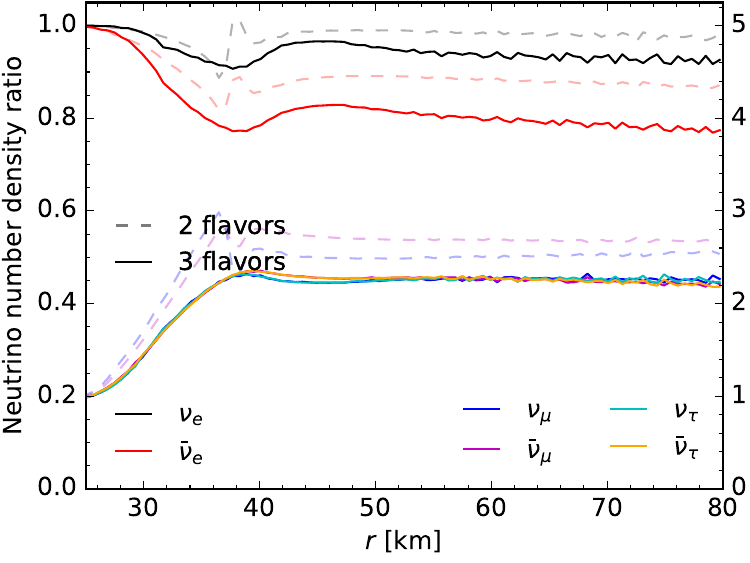}
\llap{\parbox[b]{3.8in}{\small (c)\\\rule{0ex}{1.4in}}}
\caption{\label{fig:rp_f2f3} Comparison of the radial profiles of neutrino number density ratios $n_\nu(t)/n_\nu(t=0)$ between two-flavor (dashed) and three-flavor (solid) schemes for Model II at $320~\mu$s (a), Model IV at $32~\mu$s (b) and at $320~\mu$s (c). Note that the scales for heavy-lepton neutrinos are labeled on the right axes.
}
\end{figure*}

To illustrate the effects of adopting schemes with different numbers of neutrino flavors in the simulations, we choose Models II and IV as the representative examples for cases without and with the dynamical collisional feedback.
The results from Models II and IV are compared with Models IIf3 and IVf3 in Fig.~\ref{fig:rp_f2f3}.
For Models II and IIf3, the radial profile of the number density ratios $n_\nu(t=320~\mu{\rm s})/n_\nu(t=0)$ for all flavors, including $\nu_\tau$ and $\bar\nu_e$ in the three-flavor scenario, are shown in Fig.~\ref{fig:rp_f2f3}(a).
Note that we use different scales for the electron flavors (left axis) and the heavy-lepton flavors (right axis) for better readability.
For Models IV and IVf3, the number density ratios at $t=32$s and $320~\mu$s are shown in Fig.~\ref{fig:rp_f2f3}(b) and (c), respectively, to illustrate the impact after the prompt FFC and that at the asymptotic state after the collisional feedback takes place.

These plots show that, with three flavors, the enhancement of $\nu_\mu$ and $\bar\nu_\mu$ number densities is reduced, due to the additional flavor conversion channel from electron to tauon flavors.
The enhancement in Model IIf3 is $\simeq 50\%$ of that in Model II.
For Models IVf3 and Model IV, the enhancement of heavy-lepton flavor number densities in Model IVf3 is reduced by less than $50\%$, particularly at $t=320~\mu$s.
This implies that the flavor conversions result in a larger amount of the total heavy-lepton flavor number density in Model IVf3 than in Model IV.

\begin{figure*}[!hbt]
\includegraphics[width=0.32\textwidth]{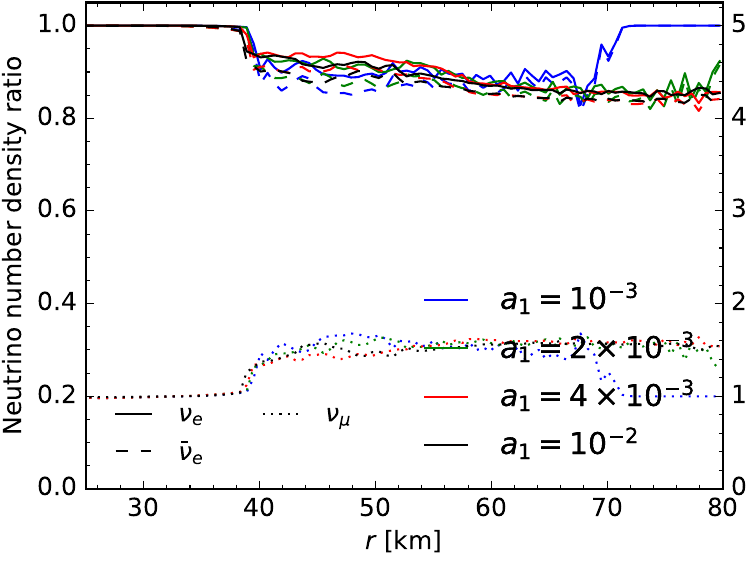}
\llap{\parbox[b]{3.8in}{\small (a)\\\rule{0ex}{1.4in}}}
\includegraphics[width=0.32\textwidth]{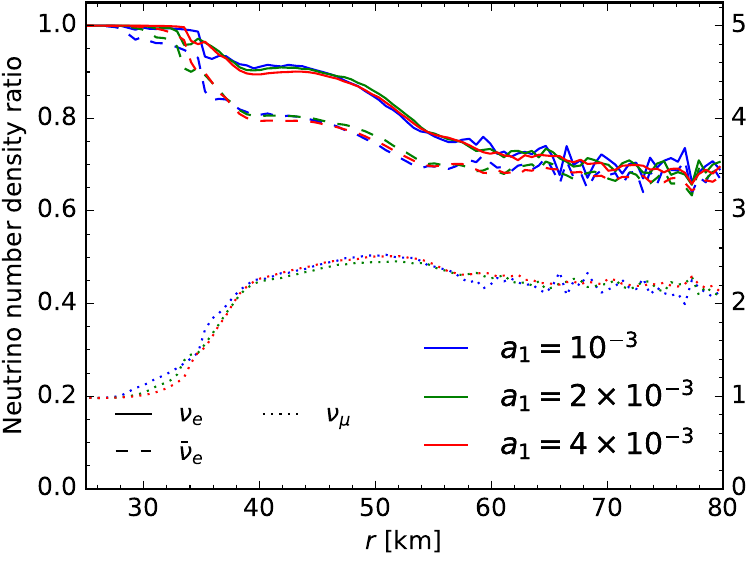}
\llap{\parbox[b]{3.8in}{\small (b)\\\rule{0ex}{1.4in}}}
\includegraphics[width=0.32\textwidth]{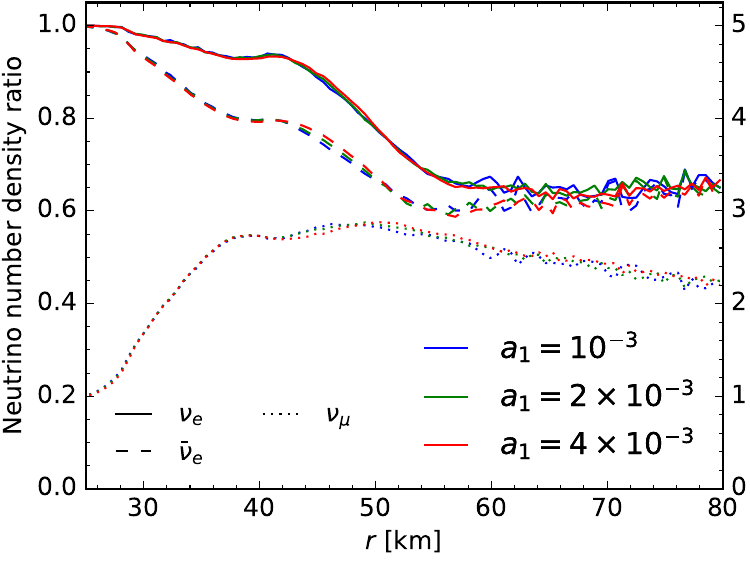}
\llap{\parbox[b]{3.8in}{\small (c)\\\rule{0ex}{1.4in}}}
\caption{\label{fig:rp_a1} Comparison of the radial profiles of neutrino number density ratios $n_\nu(t)/n_\nu(t=0)$ (blue) in Models II (a), III (b), and IV (c) at $64~\mu$s with different attenuation factors $a_1=10^{-3}$ (blue), $2\times 10^{-3}$ (green), $4\times 10^{-3}$ (red), and $1\times 10^{-2}$ (black in Model II). Note that the scales for the ratio of $\nu_\mu$ are labeled on the right axes.
}
\end{figure*}

\begin{figure}[!hbt]
\includegraphics[width=0.4\textwidth]{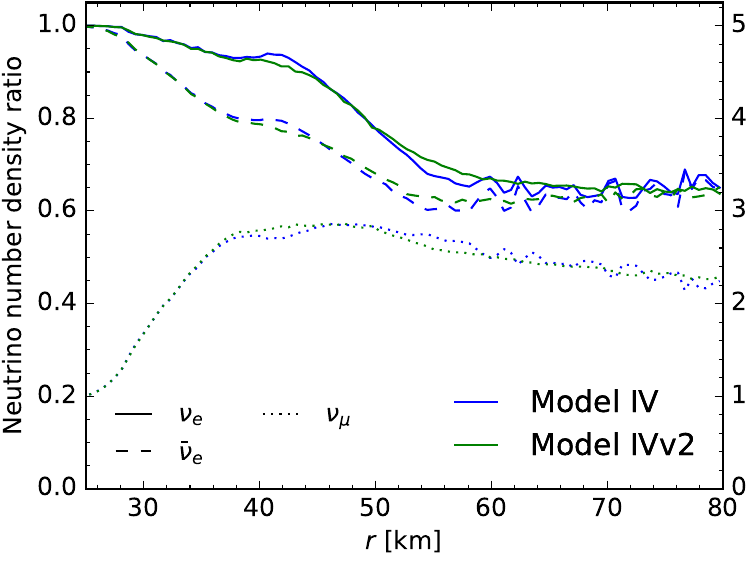}
\caption{\label{fig:rp_vac} Comparison of the radial profiles of neutrino number density ratios $n_\nu(t)/n_\nu(t=0)$ between Models IV and IVv2 at $64~\mu$s. The scale for the ratio of $\nu_\mu$ is labeled on the right axis.
}
\end{figure}

For $\nu_e$ and $\bar\nu_e$, their number densities are generally more reduced by flavor conversions, while the amount of reduction in Models IIf3 and IVf3 from the corresponding two-flavor cases differs.
In Model IIf3, the reduction is relatively minor and only appears at $r\gtrsim 50$~km.
In Model IVf3, the reduction at $t=32~\mu$s is much more significant than in Model IIf3, reflecting the shift of $\nu_e$ and $\bar\nu_e$ survival probability from $\simeq 1/2$ to $\simeq 1/3$ when reaching flavor equilibration (see, e.g. \cite{richers2021neutrino}).
At $t=320~\mu$s, although the $\nu_e$ and $\bar\nu_e$ number densities are restored back to values closer to those prior to FFCs due to the collisional feedback effect, the three-flavor case still shows more reduction than the two-flavor case.
Overall, including the tauon flavor allows more flavor conversion of $\nu_e$'s and $\bar\nu_e$'s, which is consistent with the observed increase of the total heavy-lepton flavor content.

\subsection{Attenuation on the neutrino self-induced term}
We discussed in Sec.~\ref{sec:emergence} the impact of $\mathbf H_{\nu\nu}$ attenuation on the dispersion relation and the small-scale structure.
In this subsection, we examine its impact on the flavor evolution in macroscopic scale.
Figure~\ref{fig:rp_a1}(a)--(c) shows the neutrino number density ratios in Models II--IV at $t=64~\mu$s with different attenuation factors $a_1$.
At this time, only the case in Model II with $a_1=10^{-3}$ shows substantial differences in the number density ratios at $r\gtrsim 70$~km.
The reason is that the evolution in Model II is mainly determined by the prompt FFCs.
Thus, cases with more attenuation (smaller $a_1$) evolve slower than those with less attenuation.
Another slight difference is that the starting radius of FFCs with less attenuation is slightly shifted leftward from $r\approx 39$~km to $\approx 38$~km.

For Models III and IV, the number density ratios in Fig.~\ref{fig:rp_a1}(b) and (c) exhibit a transition at $r\simeq 50$~km, within which the impact of collisional feedback appears.
Since the timescale of the collisional feedback effect does not depend on the attenuation factor, provided that the FFC timescale remains short enough, the results shown in these two panels are therefore hardly affected by the choice of different attenuation factors.

\subsection{Vacuum term}
\label{sec:vacuumterm}
We also discussed in Sec.~\ref{sec:emergence} that the general trend of the full dispersion relation is almost independent of the vacuum term when no $\mathbf H_{\nu\nu}$ attenuation is applied.
When taking $a_1=10^{-3}$, the dispersion relation in Model II, which has the smallest instability growth rate, can be affected by the adopted value of $\delta m^2$, while the other models such as IV and V remain nearly unaffected.
Here, we examine whether the simulation results at late times are affected by the strength of the vacuum term by comparing Model IV to Model IVv2 at $t=64~\mu$s in Fig.~\ref{fig:rp_vac}.
We find that taking a different $\delta m^2$ indeed only results in negligible effects, which is consistent with the findings reported in Refs.~\cite{martin2021fast,nagakura2023basic,abbar2024applications}.
For both the radial regions affected by the prompt flavor conversion and that affected by the collision feedback, a larger $\delta m^2$ only leads to insignificant deviations.

\section{Effects and observables}
\label{sec:effects}
After discussing the evolution of FFCs, we examine in this section their implications and feedback to key quantities relevant to CCSN physics, including the net neutrino heating rates and the equilibrium electron fraction.
In addition, the impact of FFCs on the free-streaming neutrino energy spectra, which are important inputs for the expected supernova neutrino signals, will be discussed.

\subsection{Heating-cooling rates}
\begin{figure*}[!hbt]
\includegraphics[width=0.32\textwidth]{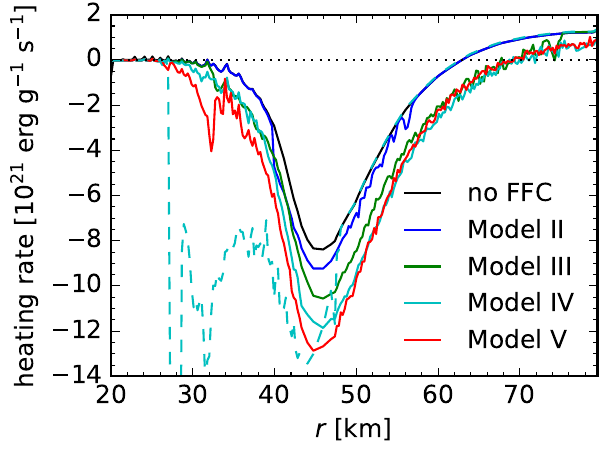}
\llap{\parbox[b]{3.5in}{\small (a)\\\rule{0ex}{1.47in}}}
\includegraphics[width=0.32\textwidth]{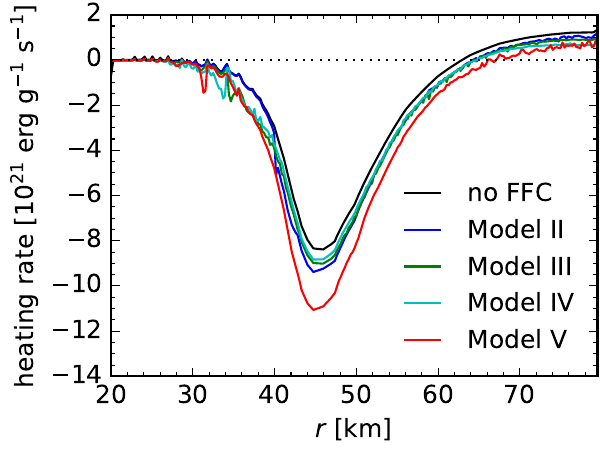}
\llap{\parbox[b]{3.5in}{\small (b)\\\rule{0ex}{1.47in}}}
\includegraphics[width=0.32\textwidth]{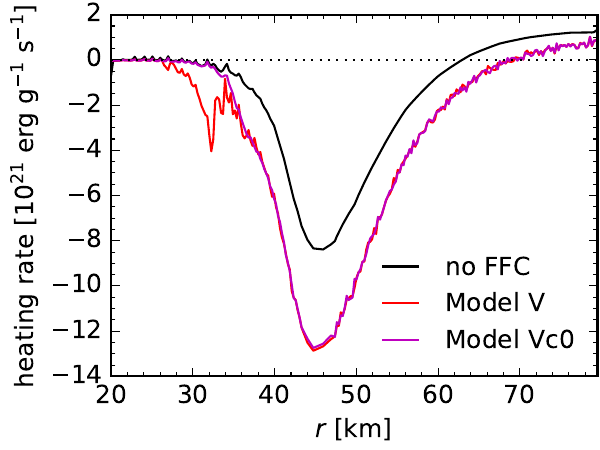}
\llap{\parbox[b]{3.5in}{\small (c)\\\rule{0ex}{1.47in}}}
\caption{\label{fig:heating} Comparison of the specific heating-cooling rate after the prompt FFCs at $t=32~\mu$s (a) and in the asymptotic state at $t=160~\mu$s (b) for Models II (blue), III (green), IV (cyan), and V (red). The original heating rates without flavor conversions are shown by the black curves and are indistinguishable for all models. The cyan dashed curve in (a) is for Model IV at $t=3.2~\mu$s. The magenta curve in (c) is for Model Vc0 at $t=32~\mu$s, compared to Model V (red).
}
\end{figure*}

The specific net heating rate dominated by the EA process of $\nu_e$ and $\bar\nu_e$, processes (1a--1c) in Table~\ref{tab:nu_process}, is given by
\begin{align}
    \dot q = \frac{1}{\rho} \int dE\, & dv_r\, E [
    \chi_{e,\rm EA}\varrho_{ee}
    - j_{e,\rm EA}(\varrho_{ee,\rm FO}-\varrho_{ee}) \nonumber\\
    & + \bar\chi_{e,\rm EA}\bar\varrho_{ee}
    - \bar j_{e,\rm EA}(\bar\varrho_{ee,\rm FO}-\bar\varrho_{ee}) ].
\end{align}
We show the above net heating rate in Models II--V for cases with and without FFCs in Fig.~\ref{fig:heating}.
Notice that the rates before FFCs (black curve) are practically indistinguishable on the plot for all models despite different attenuation strengths adopted to $Y_e$ profiles.
This is because the reduction of $\nu_e$'s and enhancement of $\bar\nu_e$'s are complementary to each other.
Without FFCs, the regions with net cooling ($\dot q<0$) and heating ($\dot q>0$) are separated by the gain radius at $r\approx 63$~km.
In the cooling region, the maximum net cooling rate is as low as $\dot q\approx -8.4 \times 10^{21}~{\rm erg}\,{\rm g}^{-1}\,{\rm s}^{-1}$ at $r\simeq 46$~km.

With FFCs, other solid curves in Fig.~\ref{fig:heating}(a) show the radial profiles of $\dot q$ at $t=32~\mu$s for Models II--V.
Since the pairwise FFCs from electron to heavy-lepton flavors lead to the reduction of heating rates in these models, $\dot q$ are reduced in the entire domain.
As a result, the gain radius shifts to larger radii around $r\simeq 69$~km, consistent with that found in \cite{nagakura2023roles,ehring2023fast2}.
At this time, $\dot q$ in models with more $Y_e$ attenuation that exhibit more flavor conversions of $\nu_e$ to $\nu_\mu$ shown in Fig.~\ref{fig:rn} are more negative, with Model V showing the most negative $\dot q \simeq -1.3 \times 10^{22}~{\rm erg}\,{\rm g}^{-1}\,{\rm s}^{-1}$.

It should also be highlighted that because the prompt flavor conversion takes place in different timescales among models and at different radii, the heating rate can be reduced more at an earlier time.
For example, the cyan dashed curve in Fig.~\ref{fig:heating}(a) demonstrates that the net cooling from $r\simeq 28$~km to $\simeq 46$~km is drastically enhanced at $t=3.2~\mu$s in Model IV.
The reduction of $\dot q$ at this time is particularly severe around $r\simeq 30$~km and $\simeq 44$~km, where the initial ELN deepness ratios are $\approx 1$ (Fig.~\ref{fig:I_ratio}), allowing nearly complete flavor equilibration or even overconversion after the prompt FFC.
In contrast, for models in which the feedback between FFCs and collisional weak processes leads to additional production of $\nu_e$'s and $\bar\nu_e$'s inside neutrinospheres, the reduction of $\dot q$ can be alleviated at later times.
This can lead to the change of ordering in terms of $\dot q$ reduction from what was discussed above at $t=32~\mu$s.
Figure~\ref{fig:heating}(b) shows that the reductions of $\dot q$ in Models III and IV are less than that in Model II around $r\simeq 46$~km at $t=160~\mu$s when the systems approach toward asymptotic states.
For Model V, although it remains to be the model affected by FFCs the most, the impact also becomes slightly smaller at this time.
As for Models VI and VII, because the impact of FFC on $\nu_e$'s and $\bar\nu_e$'s flavor content is small, $\dot q$ in these two models are only affected insignificantly, and are thus not shown in Fig.~\ref{fig:heating}.

In addition to the FFCs, the CFI that is present in Model V also contributes to the reduction of $\dot q$.
Figure~\ref{fig:heating}(c) shows that an additional $\dot q$ reduction appears at $r\simeq 32$~km and $t=32~\mu$s in Model V, which is otherwise not present in the corresponding model without the CFI (Model Vc0).
The impact of this component also largely diminishes at later times as shown in Fig.~\ref{fig:heating}(b).

We would like to caution that the impact of FFC on the net heating rate discussed here is under the assumption that background matter profiles are static.
In realistic situations, even if the prompt flavor conversions can be reset by the collisional weak processes, its imprint on the matter background may introduce further feedback to neutrinos and lead to an  outcome that sits in between the state after the prompt conversions and the  asymptotic one shown above.
This additional feedback effect may also depend on involved hydrodynamic timescales and will need to be evaluated in future work.

\subsection{Equilibrium electron fraction}
\label{sec:yeeq}
Another pertinent implication of FFC on supernova physics is related to the nucleosynthesis of elements.
Although our background matter profile is taken from a supernova model that does not explode, we expect that the analyses of the impact of FFC on the chemical composition in our models can still shed light on what may happen in successfully exploded supernovae with different progenitor masses as well as the neutrino-driven wind at a later epoch.

For this purpose, we examine the effect of FFC on the equilibrium $Y_e$, which is a key quantity relevant to supernova nucleosynthesis.
For $T\lesssim 1$~MeV, the equilibrium $Y_e$ is dominated by the neutrino absorption reactions and can be approximated by
\begin{equation}
    Y_e^{\nu,\mathrm{eq}} = \left( 1+\frac{L_{\bar\nu_e}}{L_{\nu_e}} \frac{\langle E_{\bar\nu_e} \rangle-2\Delta_{np}+1.2\Delta_{np}^2/\langle E_{\bar\nu_e} \rangle}{\langle E_{\bar\nu_e} \rangle+2\Delta_{np}+1.2\Delta_{np}^2/\langle E_{\nu_e} \rangle} \right)^{-1},
\end{equation}
where $\Delta_{np}=m_n-m_p$ is the neutron–to-proton mass difference \cite{qian1996nucleosynthesis}.
Notice that although all neutrinos propagate radially outward at the outer boundary of our simulation domain, there remains a fraction of neutrinos with non-negligible transverse velocities, i.e., $v_r\neq 1$.
Assuming that the neutrino flux is conserved and no other types of flavor oscillations occur outside the outer boundary, the neutrino spectra beyond the simulation domain for $r>r_{\rm ob}$ can then be approximated by
\begin{align}
    \langle \varrho \rangle_A(E) = & \int_0^1 dv_r\, \varrho(E,v_r,r_{\rm ob}) \times \nonumber\\
    & \left( \frac{r_{\rm ob}}{r} \right)^2 \frac{v_r}{\sqrt{1-(\frac{r_{\rm ob}}{r})^2 (1-v_r^2)}},
\end{align}
which can be further used to calculate the neutrino luminosity and mean energy relevant to nucleosynthesis.

\begin{figure*}[!hbt]
\includegraphics[width=0.49\textwidth]{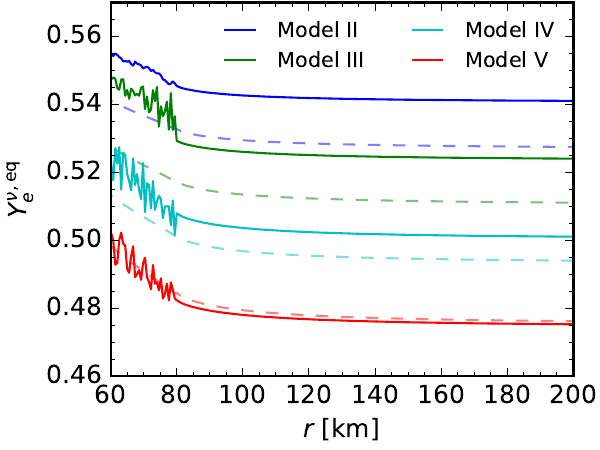}
\llap{\parbox[b]{5.0in}{\small (a)\\\rule{0ex}{2.3in}}}
\includegraphics[width=0.49\textwidth]{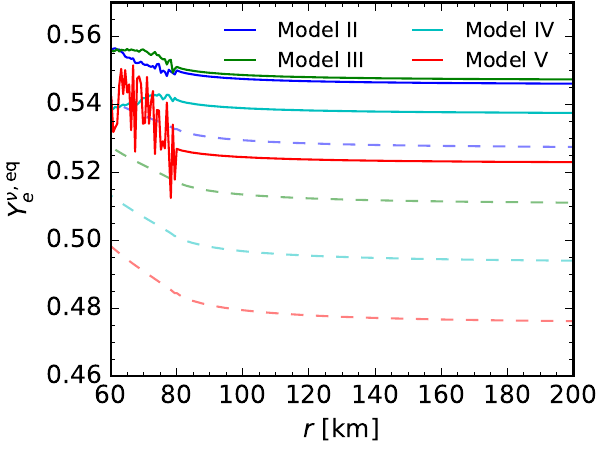}
\llap{\parbox[b]{5.0in}{\small (b)\\\rule{0ex}{2.3in}}}
\caption{\label{fig:yeeq} Comparison of the equilibrium electron fraction $Y_e^{\nu,\rm eq}$ after the prompt FFCs (a) and at a later time $t=160~\mu$s (b) for Models II (blue), III (green), IV (cyan), and V (red). The dashed curves show the corresponding profiles without flavor conversions for comparisons.
}
\end{figure*}

The equilibrium $Y_e^{\nu,\mathrm{eq}}$ in Models II--V are shown from $r=60$ to 200~km in Fig.~\ref{fig:yeeq}.
The values asymptotically approach constants at larger radii where neutrinos propagate predominately along the radial direction.
Unlike the heating rate, which is hardly affected by the $Y_e$ attenuation for cases without any flavor conversions, the $Y_e^{\nu,\mathrm{eq}}$ profiles depend on the ratio between $\nu_e$'s and $\bar\nu_e$'s and hence vary from Models II--IV without FFCs (transparent dashed curves).
At $r=200$~km, $Y_e^{\nu,\mathrm{eq}}$ are $\approx 0.527$, 0.511, 0.493, and 0.476 in Models II--V without FFCs, respectively,  which reflects the increasing dominance of $\bar\nu_e$ capture rates that result in more neutron-rich conditions.

The impact of the prompt FFCs on $Y_e^{\nu,{\rm eq}}$ is shown in Fig.~\ref{fig:yeeq}(a).\footnote{Because the timescale for prompt FFCs at $r_{\rm ob}$ varies from model to model, we show the profiles at $t=96~\mu$s for Model II, at $t=64~\mu$s for Model III, and at $t=32~\mu$s for Models IV and V, respectively.}
Since the prompt FFCs transform an approximately same amount of $\nu_e$ and $\bar\nu_e$ into the heavy-lepton flavors, it implies that for cases with an initial asymmetry between $\nu_e$ and $\bar\nu_e$, the asymmetry gets further enhanced by FFCs.
As a result, a $\nu_e$-dominated condition becomes even more dominated by $\nu_e$'s, and vice versa.
In addition, given that the mean energy of $\nu_e$ is generally lower than those of other species, FFC enhances $\langle E_{\nu_e} \rangle$ relatively more than $\langle E_{\bar\nu_e}\rangle$, which also leads to the increased $Y_e^{\nu,\mathrm{eq}}$.
Combining these two effects, Fig.~\ref{fig:yeeq}(a) shows that $Y_e^{\nu,\mathrm{eq}}$ at $r=200$~km are increased by $\approx 0.013$, 0.013, 0.008, and $-0.001$ in Models II--V, respectively.
The impact is generally larger for models whose $Y_e^{\nu,\mathrm{eq}}$ deviate more from 0.5.
Note that the impact in Model II is similar to that in Model III, due to the shallower ELN crossing in Model II that only allows a smaller amount of FFCs.

Similar to the impact on the heating rate, the impact of flavor conversions on $Y_e^{\nu,\mathrm{eq}}$ at the asymptotic times are reshaped by the feedback effect from the collisional weak processes in Models III--V.
As discussed in Sec.~\ref{sec:III_IV}, the feedback effect tends to bring the flavor content of $\nu_e$ back closer to the level prior to oscillations than for $\bar\nu_e$, this increases the ratio of the $\nu_e$ number density over $\bar\nu_e$ by more than 10\% compared to the ratio before conversions, which leads to a large change on $Y_e$.
Figure~\ref{fig:yeeq}(b) shows that, for Model II without feedback effect, $Y_e^{\nu,\mathrm{eq}}$ at $t=160~\mu$s is indeed similar to that shown in Fig.~\ref{fig:yeeq}(a).
In contrast, $Y_e^{\nu,\mathrm{eq}}$ in Models III--V are substantially increased by $\simeq 0.02$--0.04 at the asymptotic times compared to the values immediately after the FFCs, due to the increased $\nu_e$ to $\bar\nu_e$ number density ratios.
These results suggest that with the feedback effect, a neutron-rich condition may even be turned into a proton-rich condition (Model V).
Moreover, the feedback effect may further enhance the proton richness than the outcome of prompt FFCs.
All these may facilitate the production of light $p$-nuclei through the $\nu p$-process \cite{frohlich2006neutrino,xiong2020potential}.

\subsection{Free-streaming neutrino spectra}
Finally, we examine the impact of the flavor conversions on the energy spectra of neutrinos at sufficiently large radii outside the simulation boundary.
We follow the same extrapolation method in Sec.~\ref{sec:yeeq} to obtain the differential number luminosity
\begin{equation}
    \frac{dL_{\rm num}}{dE}(E) = 4\pi \int_0^1 dv_r\, \varrho(E,v_r,r_{\rm ob}) r_{\rm ob}^2 v_r
\end{equation}
at sufficiently large radii.
Note that no additional flavor conversions are included in order to focus on the impact of FFCs on the free-streaming neutrinos.

\begin{figure*}[!hbt]
\includegraphics[width=0.32\textwidth]{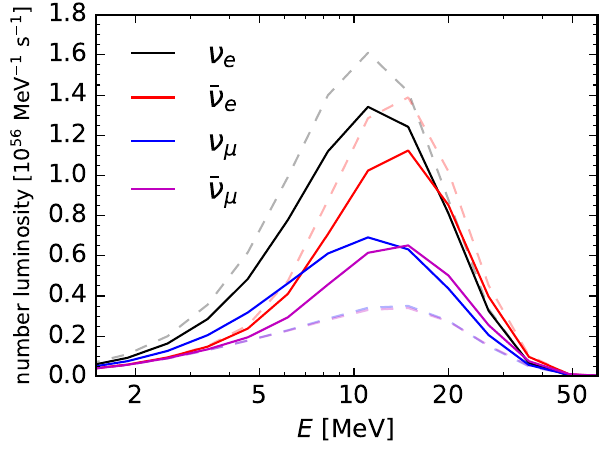}
\llap{\parbox[b]{2.0in}{\small (a) Model II\\\rule{0ex}{1.4in}}}
\includegraphics[width=0.32\textwidth]{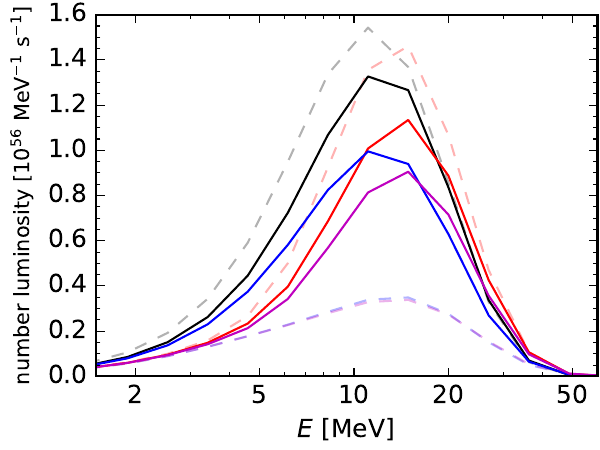}
\llap{\parbox[b]{3.0in}{\small (b) Model III\\\rule{0ex}{1.4in}}}
\includegraphics[width=0.32\textwidth]{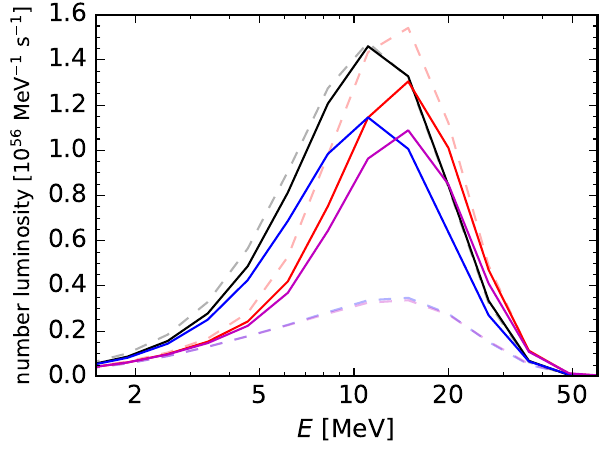}
\llap{\parbox[b]{3.0in}{\small (c) Model IV\\\rule{0ex}{1.4in}}}
\includegraphics[width=0.32\textwidth]{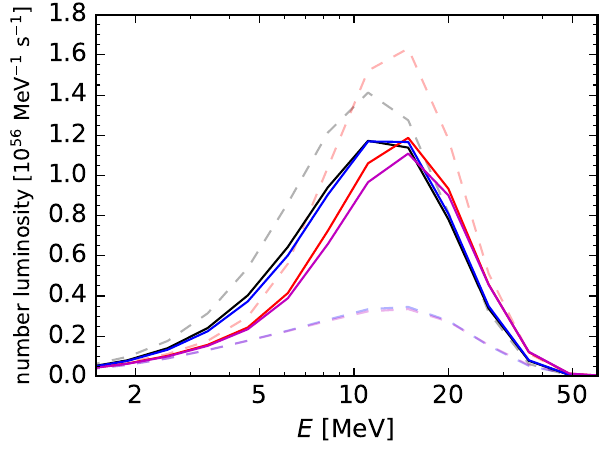}
\llap{\parbox[b]{3.0in}{\small (d) Model V\\\rule{0ex}{1.4in}}}
\includegraphics[width=0.32\textwidth]{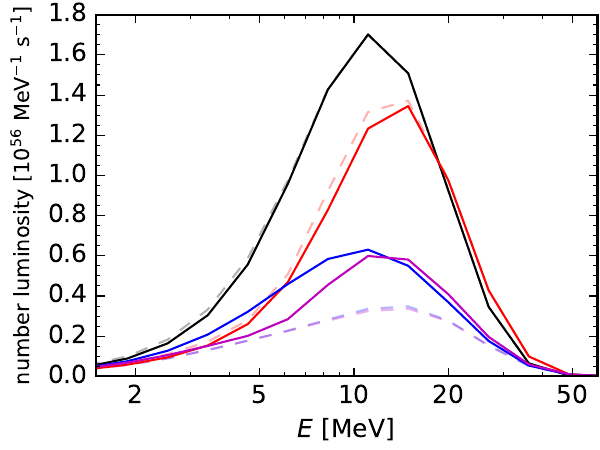}
\llap{\parbox[b]{3.0in}{\small (e) Model VI\\\rule{0ex}{1.4in}}}
\includegraphics[width=0.32\textwidth]{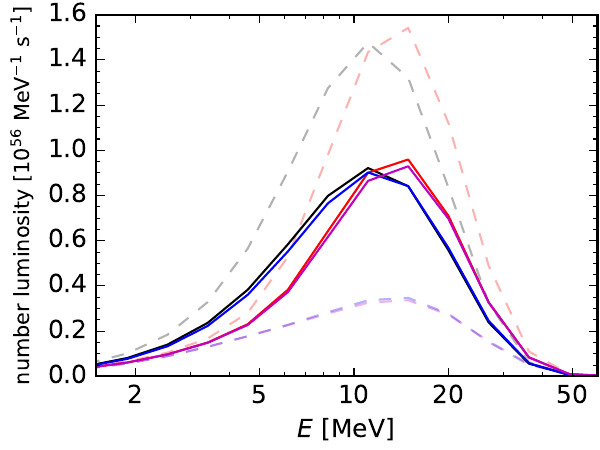}
\llap{\parbox[b]{2.6in}{\small (f) Model IV, $64~\mu$s\\\rule{0ex}{1.4in}}}
\includegraphics[width=0.32\textwidth]{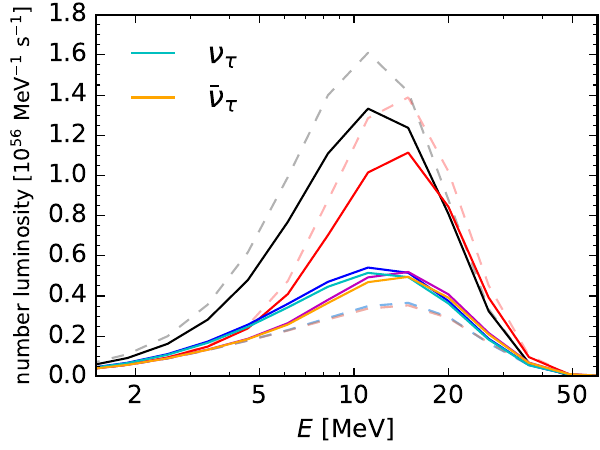}
\llap{\parbox[b]{1.9in}{\small (g) Model IIf3\\\rule{0ex}{1.4in}}}
\includegraphics[width=0.32\textwidth]{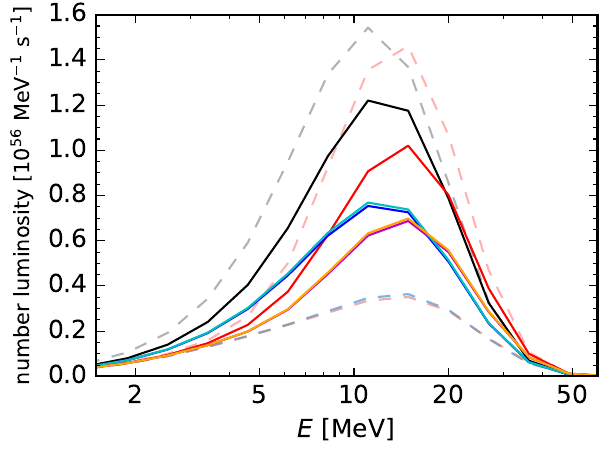}
\llap{\parbox[b]{2.9in}{\small (h) Model IIIf3\\\rule{0ex}{1.4in}}}
\includegraphics[width=0.32\textwidth]{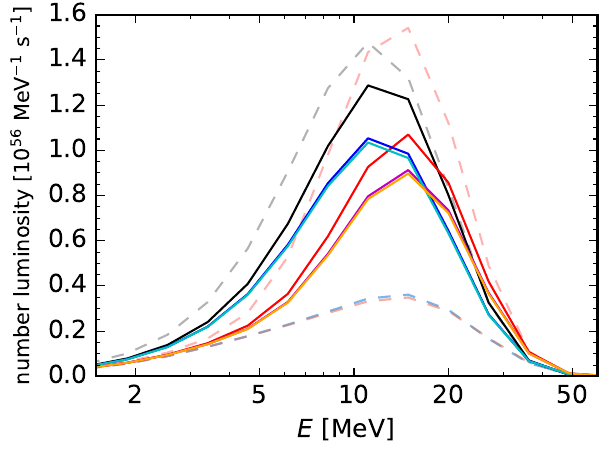}
\llap{\parbox[b]{2.9in}{\small (i) Model IVf3\\\rule{0ex}{1.4in}}}
\caption{\label{fig:en_ob} Free-streaming neutrino energy spectra at $t=320~\mu$s in Models II (a), III (b), IV (c), V (d), VI (e), IIf3 (g), IIIf3 (h), and IVf3 (i). Note that the panel (f) shows the spectrum after the prompt FFC at $t=64~\mu$s in Model IV for illustration. The dashed curves show the corresponding spectra without flavor conversions for comparisons.
}
\end{figure*}

Figure~\ref{fig:en_ob}(a--e) shows the free-streaming neutrino energy spectra at $t=320~\mu$s in Models II--VI for cases with and without flavor conversions.
In addition, Fig.~\ref{fig:en_ob}(f) shows the spectra after the prompt FFCs at $t=64~\mu$s in Model IV.
A common feature is that the degeneracy between $\nu_\mu$'s and $\bar\nu_\mu$'s are all broken after the flavor conversions given that $\nu_e$'s and $\bar\nu_e$'s have different energy spectra prior to flavor conversions.

However, these plots also clearly show that flavor equilibration is not a general outcome of FFCs.
Only the Model V [Fig.~\ref{fig:en_ob}(d)] as well as the earlier phase after FFCs in Model III (not shown) and IV [Fig.~\ref{fig:en_ob}(f)] obtain outcomes that are close to flavor equilibration.
For the rest, equilibration is not achieved due to the initial shallow ELN angular crossings in Model II, the dynamical feedback from collisions after the FFCs in Model III and IV, and a complete reset from the collisional weak processes when the angular crossing is inside neutrinospheres in Model VI.
For cases that include three flavors shown in Fig.~\ref{fig:en_ob}(g--i) for Model II--IV, the results are qualitatively similar to that in the corresponding two-flavor cases.

We also emphasize again that for models where the collisional feedback effect is important, it results in a net total amount of neutrino luminosity summing all flavors compared to the values without flavor conversions, as discussed earlier in Sec.~\ref{sec:III_IV}.
Even though the flavor equilibration is achieved in Model V, the energy spectra of all neutrino species lie above the average of the unoscillated level.
Based on this observation, we expect that the FFC can lead to a  significant enhancement for the energy loss rate due to neutrino emission, which will further affect the CCSN dynamics.
This effect, however, needs to be fully evaluated via a CCSN hydrodynamic simulation that incorporates the FFCs.

\section{Discussion and conclusions}
\label{sec:discussion}
We have utilized our multigroup and discrete-ordinate collective neutrino oscillation simulation code based on \textsc{cose$\nu$} to comprehensively study the emergence, evolution, and effects of FFCs in a spherically symmetric supernova background obtained with \textsc{agile-boltztran}.
The collisional weak processes in this study have been extended from Ref.~\cite{xiong2023evolution} to include neutrino-electron scatterings and neutrino pair reactions.
We adopted radial-dependent $Y_e$ attenuation schemes that allow us to probe various scenarios with the ELN angular crossings residing outside, inside, and across the neutrinosphere, as well as conditions with neutrino number densities dominated by $\nu_e$'s or $\bar\nu_e$'s, prior to the FFCs.
We have also investigated the impact of an assumed number of neutrino flavors, the choice of the $\mathbf{H}_{\nu\nu}$ attenuation factor, the size of the vacuum term, and the coexistent CFI.
All models examined in this work are listed in Table~\ref{tab:parameters}.

To analyze our simulation results, we introduced the spectrogram method that is suitable for the inhomogeneous neutrino gas in Sec.~\ref{sec:ana-meth-spec} to inspect the small-scale structure of the flavor coherence.
The spectrograms are compared to the dispersion relation obtained from the LSA in Sec.~\ref{sec:emergence}.
We examined in detail the evolution history in both two- and three-flavor models as well as the impact on FFCs from other flavor instabilities in Sec.~\ref{sec:evolution}.
Finally, we discussed the general effects of FFCs on CCSN physics and the free-streaming neutrino energy spectra in Sec.~\ref{sec:effects}.

We confirmed several characteristics of FFCs reported in a recent study~\cite{nagakura2023basic}.
After the FFCs take place, neutrino profiles eventually reach an asymptotic state on the coarse-grained level, with the angular crossings in the ELN distributions eliminated in general.
The asymptotic state of neutrinos shows a strong dependence on the collisional neutrino-matter interactions, particularly when the ELN angular crossings appear inside the neutrinospheres.

On top of that, we summarize our new findings as follows:
\begin{enumerate}

\item The comparison of our LSA and spectrogram results shows that the emergence of FFI and the early evolution of FFCs qualitatively follow the prediction of dispersion relation from the LSA (Fig.~\ref{fig:sp}).
The unstable modes with larger growth rates develop predominantly, which corresponds to the emergence of small-scale structure in the off-diagonal flavor coherence.
Although the dominant modes of the small-scale structure may be shifted to slightly different $K_r$ during the later evolution or even disappear for some radial ranges, they generally survive and stay rather close to the initially most unstable $K_r$ mode from LSA in Models II--IV and Vc0.
In addition, we also observed that the small-scale structure can appear in the diagonal elements, i.e., neutrino number density, with a length scale consistent with that in the off-diagonal flavor coherence (Fig.~\ref{fig:small_scale_structure}).

\item We showed that a good scalability of the dispersion relation exists even when the attenuation on $\mathbf H_{\nu\nu}$ is employed (Figs.~\ref{fig:DR_other}--\ref{fig:gr}).
Although this implies the typical length of the small-scale structure changes with the attenuation factors, the neutrino flavor evolution has rather weak dependence on the chosen attenuation factors so long as the dominant $K_r$ modes of the dispersion relation are well resolved (Fig.~\ref{fig:rp_a1}).

\item The evolution of the neutrino flavor content exhibits a large variation among different models (Fig.~\ref{fig:rn}).
The prompt flavor conversions immediately after the emergence of the local FFI can lead to nearly complete flavor equilibration for models with deep ELN angular crossings (Models III--V).
Models with shallow crossings result in incomplete flavor conversions, demonstrated by Model II.

\item We found that after the prompt FFC phase in Models III--V, an interesting feedback effect due to the neutrino-matter interaction just below the neutrinosphere continues to restore the $\nu_e$'s and $\bar\nu_e$'s, and reshapes the local ELNs as well as the angular distribution of each neutrino species (Figs.~\ref{fig:an_iz26} and \ref{fig:an_iz50}).
This effect can largely alter the flavor state of neutrinos around and outside the neutrinosphere without incurring an additional ELN angular crossing, leading to an asymptotic state much different from that obtained right after the prompt FFCs.
Although heavy-lepton flavor neutrinos do not directly experience collisional restoration, their flavor contents are also enhanced through further flavor conversions from the restored $\nu_e$'s and $\bar\nu_e$'s.
Detailed explanations on this dynamical feedback were discussed in Sec.~\ref{sec:III_IV}.

\item When the ELN angular crossings only reside inside the neutrinosphere but not outside (Models VI and VII), the impact of FFCs can affect the heavy-lepton flavor neutrinos in the free-streaming regime, as they suffer less collisional processes on diffusing outward.
However, negligible effects are observed for $\nu_e$'s and $\bar\nu_e$'s owing to the large opacity that they experience (Fig.~\ref{fig:rn}).
Overall, the asymptotic outcome of FFCs in the free-streaming regime depends on the how deep the FFI region is inside the neutrinosphere.

\item Regarding the implication of FFCs on the CCSN dynamics, we found that with the prompt FFCs, it reduces the net heating rate behind the shock due to the reduced $\nu_e$ and $\bar\nu_e$ absorption rates, resulting in more cooling below as well as less heating above the gain radius, consistent with \cite{nagakura2023basic}.
However, we also found that for Models III--V in which the collisional feedback plays an important role after the prompt FFCs, the restoration of $\nu_e$'s and $\bar\nu_e$'s at later times reduces the impact of FFCs on the enhanced cooling in the asymptotic state (Fig.~\ref{fig:heating}).

\item The feedback mechanism also leads to another interesting consequence on $Y_e^{\nu,\rm eq}$.
Right after the prompt FFCs, because similar amount of flavors is converted, the asymmetry between $\nu_e$ and $\bar\nu_e$ number densities is increased.
Hence, it enhances the proton or neutron richness depending on the initial value of $Y_e$.
However, the collisional feedback effect restores more $\nu_e$'s than $\bar\nu_e$'s due to the disparity between their collisional rates.
It systematically leads to higher $Y_e^{\nu,\rm eq}$ and makes material more proton-rich in the asymptotic state (Fig.~\ref{fig:yeeq}).

\item For the emerging neutrino energy spectra after the FFCs relevant to the neutrino detection, we found that flavor equilibration is not a general outcome.
Significant differences between neutrinos of electron and heavy-lepton flavors can exist (Fig.~\ref{fig:en_ob}).
This finding indicates that the subsequent slow flavor instability and Mikheyev–Smirnov–Wolfenstein matter effect \cite{wolfenstein1978neutrino,mikheev1985resonance} can be relevant and will need to be considered for studying the neutrino signals.

\item Although the above conclusions are based on the two-flavor scenario, we found that the general features of flavor conversions remain qualitatively similar to that with three flavors.
Nevertheless, including a third flavor can result in additional flavor conversions into the additional sector, which leads to less amount of $\nu_e$ and $\bar\nu_e$.
For the heavy-lepton sectors, the enhanced amounts for muon and tauon flavors in the three-flavor scenarios are less than that for muon flavors in the two-flavor cases.
However, the total amount of heavy-lepton flavors in three-flavor cases is in fact larger than that in the two-flavor cases (Fig.~\ref{fig:rp_f2f3}).

\item We showed that the dispersion relation in our simulation domain is barely affected by the inclusion of the vacuum term when no attenuation is applied to $\mathbf H_{\nu\nu}$.
When attenuation is applied, one may pragmatically adopt a reduced $\delta m^2$ to maintain the scalability of the dispersion relation.
In this case, the simulation results are minimally influenced by the presence of the vacuum term (Fig.~\ref{fig:rp_vac}).

\item We found that the occurrence and the strength of CFI, which is intrinsically a multienergy phenomenon due to the energy dependence of the EA collisional rates, can depend on the energy spectra of heavy-lepton neutrinos.
When the NES processes are included, the condition for the CFI in this work becomes different from that in Ref.~\cite{xiong2023evolution} and can be understood using the formula developed in Ref.~\cite{xiong2023collisional} (Sec.~\ref{sec:otherFI}).
Among all models considered in this paper, noticeable flavor conversion due to the CFI is only observed inside the FFC region in Model V when the initial condition is dominated by $\bar\nu_e$'s.
While this affects less the properties of neutrinos than the FFC does, it can lead to a transient enhancement of the net cooling rate in the corresponding region (Fig.~\ref{fig:heating}).

\end{enumerate}

In summary, our comprehensive study of neutrino flavor conversions in the postshocked region supports that it is important to include FFCs in supernova simulations, in order to fully understand the role of neutrinos in supernova dynamics and nucleosynthesis, as well as to predict the CCSN neutrino signals.
Our work also highlights the need for using energy-dependent and realistic neutrino collision rates in relevant simulations, because the condition and evolution of FFC and CFI can be affected by those collision processes either directly or through feedback.
Certain assumptions that were introduced in our work such as the spherical symmetry, the omission of the modification on the proton and neutron abundances when attenuating $Y_e$, the static supernova background without feedback on $Y_e$, and the missing coherent forward scattering with the matter remain to be further examined and will be studied in future.

We acknowledge the following software: \textsc{matplotlib}~\cite{matplotlib}, \textsc{numpy}~\cite{numpy}, and \textsc{scipy}~\cite{scipy}.

\begin{acknowledgments}
We thank Ninoy Rahman and Gang Guo for discussions on collisional weak processes, Yi-Siou Wu on the comparison of two- and three-flavor results, and Sherwood Richers on the effective method of treating neutrino pair reactions. 
ZX and GMP acknowledge support of the European Research Council (ERC) under the European Union’s Horizon 2020 research and innovation program (ERC Advanced Grant KILONOVA No. 885281), the Deutsche Forschungsgemeinschaft (DFG, German Research Foundation) -- Project-ID 279384907 -- SFB 1245, and MA 4248/3-1.
MRW and MG acknowledge support from the National Science and Techonology Council, Taiwan under Grant No.~111-2628-M-001-003-MY4, and the Academia Sinica (Project No.~AS-CDA-109-M11).
MRW also acknowledges support from the Physics Division of the National Center for Theoretical Sciences, Taiwan.
CYL acknowledges support from the National Center for High-performance Computing (NCHC).
TF and NKL acknowledge support from the Polish National Science Center (NCN) under Grant No. 2020/37/B/ST9/00691.
The core-collapse supernova simulations were performed at the Wroclaw Center for Scientific Computing and Networking (WCSS).
The neutrino quantum kinetic simulations were performed at the Virgo computing cluster of the GSI Helmholtzzentrum {f\"ur} Schwerionenforschung.
\end{acknowledgments}

\bibliographystyle{apsrev4-1}
\bibliography{references.bib}

\onecolumngrid

\appendix

\section{Equation of state used in supernova simulation}
\label{sec:sn_eos}
For the present supernova simulation, the DD2 relativistic mean-field (RMF) nuclear equation of state (EOS) is employed~\cite{typel2010composition,hempel2010statistical,hempel2012new}.
Its bulk properties agree quantitatively with nuclear physics experiments in the vicinity of nuclear saturation density (see Table~1 in Ref.~\cite{fischer2014symmetry} and references therein) as well as with chiral perturbation calculation for pure neutron matter~\cite{Tews2013PhRvL110}.
Furthermore, the DD2 EOS is in agreement with current astrophysical constraints derived from pulsar observations~\cite{Antoniadis13,Cromartie20,NICER_Miller2019,NICER_Watts2019,NICER_Miller2021,NICER_Riley2021} and from the GW170817 gravitational wave observation associated with a binary neutron-star merger event ~\cite{abbott2017gw170817,Lattimer2018PhRvL121_NSEOS,abbott2019properties}.
At densities below nuclear saturation density, the DD2 RMF EOS is coupled with the modified nuclear statistical equilibrium approach of Ref.~\cite{hempel2010statistical}, featuring several thousand nuclear species with partly measured as well as theoretical nuclear masses.
The transition to uniform nuclear matter at the saturation density is modeled via the excluded volume approach in the DD2 EOS, suppressing nuclear clusters (see Ref.~\cite{fischer2020medium} for a detailed discussion about this transition in the context of supernova simulations).
At temperatures below $T\simeq 0.45$~MeV, which correspond to low densities in the CCSN simulation, the RMF EOS is matched with the ideal silicon-sulfur gas approximation.
This matches the iron-core silicon-sulfur layer transition of the stellar progenitor~\cite{rauscher2002nucleosynthesis}.
Furthermore, the nuclear EOS is coupled with electron, positron and photons EOS of Ref.~\cite{timmes2000accuracy} at all densities.
Note that both RMF and silicon-sulfur gas EOS treat Coulomb contributions differently; however, the differences remain negligible for the matching conditions here.

\section{Neutrino-electron scattering}
\label{sec:nes}
We only include the diagonal elements in $\mathbf C_{\rm NES}$ for the heavy-lepton neutrinos as they are mainly thermalized by the NES process.
The general formula is given as follows~\cite{mezzacappa1993type,mezzacappa1993stellar}:
\begin{eqnarray}
\frac{4\pi^2}{E^2} \mathbf C_{ii, \rm NES}(E, v_r)
&=&
\left[1- f_{\nu_i}(E, v_r)\right]
\int E'^2 d E'\, d v_r'\,
R_{\rm NES}^{\rm in}(E, E', v_r, v_r') f_{\nu_i}(E', v_r')
\nonumber\\
&&
\quad - f_{\nu_i}(E, v_r)
\int E'^2 d E'\, d v_r'\,
R_{\rm NES}^{\rm out}(E, E', v_r, v_r')
\left[1- f_{\nu_i}(E', v_r')\right]~,~
\label{eq:appendix_nes_c1}
\end{eqnarray}
where $f_{\nu_i}$ is the neutrino distribution function, $i=\mu,\,\tau$, and $R_{\rm NES}^{\rm out/in}(E,E',v_r,v_r')$ are the in- and out-scattering kernels.
Unlike the IS, which only has contributions from the first two moments in $v_r$ and $v_r'$, the angular dependence of the NES scattering kernel can be rather complicated and computationally challenging, depending on the kinematics of the target electron and position as well as on the numerical realization of the neutrino phase space.
Nonetheless, it is common for neutrino transport codes to introduce the moment approximation.
Here, we choose to retain only the zeroth moment, $\Phi_{{\rm NES},0}(E,E')$, obtained by the appropriate integration of $R_{\rm NES}$ following Refs.~\cite{schinder1982neutrino,rampp2002radiation},
\begin{equation}
\Phi_{{\rm NES},0}^{\rm out}(E,E') =
\int_{-1}^1 d\omega\, R_{\rm NES}^{\rm out}(E,E',\omega)
\end{equation}
where $\omega$ is the cosine of the angle between the incident and outgoing neutrinos.
With the zeroth moment approximation, the scattering kernel is given as
\begin{equation}
    R_{\rm NES}^{\rm out}(E,E',v_r,v_r') = \pi \Phi_{{\rm NES},0}^{\rm out}(E,E'),
\end{equation}
denoted as the angle-independent $R_{\rm NES}^{\rm out}(E,E')$.
Notice that the detailed balance ensures the validity of Eq.~\eqref{eq:detailed_balance}, see also Sec.~IIa of Ref.~\cite{fischer2012neutrino}, and hence
\begin{equation}
    R_{\rm NES}^{\rm in}(E,E') = \exp \left( \frac{E'-E}{T} \right) R_{\rm NES}^{\rm out}(E,E').
\end{equation}
Consequently, Eq.~\eqref{eq:appendix_nes_c1} becomes
\begin{align}
\mathbf C_{ii, \rm NES} (E,v_r) = &
-2 \varrho_{ii}(E, v_r)\int d E'\, R_{\rm NES}^{\rm out}(E,E') E'^2  \nonumber\\
&
+ \int d E'\, R_{\rm NES}^{\rm in}(E, E')
\left\{
E^2+4\pi^2 \varrho_{ii}(E, v_r)
\left[
\exp\left(\frac{E-E'}{T}\right)-1
\right]
\right\}
\int d v_r'\, \varrho_{ii}(E', v_r').
\end{align}
The integration over $v_r'$ can be calculated separately, which significantly speeds up the simulation.

For the implementation in the code, the neutrino density matrix is discretized, and so is the kernel $\varrho_{ii}(E', v_r')$.
Given that NES is mostly efficient in exchanging energy between neutrinos with similar energies, in practice we keep the elements sufficiently close to the diagonal in $R_{\rm NES}$, i.e., $R^{i_E,i_E+\Delta i_E}_{\rm NES}$ where $i_E$ is the index of the energy grid.
We choose the range of energy exchange to be within $|\Delta i_E|\leq 3$.
This choice not only helps accelerate the computation but also ensures the thermalization of neutrinos at a given temperature without breaking the detailed balance.

\begin{figure}[!hbt]
\includegraphics[width=0.475\columnwidth]{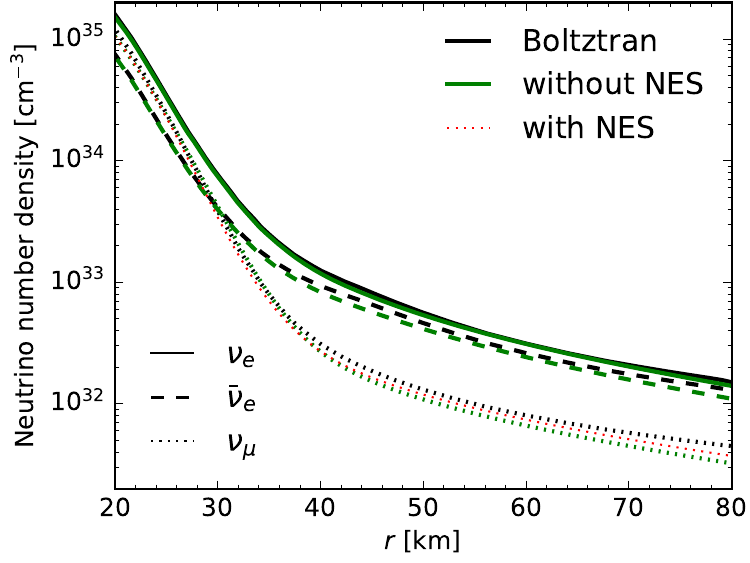}
\llap{\parbox[b]{4.9in}{\small (a)\\\rule{0ex}{2.2in}}}
\includegraphics[width=0.475\columnwidth]{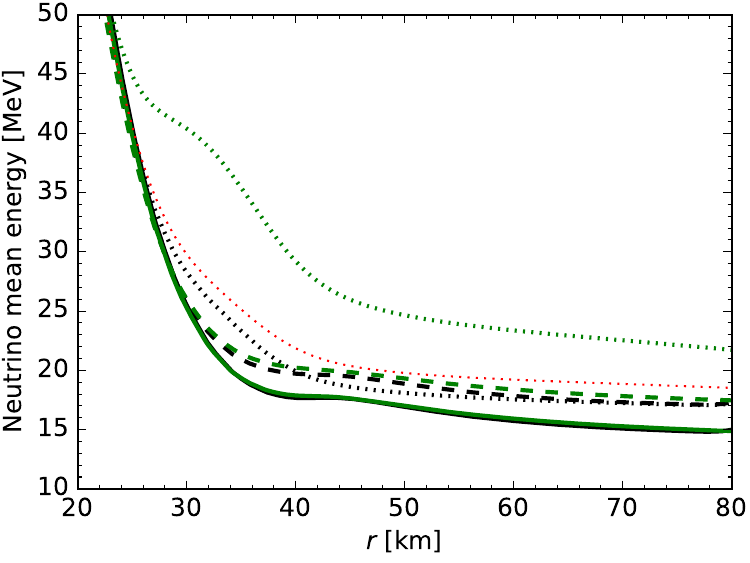}
\llap{\parbox[b]{4.9in}{\small (b)\\\rule{0ex}{2.2in}}}
\caption{\label{fig:boltztran_cosenu} Radial profiles of neutrino number density (a) and mean energy (b) based on \textsc{boltztran} (black), \textsc{cose$\nu$} without NES (green), and \textsc{cose$\nu$} with NES (red) in Model I.
}
\end{figure}

We compare the radial profiles of the neutrino number density and mean energy, obtained from \textsc{cose$\nu$} before any neutrino oscillations with the result based on the supernova simulations, denoted as \textsc{boltztran} in Fig.~\ref{fig:boltztran_cosenu}.
Both profiles of $\nu_e$ and $\bar\nu_e$ are highly consistent despite the different codes used.
For the heavy-lepton neutrinos, the most significant effect by including the NES is the reduction of the mean energy for $r\gtrsim 25$~km.
It decreases from $\simeq 40$ to $\simeq 29$~MeV at $r=30$~km, and from $\simeq 22$ to $\simeq 19$~MeV at $r=80$~km, due to the efficient thermalization through the NES~\cite{bruenn1985stellar,raffelt2002mu,keil2003monte}.
After including the NES, the deviation of $\langle E_{\nu_\mu} \rangle$ profiles between \textsc{cose$\nu$} and \textsc{boltztran} is strongly suppressed within $\simeq 2$~MeV.
It is expected that the deviation should become even smaller with a convergence between the two codes when a complete NES scheme is used.
Although the number density profile is also affected by the NES, the change is relatively small at the level of about $10\%$.
This is because the scattering process does not produce extra neutrinos unlike the processes of EA and PR.

\section{Practical method to select collective nonspurious modes}
\label{sec:selection_rule_spurious}
To preclude those spurious modes, we take the following practical selection rule based on the noncollectiveness of a given mode.
After the diagonalization of Eqs.~\eqref{eq:LEQ1} and \eqref{eq:LEQ2}, we sort the normalized modulus square of eigenmode $|Q_{i_E,i_{v_r}}|^2$ and $|\bar Q_{i_E,i_{v_r}}|^2$ for each eigenvalue $\Omega$.
Considering the number of grids that we used, there are 3000 components in each eigenmode.
To exclude those noncollective modes that mainly concentrate on one or a few angular grids, we then sum the first 30 largest components.
If the sum is more than 0.1, it implies that, on average, these components dominate the averaged contributions from the rest.
As a result, the corresponding eigenmode is likely of noncollective nature and is thus ruled out.


\end{document}